\documentclass[10pt]{amsart}
\usepackage{amsmath,amssymb,amsthm,amscd,mathrsfs}
\usepackage{graphicx,epstopdf}
\usepackage[all]{xy}
\usepackage{color}
\usepackage{dbnsymb}
\CompileMatrices
\numberwithin{equation}{section}
\newtheorem{prop}{Proposition}[section]

\newtheorem{conj}[prop]{Conjecture}
\numberwithin{equation}{section}

\newcommand{\be}{\begin{equation}}
\newcommand{\ee}{\end{equation}}

\newcommand{\IP}{\mathbb{P}}

\newcommand\IZ{\mathbb {Z}}

\newcommand{\IC}{\mathbb{C}}

\newcommand{\IR}{\mathbb{R}}

\newcommand{\ba}{\begin{array}}
\newcommand{\ea}{\end{array}}

\newcommand{\bz}{\overline{z}}
\newcommand{\bx}{\overline{x}}
\newcommand{\by}{\overline{y}}

\newcommand{\bw}{\overline{w}}

\newcommand{\oa}{{\overline a}}
\newcommand{\ob}{{\overline b}}
\newcommand{\oc}{{\overline c}}

\newcommand{\balpha}{\overline{\alpha}}
\newcommand{\bbeta}{\overline{\beta}}
\newcommand{\bgamma}{\overline{{\gamma}}}

\newcommand{\CS}{{\mathcal S}}

\newcommand{\bal}{\begin{aligned}}
\newcommand{\eal}{\end{aligned}}

\newcommand{\ch}{{\mathrm{ch}}}

\newcommand{\CO}{{\mathcal O}}

\newcommand{\CH}{{\mathcal H}}

\newcommand{\CM}{{\mathcal M}}
\newcommand{\CL}{{\mathcal L}}
\newcommand{\CJ}{{\mathcal J}}

\newcommand{\CC}{{\mathcal C}}

\newcommand{\CT}{{\mathcal T}}

\title{Large $N$ duality, lagrangian  cycles, and algebraic knots}
\author[D.-E. Diaconescu, V. Shende, C. Vafa]{D.-E. Diaconescu${}^1$, V. Shende${}^2$, 
C. Vafa${}^3$}
\address{${}^1$ NHETC, Rutgers University, Piscataway, NJ 08854-0849 USA}
\address{${}^2$ Department of Mathematics, MIT, Cambridge, MA 02139 USA}
\address{${}^3$ Jefferson Physical Laboratory, Harvard University, Cambridge MA 02138 USA}

\begin{document}

\begin{abstract}
We consider knot invariants in the context of large $N$ transitions
of topological strings.   In particular we consider aspects of Lagrangian cycles associated to knots in the conifold
geometry.   We show how these can be explicity constructed in the case of algebraic knots.  We use this
explicit construction to explain a recent conjecture relating study of stable
pairs on algebraic curves with HOMFLY polynomials.  Furthermore, for torus knots, using
the explicit construction of the Lagrangian cycle, we also give a direct A-model computation and recover
the HOMFLY polynomial for this case.
\end{abstract} 
\maketitle 

\tableofcontents

\section{Introduction}\label{sectionone}
The idea that knot invariants can be captured by physical theories 
dates back to the work
of Witten \cite{wit1} on the relation between Wilson loop observables 
of Chern-Simons quantum field theory
for $U(N)$ gauge theories and HOMFLY polynomials.  It was later noted by Witten \cite{CSstring} that the Chern-Simons theory,
in turn, describes the target space physics of A-model topological strings, in the presence of D-branes.  In particular,
if we have a stack of $N$ D-branes wrapping a three manifold $M^3\subset T^*M$ 
viewing $T^*M$ as a Calabi-Yau threefold,
the large $N$ perturbative Feynman diagrams, i.e. `t Hooft diagrams 
(known in math literature as `ribbon graphs') 
can be viewed as degenerate versions of holomorphic maps from 
Riemann surfaces with boundaries to $T^*M$ where the boundary of the 
Riemann surface
is restriced to lie on $M$.  It was later conjectured in \cite{gauge-gravity} that in the special case where $M=S^3$ at large $N$ the geometry
undergoes a transition, where $S^3$ shrinks and an $S^2$ is blown up with size equal to $Ng_s$ where $g_s$ is
the string (or Chern-Simons) coupling constant.  This is the small resolution of the conifold.  Furthermore
in this new geometry there are no more D-branes.  
In other words the partition function of the Chern-Simons theory is
equivalent to the {\it closed} topological A-model involving Riemann 
surfaces without boundaries, on the resolved conifold.
This large $N$ equivalence was checked by computating the 
partition function on both sides and observing their equality.

One can also extend this equivalence to the computation 
of the Wilson loop observables for knots, by adding to both
sides suitable `spectator D-branes' \cite{knot-top}.  
Namely for each knot $K\subset S^3$, consider the canonical 
Lagrangian $L_K\subset T^*S^3$
which intersects $S^3$ along the knot $K$.  
Note that $L_K$ has the topology $S^1\times R^2$.  
The insertion of the spectator brane leads to the insertion of Wilson loop
observables on the Chern-Simons side.  On the other 
side the original stack of $N$ D-branes has disappeared 
but the spectator D-branes $L_K$,
which have the imprint of the knot, survive.  We thus end up with the open topological A-model on the resolved conifold, in the presence
of D-branes wrapping $L_K$.   This equivalence was checked for the unknot in \cite{knot-top}.  Moreover this equivalence leads to integrality
predictions for the coefficients of HOMFLY polynomials (and their colored versions) \cite{knot-top,LMV}, which has been proven
to be true \cite{Liu1,Liu2,Liu3}.    The integrality structure follows from the fact that on the resolved side the computation
of the amplitudes captures the content of BPS particles represented by M2 branes ending on $L_K$, and one
is simply counting them.  For example,
for the unknot the partition function is captured by the fact that there are two M2 branes ending on $L_K$.

This leaves open the problem of directly computing 
the topological A-model for the resolved conifold in the 
presence of branes wrapping
$L_K$.  The difficulty in performing this task is two-fold:  first we have to 
identify the Lagrangian subspace $L_K$, and second
set up a computation for the A-model amplitudes.  The difficulty with the 
first task is that before transition $L_K$ intersects $S^3$, and thus as $S^3$
shrinks $L_K$ becomes singular and 
its continuation on the resolved side is 
delicate (though there has been progress
along these lines in \cite{Taubes,conormal-knots}.  However, it was further noted in \cite{framedknots} that to make this more well defined,
and also in order for the framing dependence to come out accurately we need to lift the original Lagrangian $L_K$, so that
it no longer touches the $S^3$, but is seperated from it by a cylinder which ends on the one hand on the knot in $S^3$
and on the other to the non-trivial circle in $L_K$.  In this way the $L_K$ is non-singular as $S^3$ shrinks and the process of
identifying it on the resolved side is more straight-forward.  We will clarify this construction later in this paper. 

 The second task is to compute the A-model amplitudes.  
When there are enough symmetries this in principle can be 
done in two ways:  Either by
direct computation using localization techniques, 
or by enumerating BPS particles ending on $L_K$.  

The enumeration of BPS particles correponding to M2 branes ending on $L_K$
is particularly simple for a special knots, including the unknot.  For example for the
 the unknot there are two BPS particles.   One corresponds
to a disc which lives on the  ${\bf C}^2$ fiber of one point in ${\bf P}^1$.  The disc intersects $L_K$ on a circle where it ends.
  The other particle is made of the
bound state of this disc with an M2 brane wrapping ${\bf P}^1$.   This follows
from the fact that the binding process is local, and we already have the disc ending on $L_K$ and the closed
M2 brane on ${\bf P}^1$ each
as BPS states, and they intersect transversally (relative codimension 4). Thus they form a unique bound state.
 From this we can recover the HOMFLY polynomial for the
unknot.    In some sense the unknot is `planar'  in that the BPS structures are captured
by objects living on the fiber or on the base independently, and simply glued together.

The question remains as for which knots are `planar' in this sense?  The natural answer ends up
being the class of knots known as algebraic knots, which can be defined by holomorphic function
of two variables.   One considers in complex dimension $2$ a holomorphic
function $f(x,y)$ with a singularity structure at the origin.  The intersection of 
$$f(x,y)=0$$
 with a large 3-sphere
$$|x|^2+|y|^2=r$$ 
for large $r$ gives a knot $K_f$ on $S^3$.  It turns out that for these knots the corresponding Lagrangian
$L_K$ can be constructed explicitly.  Moreover, just as in the case of the unknot they are `planar'.  In particular, the primitive
holomorphic curve ending on it lives on a fiber over a single point of 
$\IP^1$.  Moreover, identifying the fiber with
complex coordinates $(x,y)$ the basic holomorphic curve for the M2 brane is exactly $f(x,y)=0$ and it intersects
$L_K$ on the large three sphere along an $S^1$.  The new novely, as compared to the case of unknot, is that there could be more than one M2 brane bound state
on $f(x,y)=0$ curve.  Enumeration of such bound states turns out to map to a math problem recently studied in \cite{hilbert-links}.   However in the more general case, we have
more possibilities for forming bound states, not just the single disc as in the case of the unknot.  Furthermore,  just as in the case of the unknot, for each such disc we can form bound states
of this open $M2$ brane with a closed M2 brane wrapping ${\bf P}^1$. The number of bound states depends on the
intersection number of the ${\bf P}^1$ with the corresponding transverse bound 
state.  For each intersection point, we
get a bound state.  Considering all such BPS states wrapping the fiber and base we get the enumeration of BPS states in this geometry which leads
to the evaluation of the HOMFLY polynomial for such knots.  This turns out to explain the conjecture of \cite{hilbert-links}
relating the HOMFLY polynomial for algebraic knots with computations done for stable pairs associated
to the corresponding curve.

We explain in detail how these bound states can be evaluated for the case of the torus knots where
$$f(x,y)=x^r -y^s.$$
Furthermore, for these cases, using the explicit construction of the Lagrangian cycles $L_K$ we are able
to also directly compute the A-model amplitudes as well and rederive the HOMFLY polynomials for torus knots.

The organization of this paper is as follows. Section two is a review 
of large $N$ duality for the unknot, including the construction of 
toric lagrangian cycles on the resolved conifold. The main goal 
of this discussion is to motivate the general idea of lifting conormal 
bundle lagrangian cycles in the deformed conifold.  Section three presents an explicit construction of such a lift for algebraic knots, 
as well as the corresponding lagrangian cycles in the resolved conifold. Section four provides a physical explanation for 
the conjecture of Oblomkov and Shende \cite{hilbert-links}  
relating HOMFLY polynomials of 
of algebraic knots to certain generating functions associated 
to Hilbert schemes of plane curve singularities. 
In particular, the generating functions employed in \cite{hilbert-links} 
are identified with counting functions for open
M2-brane microstates with boundary on an M5-brane wrapping a lagrangian cycle. 
Section five is a reprise of section four in more mathematical dialect. 
Finally, section six consists of detailed computations 
of open topological {\bf A}-model amplitudes for lagrangian 
cycles corresponding to $(s,r)$-torus knots. The main result 
is a geometric derivation of the Chern-Simons $S$-matrix formula 
found in \cite{toruslinks, torusknotsmirror} by manipulations of 
open Gromov-Witten invariants. 

{\it Acknowledgments.} We thank Lev Borisov, Wu-yen Chuang, Zheng Hua, Amer Iqbal, Melissa Liu,
Sheldon Katz, Alexei Oblomkov, Andrei Okounkov, Rahul 
Pandharipande, Tony Pantev, Paul Seidel, Clifford Taubes,
Richard Thomas, Yan Soibelman and Chris Woodward for very helpful discussions. 
D.-E. D. and V.S. would especially like to thank Alexei Oblomkov 
for collaboration on related projects and many insightful 
discussions. The work of D.-E.D. was partially supported 
by NSF grant PHY-0854757-2009.  V.S. was supported by an EPSRC 
programme grant on a visit to Imperial college, and is 
currently supported by the Simons foundation. The work of C.V. 
is supported in part by NSF grant PHY-0244821. We would also 
like to thank the 2010 and 2011 Simons workshop in Mathematics 
and Physics and the Simons Center for Geometry and Physics 
for hospitality during the inception of this work. C.V. would also like to acknowledge the MIT physics department for hospitality.

\section{Large $N$ duality and lagrangian cycles for the unknot}\label{unknot}

The conifold transition is a topology changing process relating 
the smooth hypersurface $X_\mu$ 
\be\label{eq:conifoldA}
xz-yw = \mu 
\ee
in $\IC^4$ with $\mu\in \IC\setminus \{0\}$ to the small resolution 
$Y$ of the singular threefold $X_0$ obtained at $\mu=0$. In fact 
there exist two such isomorphic resolutions related by a toric flop.
For concreteness, let $Y$ be the resolution obtained 
by blowing-up the subspace $y=z=0$ in $\IC^4$. Then 
$Y$ is determined by the 
equations 
\be\label{eq:smallresA}
x\lambda = w \rho, \qquad z\rho = y\lambda 
\ee
in $\IC^4\times \IP^1$ and there is a natural map $\sigma:Y\to X_0$ 
which contracts  the rational curve $y=z=0$ 
on $Y$. It can be easily seen 
that $Y$ is isomorphic to the total space of the rank two bundle 
$\CO_{\IP^1}(-1)\oplus \CO_{\IP^1}(-1)$ and the 
curve $y=z=0$ is identified with its zero section, which is 
the only compact holomorphic curve on $Y$

The deformed conifold $X_\mu$, $\mu\neq 0$, equipped with 
the symplectic form 
\[
\omega_{X_\mu}= \omega_{\IC^4}\big|_{X_\mu}, \qquad 
\omega_{\IC^4}={i\over 2}\big(dx\wedge d\bx+ dy\wedge 
d\by+dz\wedge d\bz+dw\wedge d\bw\big)
\]
is symplectomorphic 
to the total space $X$ of the cotangent  bundle $T^*S^3$. 
For $\mu\in \IR_{>0}$, this can be seen explicitly \cite{sympsurgerysing,symp-conifold} observing that 
equation \eqref{eq:conifoldA} 
becomes 
\[
\sum_{i=1}^4 z_i^2 =\mu
\]
in the coordinates 
\[
x=z_1+iz_2, \qquad z=z_1-iz_2,\qquad y=-z_3-iz_4,\qquad w=z_3-iz_4.
\]
Writing $z_j = x_j+iy_j$ 
$j=1,\ldots,4$, with $(x_j,y_j)$ real coordinates on $\IC^4$, 
equation \eqref{eq:conifoldA} is further equivalent to 
\be\label{eq:conifoldF}
 {\vec x}\cdot {\vec y} =0, \qquad |{\vec x}|^2-|{\vec y}|^2=\mu.
\ee
Here $\cdot$ denotes the Euclidean inner product on $\IR^4$ and 
$|{}\ \, |$ the Euclidean norm. 

On the other hand, the  total space $X$ of the cotangent bundle $T^*S^3$ 
is identified with the subspace 
$\{({\vec u},{\vec v})\}\subset 
\IR^4 \times \IR^4$ satisfying 
\be\label{eq:conifoldD}
|{\vec u}|=1, \qquad {\vec u}\cdot{\vec v}=0.
\ee
The canonical symplectic form on $X=T^*S^3$ is then obtained by restriction from the ambient space, 
\be\label{eq:sympformC}
\omega_X =\big( \sum_{j=1}^4 dv_j\wedge du_j\big)\big|_{X}.
\ee

According to equation \eqref{eq:conifoldF}, ${\vec x}\neq 0$ on $X_\mu$ since $\mu\in \IR_{>0}$. Therefore there is a 
well defined map 
\be\label{eq:sympmorphismA}
\phi_\mu: X_\mu \to X
\qquad
\phi_\mu({\vec x}, {\vec y}) = \left({x_j\over |{\vec x}|}, -|{\vec x}|y_j\right).
\ee
It is straightforward to check that this map is a diffeomorphism,
its inverse being given by 
\be\label{eq:invsympmorphismA}
\phi_\mu^{-1}({\vec u}, {\vec v}) = (f_\mu({\vec v}) {\vec u}, \,
-f_\mu({\vec v})^{-1} {\vec v}),\qquad 
f_\mu({\vec v}) = \sqrt{{\mu +\sqrt{\mu^2+4|{\vec v}|^2}\over 2}}.
\ee
It is also straightforward to check that 
\[
\phi_\mu^*(\omega_X) = {i\over 2} \omega_{\IC^4}|_{X_\mu}
\]
Therefore 
$\phi_\mu$ is indeed a symplectomorphism. 

A similar construction yields a symplectomorphism 
$\phi_0: X_0\setminus \{0\}\to X\setminus \{{\vec v}=0\}$ 
between the complement of the singular point in $X_0$ and 
the complement of the zero section in $X=T^*S^3$. Observing that 
${\vec x}\neq 0$ on $X_0\setminus \{0\}$, $\phi_0$ is given 
exactly by the same formula as $\phi_\mu$, $\mu>0$. The same computation shows 
that $\phi_0$ is a symplectomorphism if $X_0\setminus \{0\}$ is equipped with the symplectic structure obtained by restriction from 
$\IC^4$. 

Note also that 
there is an antiholomorphic involution 
\be\label{eq:antiholinv}
(x,y,z,w) \mapsto ({\overline z}, -{\overline w}, {\overline x}, 
     -{\overline y})
     \ee
on $\IC^4$ which preserves $X_\mu$ with   
$\mu\in \IR_{\geq 0}$. Therefore there are 
induced 
antiholomorphic involutions $\tau_\mu :X_\mu\to X_\mu$, 
$\mu\in \IR_{\geq 0}$. 
For $\mu>0$, the 
 fixed locus $S_\mu$ of $\tau_\mu$ is isomorphic to the three-sphere  
$|x|^2+|z|^2=\mu$ in $\IC^2$.
 By construction, $S_\mu$ is a 
special lagrangian cycle on $X_\mu$ 
and the image $\phi_\mu(S_\mu)$ 
is  the zero section $S=\{{\vec v}=0\}$ of the cotangent 
bundle $T^*S^3$.

\subsection{Large $N$ duality for the unknot}\label{largeNunknot}
The primary example of large $N$ duality for topological strings 
\cite{gauge-gravity} is an equivalence between 
the large $N$ limit of 
the topological {\bf A}-model on $X_\mu$ with $N$ lagrangian branes on $S_\mu$ and 
 the topological {\bf A}-model on $Y$. The 
 partition function of the latter is given by 
\[
Z_Y(q,Q) = \prod_{n\geq 1}(1-Q(-q)^n)^n
\]
where $q$ and $Q$ are related to the string coupling constant 
$g_s$ and the symplectic area $t_{0}$ of $C_0$ by 
$q=e^{ig_s}$, $Q=e^{-t_0}$. 

According to \cite{CSstring} the topological ${\bf A}$-model on 
$X_\mu$ with $N$ lagrangian branes on the sphere $S_\mu$ is equivalent to  $U(N)$ Chern-Simons theory on $S_\mu$. The level 
$k$ of the Chern-Simons theory is related to the string coupling 
constant.
The partition function of the Chern-Simons theory on $S_\mu$
is naturally expanded in terms of the large $N$ variables  
\[
g_s={2\pi\over k+N} , \qquad \lambda = {2\pi N \over k+N}.
\]
Then large $N$ duality 
\cite{gauge-gravity} 
suggests that the theory on $S_\mu$ where there is brane, is equivalent
to the one after geometric transition where the branes have disappeared and
replaced by a blown up 2-sphere.  This duality thus
identifies the analytic part of the Chern-Simons large $N$ expansion 
with the closed topological string amplitude $Z_Y(q,Q)$ on the resolved side\footnote{The non-analytic part of the Chern-Simons function can be identified
by the same change of variables with the polynomial part of the 
$N=2$ prepotential of a IIA compactification on $Y$. }
\[
Z_{CS}(g_s,\lambda) =
Z_Y(q,Q) \big|_{\substack{q =e^{ig_s},\
Q =e^{i\lambda}.}}
\]

Large $N$ duality has been extended to Chern-Simons theory with 
Wilson loops in \cite{knot-top}. The main idea is that given a 
smooth knot $K\subset S^3$ the total space $L$ 
of the conormal 
bundle $N_{K}^*$ to $K$ in $S^3$ 
is a lagrangian cycle in $X=T^*S^3$.  
Since $\phi_\mu: X_\mu \to X$ is a symplectomorphism, the 
inverse image $L_\mu =\phi_\mu^{-1}(L)$ is a lagrangian 
cycle on $X_\mu$. 
According to \cite{knot-top},  a configuration of 
$N$ branes on $S_\mu$ and $M$ branes 
on $L_\mu$ has a complex bosonic open string  mode localized on their intersection which transforms in the bifundamental representation of $U(N)\times U(M)$.
Integrating out this 
mode yields a series of Wilson line corrections to Chern-Simons 
theory on $S$ of the form 
\be\label{eq:CSdefA}
\sum_{n\geq 1} {1\over n} \mathrm{Tr}(U^n)
\mathrm{Tr}(V^{-n}).
\ee 
Here  $U$ is the holonomy of the Chern-Simons gauge field $A$ 
on $K$ and $V$ is the holonomy on $K$ 
of an arbitrary background flat 
gauge field on $L_\mu$.  This integrating out can also
be explained in terms of the annulus contributions to the amplitudes where
one boundary of the annulus ends on $S_\mu$ and the other ends on $L_\mu$.
These are `holomorphic' annuli which have zero width, corresponding to the fact that
in the dual channel there are massless bi-fundamental particles of $U(N)\times U(M)$ going 
in the loop.

Therefore in the presence of the 
$M$ noncompact branes on $L_\mu$, the (analytical part of the) 
topological open string partition function 
becomes 
\be\label{eq:CSdefAB} 
Z_{CS}(g_s,\lambda) 
\bigg\langle \mathrm{exp}\bigg(\sum_{n\geq 1}{1\over n}\mathrm{Tr}(U^n) \mathrm{Tr}(V^{-n})\bigg) \bigg\rangle 
\ee
where $\langle{}\ \ \rangle$ denote the expectation values of Wislon 
line operators 
in $U(N)$ Chern-Simons theory on $S^3$.
The main question is then to construct a dual topological string model on the resolution $Y$, extending the results of \cite{gauge-gravity}. 

This problem was solved in \cite{knot-top} for the case when 
$K$ is the unknot.  
 For concreteness let  $K\subset S^3$ be determined by the equations 
\be\label{eq:unknotA}
y=w=0, \qquad |x|=|z|=\sqrt{\mu}
\ee
on $X_\mu$. Omitting the details, a straightforward computation 
shows that the inverse image $\phi_\mu^{-1}(N^*_K)$ 
is the lagrangian cycle $L_\mu$ in $X_\mu$
determined by the equations 
\be\label{eq:unknotB} 
y=\bw, \qquad |x|=|z|.
\ee
Assuming $K$ 
to be trivially framed, the large $N$ expansion of the partition function \eqref{eq:CSdefAB} is in this case 
\be\label{eq:CSdefB} 
Z_{CS}(g_s,\lambda) \mathrm{exp}\left[-i\sum_{n\geq 1} 
{e^{in\lambda/2} - e^{-in\lambda/2} 
\over 2n\mathrm{sin}(ng_s/2)}{\mathrm{Tr}} (V^{-n})\right].
\ee
In order to find a large $N$ duality interpretation, note that 
the above partition function is related by analytic continuation 
to 
\be\label{eq:CSdefC}
Z_{CS}(g_s,\lambda) \mathrm{exp}\left[-i\sum_{n\geq 1} 
{ {\mathrm{Tr}} (V^{n})+{\mathrm{Tr}} (V^{-n})
\over 2n\mathrm{sin}(ng_s/2)}e^{in\lambda/2}\right].
\ee
This expression is then identified with a series of open 
Gromov-Witten invariants of a lagrangian cycle $M$ in $Y$ 
determined by the equations 
\be\label{eq:unknotC}
|\lambda|=|\rho|, \qquad x\lambda = {\overline y}\rho. 
\ee
By construction, $M$ intersects the zero section $C_0$ along the 
circle $|\lambda|=|\rho|$, dividing it into two discs 
${\sf D}_\pm$ with common boundary. The terms weighted by 
$\mathrm{Tr}(V^{n})$, $\mathrm{Tr}(V^{-n})$ in the 
in the exponent of \eqref{eq:CSdefC} represent open Gromov-Witten invariants with positive, respectively negative winding 
numbers along the circle $|\lambda|=|\rho|$. This was confirmed 
by virtual localization computations in \cite{KL,LS-open}. In particular, the terms with positive winding numbers 
are obtained by summing over multicovers of ${\sf D}_+$ while  
those with negative winding numbers are obtained from multicovers 
of ${\sf D}_-$. 

The main difficulty in extending the above results to 
 more general knots in $S^3$ resides in the identification 
 of the
 lagrangian cycle $M$ in $Y$ associated to a given knot $K$.
 Ideally there should be a natural geometric relation 
 between the cycle $M\subset Y$ and the specialization 
 $L_0\subset X_0$ of 
 $L_\mu\subset X_\mu$ as $\mu\to 0$, exploiting the fact that the conifold 
 transition is a  basic example of symplectic surgery 
 \cite{symp-conifold,sympsurgerysing}. In symplectic 
 geometry the blow-up of $X_0$ as a symplectic manifold 
 depends on a positive real parameter $\epsilon\in \IR_{>0}$
 which measures the symplectic area of the exceptional curve 
 $C_0\subset Y$.
  More precisely let $\omega_0$ denote 
 the symplectic form $\omega_{\IC^4}\big|_{X_0\setminus \{0\}}$ 
 on the complement of the conifold singularity in $X_0$. 
 Then the blow-up of $X_0$ is a family of symplectic K\"ahler manifolds 
 $Y_\epsilon=(Y,\omega_{Y,\epsilon})$ 
 such that the resulting family of symplectic K\"ahler  forms 
 $\omega_{Y,\epsilon}\big|_{Y\setminus \{C_0\}}$ 
 on the complement of $C_0$ 
 degenerates to $\sigma^*\omega_0$ at $\epsilon=0$. 
  This yields a more symmetric picture of the conifold transition 
 transition, involving two families of symplectic manifolds 
 $X_\mu$, $Y_\epsilon$ satisfying a natural compatibility 
 condition at $\mu=0$, $\epsilon=0$ respectively. 
 This process is schematically summarized by the following diagram. 
 \be\label{eq:coniftransA} 
 \xymatrix{ 
& Y_0 \ar[d]_-{\sigma} & Y_\epsilon\ar@{~>}[l]  \\ 
 X_\mu \ar@{~>}[r] & X_0 & \\ }
 \ee
 where $\sigma:Y_0\to X_0$ is the blow-up map. 
 Note that all $Y_\epsilon$ with $\epsilon\geq 0$ are 
 identical as complex manifolds, but not as symplectic manifolds. 
 The symplectic structure is degenerate at $\epsilon=0$ since 
 $C_0$ has zero symplectic area with respect to $\omega_0$.

 In this framework, a natural formulation of large 
 $N$ duality for knots requires two families of 
 lagrangian cycles 
 $X_\mu\subset X_\mu$, $M_\epsilon \subset Y_\epsilon$ 
 such that the degenerations $L_0$, $M_0$ are related by 
 $M_0=\sigma^*L_0$, at least on the complement of the 
 exceptional curve $C_0$. 
  Schematically, such a process would be
  captured by an enhanced diagram 
   \be\label{eq:coniftransB}
  \xymatrix{ 
& Y_0 \ar[d]_-{\sigma} & Y_\epsilon\ar@{~>}[l]  & & & \\ 
 X_\mu \ar@{~>}[r] & X_0 & M_0 \ar[dl]_-{\sigma}
 \ar@{^{(}->}[lu] & M_\epsilon \ar@{~>}[l] \ar@{^{(}->}[lu]\\ 
 L_\mu \ar@{~>}[r] \ar@{^{(}->}[u]& L_0 
 \ar@{^{(}->}[u]& & & \\} 
    \ee
 
\begin{figure}
\setlength{\unitlength}{1mm}
\hspace{-130pt}
\begin{picture}(80,45)
\qbezier(30,0)(20,20)(30,40)
\qbezier(0,0)(10,20)(0,40)
{\color{cyan}\put(8,20){\oval(6,7)[l]}
\put(22,20){\oval(6,7)[r]}
\put(8,16.4){\line(1,0){14}}
\multiput(9.5,23.5)(2,0){6}{.}}
{\color{red}
\thicklines\qbezier(7,5)(11,20)(7,35)
\thicklines\qbezier(21,5)(17,20)(21,35)}
{\color{green} \thicklines\put(13,20){\oval(9.5,4)}}
\put(-2,19){$S^3$}
\put(13,35){$L$}
\put(12,20){${}_{K}$}
\thinlines\put(90,20){\line(-1,2){10}}
\put(90,20){\line(-1,-2){10}}
\put(110,20){\line(1,2){10}}
\put(110,20){\line(1,-2){10}}
{\color{cyan}\put(94,20){\oval(8,8)[l]}
\put(106,20){\oval(8,8)[r]}
\put(94,16){\line(1,0){12}}
\multiput(94,24)(2,0){6}{.}}
{\color{blue}\thicklines\qbezier(96,3)(98,20)(91,34)
\qbezier(104,10)(103,30)(109,41)}
{\color{green}\qbezier(95,16)(97,24)(103,24)
\qbezier(95,16)(102,18)(103,24)}
\put(97,35){$M$}\put(112,19){${\mathbb P}^1$}
\put(87,25){${}_{{\sf D}_-}$}\put(106,14){${}_{{\sf D}_+}$}
\end{picture}
\[
\xymatrix{ {} \ar@{~>}[drr]^-{\mu\to 0} & & {} \\ 
 & & \\}\qquad \qquad 
\qquad \qquad \qquad \qquad 
\xymatrix{ {} & & \ar[dll]_-{\sigma} {}  \\ & & \\}
\]
\bigskip\bigskip\bigskip 

\hspace{-290pt}
\begin{picture}(30,45)
\put(47,54){\line(3,-4){36}}\put(83,54){\line(-3,-4){36}}
\put(64.2,29.4){{\color{cyan}$\bullet$}}\put(67,29.4){conifold}
{\color{magenta}\thicklines\qbezier(58,48)(72,30)(58,12)
\qbezier(72,48)(58,30)(72,12)}
\put(56,50){$L_0=\sigma(M)$}
\end{picture}
\caption{Conifold transition for unlifted lagrangian cycles.}
\label{unliftedtransition}
\end{figure}
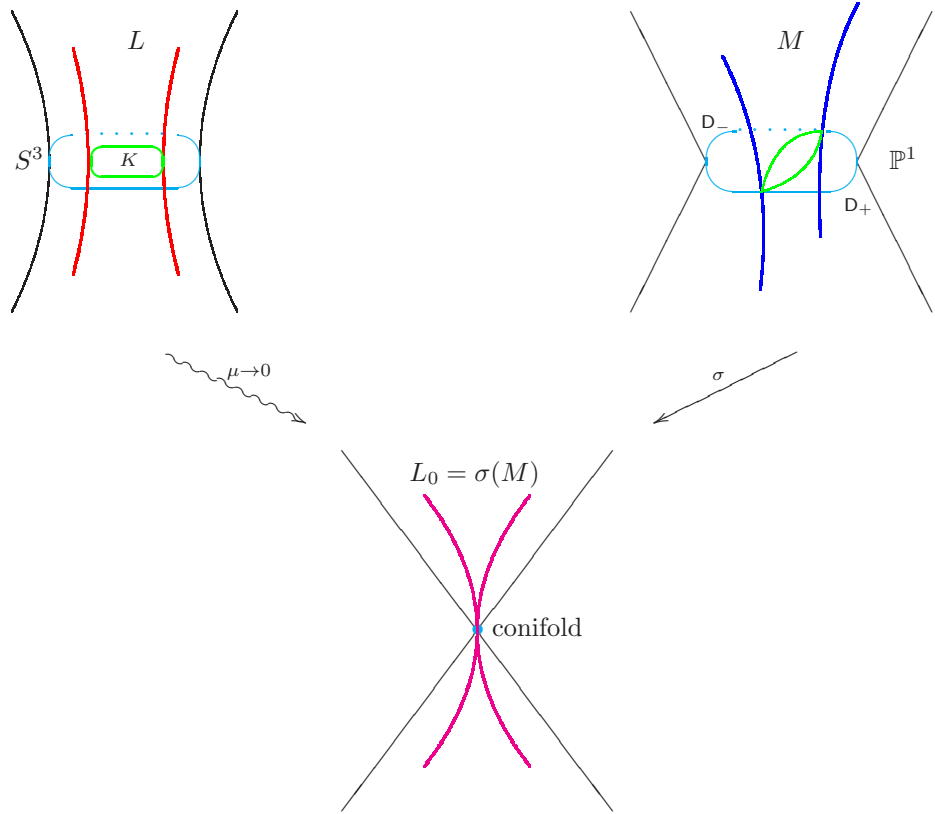

 In the case of the unknot reviewed above,  the specialization  
 of the cycle $L_\mu$ in equation \eqref{eq:unknotB}
at $\mu=0$  is the singular lagrangian cone $L_0\subset X_0$ 
determined by 
\be\label{eq:unknotD}
y=\bw, \qquad |x|=|y|.
\ee
At the same time, the cycle $M$ constructed in equation \eqref{eq:unknotC} is lagrangian 
 with respect to any symplectic K\"ahler form $\omega_{Y,\epsilon}$ 
 because it is the fixed point set of an antiholomorphic involution. 
 The image of $M$ via the blow-up map $\sigma$ is
 precisely the singular lagrangian cycle $L_0$ 
 determined by the same equations \eqref{eq:unknotD}. 
 Therefore the compatibility condition at $\mu=0$, $\epsilon=0$ is 
 satisfied.  
  For illustration, the resulting geometric picture is represented 
  in figure (\ref{unliftedtransition}).

Since the knot $K$ is contracted in this process 
 it is not clear how such a construction can be extended to more 
 general knots especially such that the resulting open string Gromov-Witten theory on $Y$ is tractable. 
 A related problem is that the analytic continuation required 
 by a proper enumerative  interpretation of 
 the partition function does not have a direct geometric interpretation.

 Both these problems
  lead to the idea \cite{framedknots,all-loop} that a better formulation 
 of large $N$ duality would be obtained using lagrangian cycles 
 supported in the complement of the zero section $C_0\subset Y$, respectively  $S^3\subset X_\mu$. 
 Said differently, this means that the lagrangian cycle $L=N_{K}^*$ must be lifted to a lagrangian cycle disjoint from 
 the zero section prior to the transition. Accordingly, the corresponding 
 lagrangian cycle in $Y$ will be lifted to a cycle 
 disjoint from the zero section 
 $C_0$.  Moreover, once properly lifted, these cycles 
 should form families 
 naturally related by symplectic surgery as explained above.   That there is such a lift
can be argued as follows:  Assume that with the proper choice of metric, the lagrangian
cycle $L$ is actually {\it special lagrangian} \cite{SYZ}.  
In this case it is known that the dimension
of moduli of $L$ is equal to the dimension of $H_1$.  This is rigorously the case
for compact lagrangians, and we assume it to hold for non-compact ones as well where
we have imposed sutiable finiteness conditions on the norm of deformations of the lagrangian.
Since the topology of $L$ is $R^2\times S^1$, there is exactly 1 deformation.  This corresponds
to moving the special lagrangian in the 1-form dual the $S^1$, by identifying the infinitesimal normal deformation
to the lagrangian with its cotangent space.  It is this deformation that lifts the $L$ off of $S_\mu$.  Moreover
it suggests that there is a unique such canonical lift for special lagrangian cycles.  Even though
we will mainly deal with just lagrangian ones, this suggests that the choice of the special
lagrangian ones make the constructions more `canonical'.
  
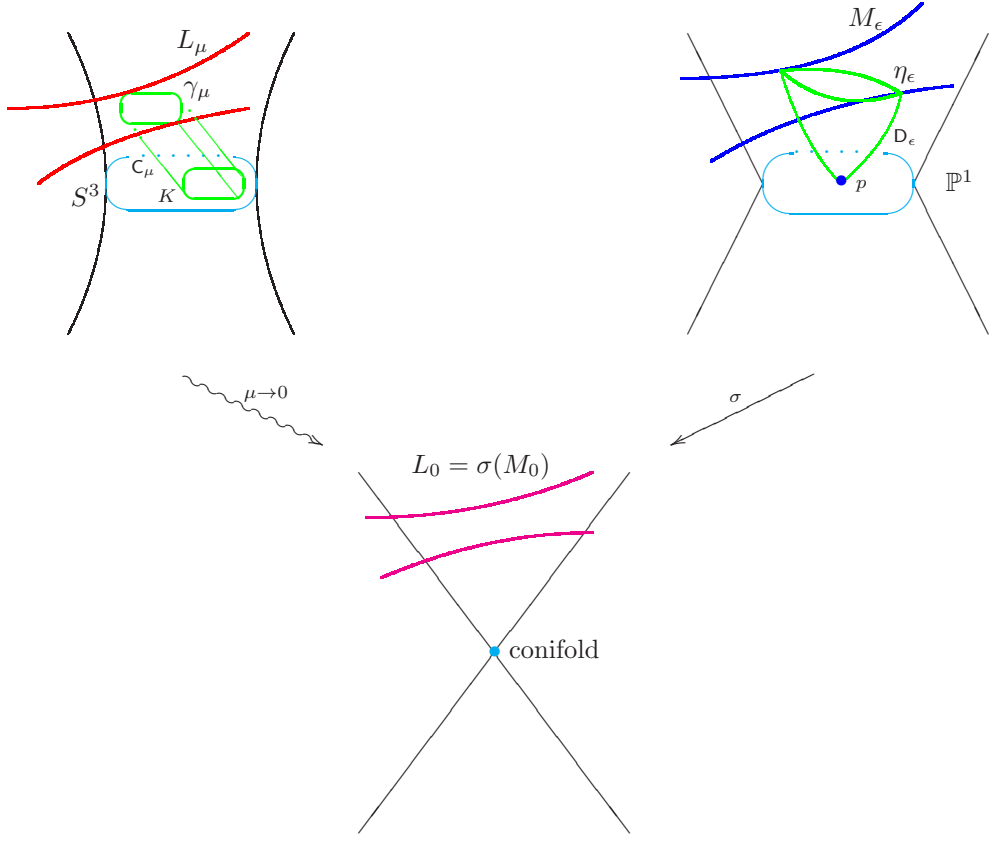
\begin{figure}
\setlength{\unitlength}{1mm}
\hspace{-100pt}
\begin{picture}(80,45)
\qbezier(30,0)(20,20)(30,40)
\qbezier(0,0)(10,20)(0,40)
{\color{cyan}\put(8,20){\oval(6,7)[l]}
\put(22,20){\oval(6,7)[r]}
\put(8,16.4){\line(1,0){14}}
\multiput(9.5,23.5)(2,0){6}{.}}
\thicklines\put(11,30){\color{green}{\oval(8,4)}}
{\color{red}\qbezier(-8,30)(10,30)(24,40)
\qbezier(-4,20)(5,27)(24,30)}
\thinlines{\color{green}\put(13.5,28){\line(4,-5){8}}
\put(8.5,26){\line(4,-5){6}}
\put(15.9,28.5){\line(4,-5){6}}
\multiput(8.0,26)(-0.8,1){2}{.}
\multiput(15.3,28.5)(-0.8,1){2}{.}}
\thicklines{\color{green}\put(17,20){\oval(8,4)}}
\put(-2,17){$S^3$}
\put(12,38){$L_\mu$}
\put(6,22){${}_{{\sf C}_\mu}$}
\put(13,32){${}\gamma_\mu$}
\put(9.5,18){${}{{}_K}$}
\thinlines\put(90,20){\line(-1,2){10}}
\put(90,20){\line(-1,-2){10}}
\put(110,20){\line(1,2){10}}
\put(110,20){\line(1,-2){10}}
{\color{cyan}\put(94,20){\oval(8,8)[l]}
\put(106,20){\oval(8,8)[r]}
\put(94,16){\line(1,0){12}}
\multiput(94,24)(2,0){6}{.}}
\thicklines
{\color{blue}\qbezier(78,34)(100,34)(110,44)
\qbezier(82,23)(95,31)(114,33)}
{\color{green}\qbezier(90,35)(100,36)(106,32)
\qbezier(90,35)(97,29)(106,32)
\thinlines\qbezier(98,20)(93,25)(90,35)
\qbezier(98,20)(105,26)(106,32)}
\put(99,41){$M_\epsilon$}\put(112,19){${\mathbb P}^1$}
\put(105,26){${}_{{\sf D}_{\epsilon}}$}
\put(105,34){${\eta_{\epsilon}}$}
\put(97.2,19.5){\color{blue}$\bullet$} \put(100,20){${}_p$}
\end{picture}
\[
\xymatrix{ {} \ar@{~>}[drr]^-{\mu\to 0} & & {}  \\ & & \\}\qquad \qquad 
\qquad \qquad \qquad \qquad 
\xymatrix{ {} & & \ar[dll]_-{\sigma} {} \\ & & \\  }
\]
\bigskip\bigskip\bigskip

\hspace{-290pt}
\begin{picture}(30,45)
\put(47,54){\line(3,-4){36}}\put(83,54){\line(-3,-4){36}}
\put(64.2,29.4){\color{cyan}$\bullet$}\put(67,29.4){conifold}
\thicklines
{\color{magenta}\qbezier(48,48)(64,48)(78,54)
\qbezier(50,40)(64,46)(78,46)}
\put(54,54){$L_0=\sigma(M_0)$}
\end{picture}
\caption{Conifold transition for lifted lagrangian cycles.}
\label{liftedtransition}
\end{figure}

Accepting the idea of lifting lagrangian cycles, a legitimate question 
is how can one then obtain the Wilson loop corrections 
 \eqref{eq:CSdefA}, given that $L_\mu$ and $S_\mu$ do not intersect. 
This is also natural.  Lifting off of the lagrangian brane off of $S_\mu$ is simply giving
the bi-fundamental particles a mass given by the amount of lifting.  In other words,
the annuli which whose dual channel corresponded to bi-fundamental strings, now have
a finite width depending on the amount of lift.  These corrections can now be interpreted as `honest'
intantons, i.e. holomorphic cylinders which on the one hand end on $S_\mu$ and on the other hand on $L_\mu$.
Such corrections were predicted in \cite{CSstring} assuming that 
there are finitely many rigid holomorphic Riemann surfaces 
${\sf C}_{\mu}^{(\alpha)}$ in 
$X_\mu$ with boundary components on $S_\mu,L_\mu$. 
 Each such surface gives rise to a series of Wilson loop 
 corrections by summing over multicovers. In particular, 
 a rigid holomorphic cylinder ${\sf C}_\mu$ in $X_\mu$ with 
 boundary components in $S_\mu,L_\mu$ yields a series of instanton corrections 
 \be\label{eq:opinstA} 
\sum_{n\geq 1} {e^{-t_{\sf C}}\over n} 
\mathrm{Tr}(U^n) \mathrm{Tr}(V^{n}) 
\ee
where $t_{\sf C}$ is the symplectic area of the cylinder 
${\sf C}_\mu$, and can be interpreted as the mass of the bi-fundamental state (where we have
changed the variables by $V\mapsto V^{-1}$). 
Note that the factor $e^{-t_{\sf C}}$ can be absorbed 
by a redefinition of the holonomy variable $V$, hence it will be 
omitted from now on. 
Figure (\ref{liftedtransition})
is a schematic representation of the surgery process 
in terms of lifted lagrangian cycles. 

To summarize, {\it for a given knot}  $K\in S^3$, {\it  large} $N$ 
{\it duality requires a family of lagrangian cycles} $L_\mu\subset X_\mu$,
{\it disjoint from} $S_\mu$, {\it such that there is a 
unique rigid holomorphic holomorphic cylinder} ${\sf C}_\mu$ 
{\it in} $X_\mu $ {\it with 
 boundary components in} $S_\mu,L_\mu$. 
 Moreover, the boundary component in $S_\mu$ must be isotopic to the given knot $K$. 
Note that the rigidity assumption on ${\sf C}_\mu$ is not needed 
if there exists a torus action on $X_\mu$ preserving $L_\mu$. In this case
it suffices to require ${\sf C}_\mu$ to be the unique torus invariant 
holomorphic cylinder satisfying these boundary conditions. 
Then the series \eqref{eq:opinstA} follows by a virtual localization 
computation analogous to \cite{KL}, as shown for example 
in  \cite{geom-delpezzo}. A concrete construction 
of such families of lagrangian cycles for algebraic 
knots is presented in section (\ref{lagtrans}). By analogy with the 
unknot, the cycles $L_\mu$ will be obtained by taking inverse 
images $\phi^{-1}_\mu(L)$ of a fixed lift 
$L\subset X$ of $N_K^*$ 
in $T^*S^3$. Uniqueness and rigidity of the associated 
holomorphic cylinders will be proven 
only for torus knots in section (\ref{torusknots}) and conjectured 
to hold for all algebraic knots. 

The family of lagrangian cycles $M_\epsilon\subset Y$ 
related to $L_\mu$ by geometric 
transition is expected to have a similar property. Namely there should exist a unique holomorphic disc ${\sf D}_\epsilon$ in $Y$ with boundary $\eta_\epsilon\subset M_\epsilon$. Note that ${\sf D}_\epsilon$ may have isolated singularities 
away from the boundary. Again, if there is a torus action on $Y$
preserving $M_\epsilon$, it
 suffices for ${\sf D}_\epsilon$ to be the unique 
torus invariant 
disc with boundary on $M_\epsilon$. Then large $N$ duality predicts 
an identification between the Chern-Simons partition function 
on $S^3$, including the instanton corrections \eqref{eq:opinstA}, and the partition function of Gromov-Witten theory on $Y$
with lagrangian boundary conditions on $M_\epsilon$. 

As a first example, the above program will 
 be carried out in detail in the next subsection for an unknot 
of the form \eqref{eq:unknotA}.  In this case 
the cycles $M_\epsilon, L_\mu$ will be
 explicitly constructed employing 
toric methods \cite{disc-mirror}
It will be shown that both cycles are preserved by 
a circle action determined by an action on $\IC^4$ of the form 
\be\label{eq:gentoractB}
(x,y,z,w) \mapsto \big(e^{-in_1\varphi}x, e^{-in_2\varphi}y, 
e^{in_1\varphi}z, e^{in_2\varphi}w\big).
\ee
Note that the action on $Y$ is uniquely determined by 
the condition that the blow-up equations \eqref{eq:smallresA} 
be left invariant. In particular it yields the circle action 
\[
[\lambda,\rho] \mapsto [ \lambda, e^{i(n_1+n_2)\varphi} 
\rho]
\]
on $\IP^1$. 
Assuming the unknot 
trivially framed, the Chern-Simons expectation value of the 
instanton corrections \eqref{eq:opinstA}  is
\be\label{eq:defCSE}
\mathrm{exp}\left[i\sum_{n\geq 1} 
{(1 - e^{in\lambda})\over 2n\mathrm{sin}(ng_s/2)} 
{\mathrm{Tr}} (V^{-n})\right]
\ee

The open Gromov-Witten invariants with boundary 
condition on $M_\epsilon$ can be computed in close analogy with 
\cite{KL}. As explained in \cite{KL}, 
the result depends on the choice of a torus action, reflecting the 
fact that the moduli space of stable maps with lagrangian boundary conditions is non-compact. This dependence is related by large 
$N$ duality  to the 
framing dependence of knot invariants in Chern-Simons theory 
\cite{framedknots}. 
Choosing the torus action $(n_1,n_2) =(1,0)$, which corresponds 
to the trivial framing, the result takes the simple form 
\be\label{eq:openGWunknot} 
\mathrm{exp}\left[i\sum_{n\geq 1} 
{ (1- Q^n)
\over 2n\mathrm{sin}(ng_s/2)} \mathrm{Tr}(V^{-n})\right]
\ee
This is in agreement with equation \eqref{eq:defCSE} 
via the change of variable $Q=e^{i\lambda}$.
Note that the term involving a single $\mathrm{Tr}(V^{-1})$ in the exponent has the form (up to an
overall factor of $q^{1/2}$)
$$(1-Q)/(1-q)={1\over (1-q)}-{Q\over (1-q)}$$
where $q=exp(ig_s)$.  Each of these two terms was interpreted in \cite{knot-top} 
as the contribution of an M2 brane ending on the Lagrangian brane corresponding
to the unknot.    The two term differ by a factor of $Q$ indicating that one of the two M2 branes
is in addition wrapped around the ${\bf P}^1$.  The minus sign in front of the second term can be interpreted as the fermion number
associated with the M2 wrapped around ${\bf P}^1$.   Moreover the term 
$${1\over (1-q)}=1+q+q^2+...$$
signifies the fact that an M2 brane particle has one mode for each  positive integer $n>0$.
Each such $n$ corresponds to the spin of the M2 brane on a plane, in the presence of a magnetic flux.
Moreover in the type IIA perspective since the rotation around the 11-th circle is identified with
the rotation on the 2-plane, $n$ can also be identified with the D0 brane charge \cite{Mtopstringdual,Mwallcrossing,CNV}.
The fact that there are two BPS states for the unknot will be explained in the next subsection.

\subsection{Toric lagrangian cycles in the resolved conifold} 
The construction of the lifted lagrangian cycles $M_\epsilon, L_\mu$ 
will be carried out in detail below for the unknot using toric geometry 
as in \cite{disc-mirror}. 
The gauged linear sigma model which flows to $Y$ 
 is a two dimensional $U(1)$ gauge theory containing 
  four chiral superfields 
$Z_1,\ldots, Z_4$ with charges  
\[
\begin{array}{ccccc}
 &Z_1 & Z_2 & Z_3 & Z_4 \\
U(1) & 1 & 1 & -1 & -1. \\
\end{array}
\]
and trivial superpotential. The D-term equation is 
\be\label{eq:moment}
|Z_1|^2+|Z_2|^2-|Z_3|^2-|Z_4|^2=\epsilon,
\ee
where
$\epsilon \in \IR_{>0}$ is an FI parameter. 
The symplectic quotient construction yields a family of 
symplectic 
K\"ahler manifolds $Y_\epsilon=(Y, \omega_{Y,\epsilon})$. 
The exceptional curve $C_0$ is given by $Z_3=Z_4=0$, 
and has symplectic area proportional to $\epsilon$. 
The contraction map $\sigma:Y\to X_0$ is 
determined by the
$U(1)$-invariant monomials 
\[
 x= Z_3Z_1, \qquad y= Z_4Z_1, \qquad z= Z_4Z_2, \qquad 
 w=Z_3Z_2
 \]
which satisfy the relation $xz=yw$.  

Lagrangian cycles in $Y$ 
are constructed by a linear gauged linear sigma model with boundary, which is expected to flow to a boundary 
conformal field theory in the infrared. In particular consider the cycles 
$M_\epsilon$ be defined by  
the boundary D-term equations 
\be\label{eq:slagA}
|Z_2|^2-|Z_4|=0,\qquad  |Z_3|^2-|Z_4|^2=c, 
\ee
where $c\in \IR_{>0}$ is a boundary FI parameter,
and the phase condition 
\be\label{eq:slagB}
Z_1\cdots Z_4 = |Z_1\cdots Z_4|.
\ee
On the open subset $Z_i\neq 0$, 
 where all angular coordinates $\theta_i$, $i=1,\ldots, 4$,
  are well defined this condition is equivalent 
to $\theta_1+\cdots + \theta_4=0$. 
A detailed construction of the boundary gauged linear sigma 
models has been carried out in \cite{boundary-glsm-I,
boundary-glsm-II,boundary-glsm-III,boundary-glsm-IV}.
The boundary FI parameter $c>0$ will be kept fixed throughout 
this discussion. 

In order to understand the geometry of $M_\epsilon$, note that 
equations \eqref{eq:moment}, \eqref{eq:slagA} imply 
\[
 |Z_1|^2-|Z_3|^2=\epsilon.
 \]
 Since $\xi,c>0$, it follows that  
 $Z_1, Z_3$ cannot vanish on $M_\epsilon$. Then the phase 
 $\theta_1$ can be set to $0$ by $U(1)$ gauge transformations, 
 and the phase relation 
 \eqref{eq:slagB} reduces to 
 \[
 Z_2\cdots Z_4 = |Z_2\cdots Z_4|.
 \]
   As emphasized in the previous subsection, 
 it is important to note that $M_\epsilon$ 
 is preserved by any  circle action $S^1
 \times Y \to Y$ of the form 
 \be\label{eq:gentoract}
 (Z_1,\ldots, Z_4) \mapsto (Z_1, e^{i(n_1+n_2)\varphi} Z_2, 
 e^{-in_1\varphi}Z_3, 
 e^{-in_2\varphi} Z_4 )
 \ee
 with $n_1,n_2\in \IZ$. It is straightforward to check that this
 is in agreement with the action \eqref{eq:gentoractB} 
  on the invariant monomials.
  It is also important to note that $M_\epsilon$ intersects the plane 
$Z_2=0$ along a circle $S^1_{c}$ given by 
\[
 |Z_2|=|Z_4|=0, \qquad |Z_3|^2 =c, \qquad |Z_1|^2 = \epsilon+c.
 \]
Since $Z_2, Z_4$ are set to $0$, the intersection 
is indeed 
a circle parameterized by the angular variable $\theta_3$. 
Moreover, there is a holomorphic disc ${\sf D}_\epsilon$ 
with boundary on $M_\epsilon$ defined by 
 \[
  |Z_2|=|Z_4|=0,\qquad |Z_3|^2 \leq c, \qquad |Z_1|^2 = \epsilon+c.
  \]
 Reasoning by analogy with \cite{geom-open,geom-delpezzo} 
 it can be checked that ${\sf D}_\epsilon$ is the only Riemann surface in 
 $Y$   with boundary on $M_\epsilon$ preserved by a torus action of the form 
 \eqref{eq:gentoract} with $n_1\neq 0$.
 
Next note that setting $\epsilon=0$ in the above construction 
yields a lagrangian cycle $M_0$ on the singular conifold $X_0$. 
In terms of the invariant monomials $(x,y,z,w)$,
the defining equations of $M_0$ in $X_0$ are 
\be\label{eq:slagC}
y-{\overline w}=0, \qquad |x|-|z|=c.
\ee
Since $c>0$, $x$ cannot vanish, hence $M_0$ 
 is contained in the complement of the singular 
point $x=y=z=w=0$. Moreover, it is easy to check 
that $M_0$ is lagrangian with respect to 
the symplectic form $\omega_0$ obtained by restricting the 
standard symplectic form $\omega_{\IC^4}$ to the complement of 
the singular point in $X_0$. Note also that equation \eqref{eq:slagC} 
yields equation \eqref{eq:unknotC} at $c=0$, 
confirming that the present construction is a lifted version of 
the previous one.

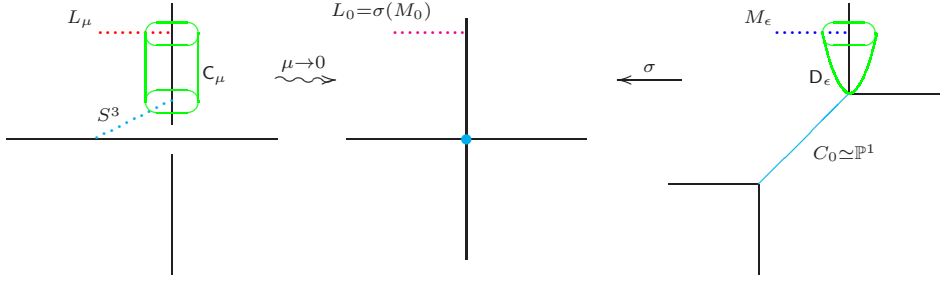
\begin{figure}[h]
\setlength{\unitlength}{1mm}
\hspace{-140pt}
\begin{picture}(80,40)
\put(102,12){\line(0,-1){12}}
\put(102,12){\line(-1,0){12}}
\put(102,12){\color{cyan}\line(1,1){12}}
\put(114,24){\line(1,0){12}}
\put(114,24){\line(0,1){12}}
\multiput(104,32)(1,0){10}{\color{blue}.}
\put(100,34){${}_{M_\epsilon}$}
\put(2,18){\line(1,0){36}}
\put(24,16){\line(0,-1){16}}
\put(24,20){\line(0,1){16}}
\multiput(13.5,18)(1,0.5){11}{\color{cyan}.}
\multiput(14,32)(1,0){10}{\color{red}.}
\put(10,34){${}_{L_\mu}$}
\put(14,21){${}_{S^3}$}
\put(24,23){\color{green}\oval(7,3)}
\put(24,32){\color{green}\oval(7,3)}
\put(20.5,23){\color{green}\line(0,1){9}}
\put(27.5,23){\color{green}\line(0,1){9}}
\put(28,26.5){${}_{{\sf C}_\mu}$}
\put(114,32){\color{green}\oval(7,3)}
{\color{green}\qbezier(114,24)(112,25)(110.5,32)}
{\color{green}\qbezier(113,24)(115,25)(116.5,32)}
\put(107.5,26){${}_{{\sf D}_\epsilon}$}
\put(108,16){${}_{C_0\simeq \IP^1}$}
\put(46,18){\line(1,0){32}}
\put(62,34){\line(0,-1){32}}
\put(61.1,17.1){\color{cyan}$\bullet$}
\multiput(52,32)(1,0){10}{\color{magenta}.}
\put(44,35){${}_{L_0=\sigma(M_0)}$}
\put(34,25){$\xymatrix{{}\ar@{~>}[r]^-{\mu\to 0}& \\}$}
\put(80,25){$\xymatrix{&\ar[l]_-{\sigma} \\}$}
\end{picture}
\caption{Conifold transition for toric lagrangian cycles.}
\label{torictrans}
\end{figure}

The family of 
 lagrangian cycles $L_\mu\subset X_\mu$, 
 $\mu>0$ is defined by the same equations, 
 \eqref{eq:slagC}, now interpreted as equations 
 on the deformation $X_\mu$. 
It is straightforward to check that $L_\mu$ is 
lagrangian with respect to the symplectic form 
$\omega_{\IC^4}|_{X_\mu}$ and it is preserved by 
the torus action 
\eqref{eq:gentoractB}. The resulting transition between lagrangian 
cycles is schematically represented in figure (\ref{torictrans}). 

Again, comparison with equation 
\eqref{eq:unknotB} shows that the cycle $L_\mu$
is a lift of the (inverse image of the) 
conormal bundle $\phi_\mu^{-1}(N_K^*)$. 
Moreover there is a  unique 
torus invariant holomorphic cylinder ${\sf C}_\mu$ 
in $X_\mu$ 
with one boundary component in $L_\mu$ and the second contained 
in the vanishing cycle $S_\mu$. This is obtained intersecting the two 
lagrangian cycles, $L_\mu, S_\mu$ with the holomorphic curve 
$C_\mu\subset X_\mu$ 
given by 
\[
y=0, \qquad xz=\mu.
\]
One then finds two circles determined by the equations 
\[
\bal
C_\mu \cap L_\mu:& \qquad y=w=0, \qquad xz=\mu, \qquad |x|={c+\sqrt{c^2+4\mu^2}\over 2},\\ 
C_\mu\cap S_\mu:& \qquad |x|=|z|=\sqrt{\mu}, \qquad y=w=0.
\eal 
\]
The cylinder ${\sf C}_\mu$ is given by
\[
y=w=0, \qquad xz=\mu, \qquad \sqrt{\mu} \leq |x| 
\leq {c+ \sqrt{c^2+4\mu^2}\over 2}.
\]
A different construction of lagrangian cycles for more general knots will be presented 
in the next section.

\section{Algebraic knots, lagrangian cycles and 
conifold transitions}\label{lagtrans}

The goal of this section is to present a construction of lagrangian 
cycles in $T^*S^3$ lifting the conormal bundle $N^*_{K}$ 
of any knot $K\subset S^3$.  
Note that such a construction was
previously carried out in \cite{conormal-knots}, where it was also 
proven that the resulting lagrangian cycles are related to totally 
real cycles on the resolved conifold via the conifold transition. 
Moerover, there is a well defined Gromov-Witten theory with 
boundary conditions on the totally real cycles, constructed in 
\cite{conormal-knots} via symplectic methods. As discussed in more 
detail below, the construction employed in this paper is a generalization of \cite{conormal-knots} motivated by the large N 
duality considerations explained in section (\ref{largeNunknot}). 
In particular, in this approach the lagrangian cycles associated to 
algebraic knots are naturally equipped with 
holomorphic cylinders  with one boundary 
component in the lifted conormal bundle, the second boundary 
component being a knot in $S^3$ in the 
isotopy class of $K$.
It will also be shown that these cycles are related by the conifold transition to lagrangian cycles 
in the small resolution of the 
conifold. For $K$ algebraic, the construction also yields a  singular 
holomorphic discs ${\sf D}_\epsilon$ in the resolved conifold with boundary 
on the corresponding lagrangian cycles. 
Furthermore, if $K$ is a torus knot, the resulting Gromov-Witten theory on the 
resolution with lagrangian boundary conditions turns out to be 
computable using a virtual localization approach similar 
to \cite{KL} and \cite{LS-open}.

The notation and geometric set-up is as in the previous section. 
The total space of the cotangent bundle $T^*S^3$ is denoted 
by $X$ and will be identified with the subspace of 
$\IR^4\times \IR^4$ determined by equations \eqref{eq:conifoldD}. In this presentation, 
the canonical symplectic form $\omega_X$ is given by 
equation \eqref{eq:sympformC}.
The natural projection map $X\to S^3$ is denoted by $\pi$ and the zero section is denoted by $S$.

\subsection{Knots and lagrangian cycles in 
$T^*S^3$}\label{lagsect} 
Consider a smooth closed curve $\gamma:S^1 \to X$ 
such that the projection $\pi\circ \gamma :S^1 \to S^3$ is a smooth knot $K$
in $S^3$. In particular, $\gamma$ intersects each fiber 
of $X\to S^3$ at most once, otherwise its projection to $S^3$ 
would have self-intersection points. 
Suppose  the map $\gamma$ is given by 
\[
\theta \in S^1 \to ({\vec u},{\vec v})=({\vec f}(\theta), 
{\vec g}(\theta)), 
\]
where ${\vec f}(\theta)=(f_j(\theta))$,  
${\vec g}(\theta)=(g_j(\theta))$, $j=1,\ldots, 4$, are smooth periodic functions of $\theta$. 

The total space of the conormal bundle $N^*_{K}$ to $K$ in $S^3$ is defined by the equations 
\[
{\vec u}={\vec f}(\theta), \qquad  {\dot {\vec f}}(\theta)\cdot 
{\vec v}=0. 
\]
where ${\dot {\vec f}}(\theta) = d{\vec f}(\theta)/d\theta$. Then a straightforward computation yields 
\[ 
\omega_X|_{N_K} = \big(d\sum_{j=1}^4 v_jdu_j\big)|_{N_K} 
= d \big(\sum_{j=1}^4 v_j {\dot f}_j d\theta\big) =0, 
\]
confirming that $N_K^*$ is a lagrangian cycle in $X$. 

Now consider the three-cycle $L_\gamma\subset T^*S^3$ determined by the equations 
\be\label{eq:lagcycleA}
\bal 
& {\vec u}={\vec f}(\theta),\qquad 
{\dot {\vec f}}(\theta)\cdot ({\vec v}-{\vec g}(\theta))
=0.\\
\eal 
\ee
By construction $L_\gamma$ is a cycle in the total space of 
the restriction $T^*S^3|_K$. The restriction of the canonical 
projection $\pi:T^*S^3\to S^3$ yields a projection 
$\pi_{L_\gamma}:L_\gamma \to K$.  The fiber of $\pi_{L_\gamma}$ over a point $p\in K$ 
is the two plane in $T^*_pS^3$ determined by the second equation 
in \eqref{eq:lagcycleA}, which is linear in $v_j$. Basically, $L_\gamma$ is obtained by a fiberwise translation of $N_K$  by a  
translation vector depending on the point $p\in K$. 
The restriction of the canonical symplectic form to $L_\gamma$ 
is given by  
\[
\bal 
\omega_X|_{L_\gamma} & = \big( d\sum_{j=1}^4 v_j du_j\big)\big|_{L_\gamma} 
=  d\big( \sum_{j=1}^4 v_j {\dot f}_j(\theta) d\theta \big) \\
\eal 
\]
Using the second equation in \eqref{eq:lagcycleA},
\[
 \sum_{j=1}^4 v_j {\dot f}_j(\theta) d\theta = \sum_{j=1}^4 g_j(\theta)
 {\dot f}_j(\theta)  
 d\theta
 \]
 on $L_\gamma$. Therefore 
 \[
 \omega_X|_{L_\gamma}= d\big(\sum_{j=1}^4 g_j(\theta)
 {\dot f}_j(\theta) d\theta \big)=0.
 \]
 In conclusion, $L_\gamma$ is a lagrangian cycle on $T^*S^3$. 
  Note that the intersection of $L_\gamma$ with the zero section ${\vec v}=0$ is determined by the equations 
 \[
 {\vec u}= {\vec f}(\theta), \qquad {\dot{\vec f}}(\theta)\cdot {\vec g}(\theta)=0.
 \]
 For sufficiently generic ${\vec f}(\theta)$, ${\vec g}(\theta)$ 
 this intersection will be empty, such that $L_\gamma$ is a 
 lift of the conormal bundle $N^*_K$ off the zero section. 
 
 Note also that the lift  constructed in 
 \cite{conormal-knots} is a special case of the above construction 
 obtained by setting ${\vec g}(\theta)={\dot{\vec f}}(\theta)$. 
 The main reason for the above generalization is that at least 
 for algebraic cycles it also yields specific holomorphic open string 
 instantons interpolating between the lifted conormal bundle and the vanishing cycle $S^3$ in the deformed conifold. This will 
 be explained next.

 \subsection{Lagrangian cycles for algebraic 
 knots}\label{lagalgknots}
 So far this construction is fairly general and can be applied to any knot in 
 $K\subset S^3$, for any lift $\gamma:S^1 \to X$ satisfying the above conditions. 
 In the special case when $K$ is an algebraic 
 knot there is a preferred construction of the 
 lift $\gamma$ motivated by AdS/CFT correspondence. The main idea 
 is to obtain a one-cycle $\gamma$ as in section (\ref{lagsect}) 
 by intersecting an $S^2$-bundle $P_a\subset T^*S^3$ of radius 
 $a>0$ with the image $\phi_\mu(C_\mu)$ of a certain holomorphic 
 curve $C_\mu\subset X_\mu$ associated to $K$ as explained 
 below. Here  $\phi_\mu:X_\mu\to X$ is the symplectomorphism
  given in equation \eqref{eq:sympmorphismA}. 
  
 Suppose $K$ is the link of the 
plane curve singularity $f(x,y)=0$ in $\IC^2$. 
For simplicity assume that the curve $f(x,y)=0$ 
is irreducible and smooth away from $x=y=0$, 
and $K$ is connected. 
Consider the complete intersection $Z_\mu\subset 
X_\mu$ determined by 
\be\label{eq:holcurveA}
f(x,y)=0, \qquad f(z,-w)=0.
\ee
Suppose that $f(x,y)$ is sufficiently generic such that 
$Z_\mu$ is smooth for generic $\mu>0$. Note that $Z_\mu$ may have several distinct 
connected components even though 
the plane curve $f(x,y)=0$ is assumed irreducible. 
For example consider the case of torus knots, 
$f(x,y)= 
x^r-y^s$ with $(r,s)$ coprime positive integers.
Then equations \eqref{eq:holcurveA} imply 
\[
(xz)^r - (-yw)^s=0,
\]
and substitution in the deformed conifold equation, $xz-yw=\mu$, yields 
\[
(xz)^r -(\mu-xz)^s =0.
\]
Therefore $xz=\eta$, where $\eta$ is a solution of the 
polynomial equation $t^r-(\mu-t)^s=0$. 
 Each such solution $\eta$ determines a connected component of $Z_\mu$ 
of the form 
\[
(x,y,z,w)=(t^s,t^r,\eta t^{-s}, (\mu-\eta)t^{-r})
\]
with $t\in\IC\setminus \{0\}$.

Obviously, if $f(x,y)$ is a polynomial 
with real coefficients, $Z_\mu$ is preserved by the 
antiholomorphic involution \eqref{eq:antiholinv}.
This will be assumed to be the case 
from now on. Then each  connected component 
of the intersection of $Z_\mu$ 
with the fixed point locus  $S_\mu=X_\mu^{\tau_\mu}$ 
is isomorphic to the one-cycle 
\[ 
|x|^2 +|y|^2 =\mu, \qquad f(x,y)=0
\]
in $\IC^2$. 
For sufficiently small $\mu>0$, this is the
link of the plane curve singularity $f(x,y)=0$ in $\IC^2$. 
Note that the symplectomorphism $\phi_\mu$ maps $S_\mu$ 
to the zero section $S=\{{\vec v}=0\}$ in $X=T^*S^3$.

Now let $P_a=\{|{\vec v}|=a\}$, $a>0$, be the sphere bundle of radius $a$ in $X=T^*S^3$, and 
$B_a\subset X$ be the bounding  disc bundle,
\[
B_a=\{({\vec u}, {\vec v})\, |\, |{\vec v}|\leq a\}.
\]
Suppose there is a connected component $C_\mu$ of $Z_\mu$
with nontrivial intersection with the vanishing cycle $S_\mu$. 
As observed above each connected component
of the intersection 
must be isomorphic to the link of the plane curve singularity 
$f(x,y)=0$.  Since $\phi_\mu(C_\mu)$ has nontrivial 
intersection with the zero section $S\subset X$, it will 
also intersect all  sphere bundles $P_a\subset X$ for sufficiently 
small values of 
$a\in \IR_{>0}$. In fact for sufficiently small $a>0$ the intersection 
$\phi_\mu(C_\mu)\cap B_a$ will be foliated by disjoint connected one-cycles $\gamma_{\mu,a'}=\phi_\mu(C_\mu)\cap P_{a'}$, $0\leq a'\leq a$. Then applying the construction in section (\ref{lagsect}) 
to $\gamma_{\mu,a}$ yields a lagrangian cycle  $L_{\gamma_{\mu,a}}\subset X$. 
The inverse image  $L_{\mu,a}\subset X_\mu=\phi_{\mu}^{-1}(L_{\gamma_{\mu,a}})$ is a lagrangian cycle in $X_\mu$ 
intersecting $C_\mu$ along the one-cycle 
$\phi_\mu^{-1}(\gamma_{\mu,a})$. Moreover, by construction 
there is a  holomorphic cylinder in ${\sf C}_{\mu,a} \subset X_\mu$
contained in $C_\mu$, with one boundary component in $S_\mu$
and the second boundary component in $L_{\mu,a}$. This is precisely 
the basic set-up of large $N$ duality in terms of lifted lagrangian 
cycles described in section (\ref{largeNunknot}), above 
equation \eqref{eq:opinstA}. In order to keep the notation 
simple $L_{\mu,a}$, ${\sf C}_{\mu,a}$ will be simply denoted 
by $L_\mu$, ${\sf C}_\mu$ the $a$-dependence being implicitly 
understood. 

Two important questions must be addressed at this point. 
The first issue is whether
 the holomorphic cylinder ${\sf C}_\mu$ with lagrangian boundary 
 conditions on $S_\mu$, $L_\mu$ is unique and rigid, 
 at least up to a 
 torus action. This question will be answered affirmatively 
 for torus knots in section (\ref{torusknots}), being left open 
 at the moment for more general algebraic knots. 
 
  The second problem is whether one can construct a 
  family of lagrangian cycles $M_\epsilon$ on $Y$ 
  completing the geometric transition picture 
  represented in \eqref{eq:coniftransB}. 
  This will be shown to be the case 
 for any algebraic knot in the next subsection, with the caveat that the resulting Gromov-Witten theory with lagrangian 
 boundary conditions on $M_\epsilon$ 
 is again tractable only for torus knots. 

A first step towards completing the diagram \eqref{eq:coniftransB}
is to understand the specialization of the above construction 
at $\mu=0$. The specialization of $Z_\mu$ 
 is a reducible 
curve $Z_0$ in the singular conifold $X_0$ with at least 
two irreducible components
$C^\pm$ given by 
\[
f(x,y)=0, \qquad z=w=0,
\]
respectively 
\[
f(z,-w)=0, \qquad x=y=0.
\]
These components meet at the conifold singularity, which is 
also a singular point of $Z_0$. 
Since $f(x,y)$ is assumed real, 
the antiholomorphic involution $\tau_0:X_0\to X_0$ exchanges 
 $C^\pm$. 
 
 For concreteness, consider again the example of torus knots, 
 $f(x,y)=x^r-y^s$. 
 In this case the defining equations of $Z_0$ 
 imply that $t=xz$ must be a solution of the polynomial 
 equation $t^r-(-t)^s=0$. Therefore $xz=0$ or $xz=\eta$ with 
$\eta^{r-s} = (-1)^{s+1}$. This implies that $Z_0$ has $r-s+1$ 
connected components. The connected component corresponding to 
$xz=0$ is the union of the two irreducible components $C^\pm$ 
defined above, which intersect at the singular point $x=y=z=w=0$. 
Each connected component corresponding to $xz=\eta$ is determined 
by the equations 
\[
xz =yw=\eta, \qquad x^r=y^s.
\]
Since these equations are invariant under the $\IC^\times$-action 
\[
(x,y,z,w) \mapsto (\alpha^sx,\alpha^ry,\alpha^{-s}z, \alpha^{-r}w)
\]
and $x,y,z,w$ cannot vanish, each such component is isomorphic to $\IC^\times$.
 
Returning to the general case, let 
 $\gamma^\pm$ be the one-cycles obtained by intersecting the inverse images 
$\phi_0(C^\pm\setminus\{0\})$ with the sphere bundle 
$P_a$. 
It is straightforward to check that $\tau_0$ exchanges the 
image cycles $\phi_0(\gamma^\pm)$.
Applying the construction of section (\ref{lagsect}) to the 
cycle $\gamma^+$, one obtains 
a lagrangian cycle $L_{\gamma^+}$ in $X$. 
The inverse image $L_0=\phi_0^{-1}
(L_{\gamma^+})$ is a lagrangian cycle in $X_0$. 
For sufficiently small 
$\mu\in \IR_{>0}$ 
there exists an irreducible  component $C_\mu$ of $Z_\mu$ such that the 
 intersection $\phi_\mu(C_\mu )\cap P_a$ has a connected 
 component $\gamma_{\mu}$ which 
specializes to $\gamma^+$ at $\mu=0$. The resulting 
family of lifted lagrangian cycles $L_\mu$ specializes to 
$L_0$ at $\mu=0$. This completes the bottom part of
diagram \eqref{eq:coniftransB}. The remaining part will 
be constructed in the next subsection. 

\subsection{Lagrangian cycles in the resolved 
conifold}\label{topres} 
Recall the resolved conifold $Y$ is determined by equations 
\eqref{eq:smallresA} in $\IC^4\times \IP^1$
and $\sigma:Y\to X_0$ denotes the natural contraction map 
to the singular conifold. The family of symplectic manifolds $Y_\epsilon$ in 
diagram \eqref{eq:coniftransB} is determined by the 
symplectic forms 
$$\omega_{Y,\epsilon}
= \big(\omega_{\IC^4}+\epsilon^2 \omega_{\IP^1}\big)|_{Y}$$
where $\omega_{\IC^4}$ is the standard symplectic form on 
$\IC^4$ and $\omega_{\IP^1}$ is the Fubini-Study form on 
$\IP^1$. 

The family of lagrangian cycles $M_\epsilon\subset Y_\epsilon$ will be constructed using 
\cite[Lemm. 7.11]{introsymptop}, which provides a geometric relation between 
the symplectic structures on $Y_\epsilon$, $X_0$. 
First it will be helpful to recall the statement of \cite[Lemm. 7.11]{introsymptop} for 
the one-point blow-up 
$\eta:{\widetilde \IC}^2\to \IC^2$ at the origin. 
Consider the following one parameter family of symplectic forms on the blow-up
\[
\omega_{{\widetilde \IC}^2,\epsilon} = 
\big(\omega_{\IC^2}+\epsilon^2
\omega_{\IP^1}\big)|_{{\widetilde \IC}^2}.
\]
For any $\epsilon\in \IR_{>0}$ let $B(\epsilon)\subset \IC^2$ be the 
ball $|z|^2 +|y|^2 \leq \epsilon^2$  and 
${\widetilde B}(\epsilon)=\eta^{-1}(B(\epsilon))$
 be its inverse image in ${\widetilde \IC}^2$. Note that there is a radial 
 map $\rho_\epsilon: \IC^2\setminus \{0\} \to \IC^2 \setminus B(\epsilon)$, 
 \[
 \rho_\epsilon(y,z) = {\sqrt{|z|^2 +|y|^2 +\epsilon^2}\over 
 \sqrt{ |z|^2 +|y|^2}} (y,z)
 \]
Then 
\cite[Lemm. 7.11]{introsymptop} proves that the 
map $\psi_{\epsilon}: {\widetilde \IC}^2\setminus E
\to  \IC^2\setminus B(\epsilon)$, 
\[
\psi_{\epsilon} = \rho_\epsilon
 \circ \eta |_{{\widetilde \IC}^2\setminus E}
\]
is a symplectomorphism for any $\epsilon\in \IR_{>0}$, where $E\subset 
{\widetilde \IC}^2$ denotes the exceptional 
curve.

In order to apply \cite[Lemm. 7.11]{introsymptop} to the present situation, note that $Y$ 
can be regarded as the quadric hypersurface $x\lambda =w\rho$, 
in the fourfold $Z$ determined by $z\rho = y\lambda$ in 
$\IC^4\times \IP^1$. Obviously, 
$Z\simeq \IC^2\times {\widetilde \IC}^2$ where 
${\widetilde \IC}^2$
is the one point blow-up of $\IC^2$ at the origin. 
Next note that the map 
$\varrho_\epsilon=1_{\IC^2}\times \rho_{\epsilon}: \IC^2 \times (\IC^2 \setminus \{0\}) \to {\IC^2}\times(\IC^2 \setminus B(\epsilon))$ preserves the nodal threefold $X_0\subset \IC^2\times \IC^2$, 
 mapping 
 $X_0\setminus \{0\}$ to the open 
 subset $X_0(\epsilon)=X_0\setminus X_0 \cap (\IC^2 \times 
B(\epsilon))$.  Note also 
that the exceptional $(-1,-1)$ curve $C_0\subset Y$ 
coincides with the curve $\{0\}\times E \subset \IC^2\times {\widetilde \IC}^2$.
This implies that the complement of the zero section $Y\setminus C_0$ 
coincides with the open subset 
$Y\cap \IC^2\times ({\widetilde {\IC}}^2 \setminus E)$. 
Then the map 
\be\label{eq:sympmorphismB}
\phi_{\epsilon}: 
Y\setminus C_0
\to X_0(\epsilon),\qquad 
\phi_{\epsilon}= (\varrho_\epsilon \circ \sigma)|_{Y\setminus C_0}.
\ee
 is a symplectomorphism.
 
 Returning to the construction of the lagrangian cycle 
$M_\epsilon\subset Y_\epsilon$, recall that the family of complete 
intersection curves $C_\mu\subset X_\mu$ given by 
\[
f(x,y)=0, \qquad f(z,-w)=0
\]
specializes to a reducible curve at $\mu=0$ with two components
\[
\bal
C^+:&\qquad f(x,y)=0, \qquad z=w=0\\
C^-:&\qquad f(z,-w)=0, \qquad x=y=0. 
\eal 
\]
The intersection of $\phi_0(C^+)$ with the sphere bundle 
$P_a$ yields a one-cycle $\gamma^+$, the limit of 
the cycles $\gamma_\mu^+$ as $\mu\to 0$.
The corresponding lagrangian cycle $L_{\gamma^+}$ 
is the limit of $L_\mu^+$ as $\mu\to 0$. 
 
Now consider the one-cycle 
$$\gamma^+_\epsilon = \phi_0\circ \varrho_\epsilon \circ  \phi_0^{-1}\circ \gamma^+: S^1 \to 
X$$ 
on $X$ obtained by 
applying the radial map to 
 the inverse image $\phi_0^{-1}\circ \gamma^+$ of the 
 path $\gamma^+$. 
 Then set 
\be\label{eq:lagcycleres}
M_\epsilon=\phi_\epsilon^{-1}( \phi_0^{-1}(L_{\gamma^+_\epsilon})) = 
\sigma^{-1} (\varrho_\epsilon^{-1}(\phi_0^{-1}(L_\epsilon))),
\ee
where $L_{\gamma^+_\epsilon}\subset X$ is the 
lagrangian cycle obtained by applying the construction of section 
(\ref{lagsect}) to $\gamma^+_\epsilon$. 
By construction $L_{\gamma^+_\epsilon}$ intersects the dilation $\varrho_\epsilon(\phi_0(C^+)))$ 
of the curve $\phi_0(C^+)$ along the cycle 
$\gamma^+_\epsilon$. Therefore the inverse 
image $\varrho_\epsilon^{-1}
(\phi_0^{-1}(L_{\gamma^+_\epsilon}))$ intersects the plane curve 
$C^+\subset X_0$
along the cycle $\phi_0^{-1}\circ \gamma^+_\epsilon$. Since 
$\sigma:Y\setminus C_0\to X_0\setminus\{0\}$ is an isomorphism of complex 
manifolds, it follows that ${M_\epsilon}$ intersects the strict transform $C\subset Y$ of $C^+$ along the 
cycle ${\eta}_\epsilon=\sigma^{-1}\circ \phi_0^{-1}\circ 
\gamma^+_\epsilon$. 
The strict transform $C$ is the plane singular curve 
cut by the equations 
\[
f(x,y)=0, \qquad \lambda=0
\]
on $Y$. Therefore it is a singular plane curve isomorphic to $C^+$, 
contained in the 
fiber $\lambda=0$ of $Y$ over $\IP^1$. 
The singular point $p\in C$ is the unique point of intersection with the zero section, $x=y=0$, $\lambda=0$. 
 The cycle $\eta_\epsilon$ divides $C$ into two connected components, 
 the component containing $p$ being a singular holomorphic 
 disc ${\sf D}_\epsilon$ in $Y_\epsilon$ with boundary  
 $\eta_\epsilon\subset M_\epsilon$. 
 This is precisely the geometric set-up outlined in 
 diagram \eqref{eq:coniftransB}. 
 In order to obtain a complete large $N$ duality picture,  one should prove that the holomorphic disc ${\sf D}_\epsilon$ is rigid, which is a 
 difficult technical question for general algebraic knots.
  Section (\ref{torusknots}) will provide an affirmative 
 answer for torus knots, leaving the general case for future work.  
 
Assuming that ${\sf D}_\epsilon$ is rigid, the next problem is 
the computation of its multicover contributions to the 
Gromov-Witten theory with lagrangian boundary conditions on $M_\epsilon$. 
One angle on this problem is to try to generalize the computations 
of \cite{KL} based on stable maps with lagrangian boundary 
conditions to the present case. This approach requires a 
torus action preserving $M_\epsilon$, 
${\sf D}_\epsilon$, which is the 
case only for torus knots. In this case, the details of the virtual 
localization computation are presented in section (\ref{torusknots}), the resulting formulas being in agreement 
with large $N$ duality predictions. 

A second approach follows from string duality considerations 
as in \cite{Mtop,knot-top}, converting the calculation of 
of topological open {\bf A}-model amplitudes to D-brane 
bound state counting. In this framework, the topological amplitudes 
are expressed in terms of BPS states as in Donaldson-Thomas type invariants, making a 
direct connection with the \cite{hilbert-links}. This will be discussed next.

\section{D-brane bound states and the Hilbert 
scheme}\label{DbranesHilbert}
 
 The goal of this section is to provide a physical explanation 
 for the  work of Oblomkov and Shende \cite{hilbert-links}
  on plane curve singularities in the framework of large 
  $N$ duality. 
    The geometric set-up will be the same as in section 
  (\ref{topres}), namely a lagrangian cycle $M_\epsilon\subset Y_\epsilon$ 
  intersecting a singular plane curve $C\subset Y$ 
  along a smooth connected one-cycle $\eta:S^1 \to M_\epsilon$. 
  The curve $C$ is given by 
 \[
  f(x,y)=0, \qquad \lambda=0,
\]
on $Y$ and it will be assumed that it has only one 
singular point $p$, given by $x=y=0$, $\lambda=0$.   
The cycle $\eta_\epsilon$ divides $C$ into two connected components,
the component containing $p$ being a holomorphic disc 
${\sf D}_\epsilon$ with boundary on $M_\epsilon$. 
Note that $M_\epsilon\simeq \IR^2 \times S^1$ 
and the cycle $\eta_\epsilon$ is a generator of $H_1(M_\epsilon)\simeq \IZ$. 
It will be  
assumed that ${\sf D}_\epsilon$ 
is rigid, which is in fact proven in section 
(\ref{torusknots}) 
for curves of the form $f(x,y)=x^r-y^s$.
The subscript $\epsilon$ will be dropped in this section because all considerations 
below hold for any fixed arbitrary value of $\epsilon>0$.

According to cite \cite{knot-top,LMV,framedknots}, string duality transformations 
show that open topological string amplitudes 
with lagrangian boundary conditions on $M$ 
 are determined by counting supersymmetric 
 M2-brane or D2-D0 bound states in different duality frames. 
 This is achieved by studying the low-enery effective 
action for type IIA D4-brane wrapped on the lagrangian 
cycle $M$, resulting in a string-like object in the 
four transverse dimensions. Open topological string 
amplitudes with boundary conditions 
on $M$ determine certain holomorphic couplings 
in the low energy effective action of this string. 
Since $H_1(M)\simeq \IZ$ is generated by $\eta$,  
open string instantons with 
fixed genus $g\in \IZ_{\geq 0}$ 
and $h=1$ boundary components are 
 topologically classified by the 
wrapping 
number $d\in \IZ_{\geq 0}$ on the holomorphic curve $C_0$ and the winding number $k\in \IZ_{\geq 1}$ 
about the cycle $\eta$. 
The corresponding Gromov-Witten invariants with 
lagrangian boundary  conditions will be denoted 
by $GW_{g,1}(d,k)$. 
Only topological 
open string amplitudes with winding number $k=1$
 will be considered 
in the following, because we are interested in the Wilson loop 
observables in the fundamental representation.  Moreover we can assume,
with no loss of generality that we have only 1 spectator lagrangian A-brane and
we replace ${\rm Tr}V$ with $V$.
According to \cite{knot-top}, these amplitudes determine 
terms of the form 
\[
\bal
 & \int d^4x d^4\theta  \delta^{(2)}(x) \delta^{(2)}(\theta) 
F_{g,1}(t,V)(W^2)^g\\
\eal \]
in the effective action of the string,  where
\[
F_{g,1}(t,V) = \sum_{d\geq 0} g_s^{2g-1} e^{-dt} GW_{g,1}(d,1)
V\\
\]
 Here $W_{\alpha\beta}$, 
 where $\alpha, \beta$ are symmetric spinor indices, 
 denotes the four dimensional graviphoton multiplet,
 and $t$ denotes the vector multiplet whose 
top component is the K\"ahler modulus of the zero section 
$C_0\subset Y$. As in section (\ref{largeNunknot}),  
$V$ 
is the holonomy of a background flat $U(1)$ gauge field on 
the D4-brane. The four dimensional superspace integral 
is restricted to the string world-sheet by the $\delta$-functions 
$\delta^{(2)}(x), \delta^{(2)}(\theta)$.

The M-theory lift of this configuration is an M5-brane wrapping 
the same lagrangian cycle $M$. Holomorphic IIA
world-sheet instantons lift to open M2-branes with boundary 
on $M$ wrapping the disc  ${\sf D}$ and the zero section 
$C_0$.
The low-energy effective theory of the M5-brane is now a 
three dimensional theory containing a spectrum of 
supersymmetric particles corresponding to bound states of 
open  M2-branes. The low energy degrees of freedom 
include a three-dimensional $N=2$ 
$U(1)$ vector multiplet, the reduction
of the M5-brane self-dual tensor multiplet
 on a harmonic generator of $H^1(M_\epsilon)\simeq \IZ$. In addition the space-time effective action includes 
 a $U(1)$ gauge field 
field in the  supergravity multiplet. 
The three-dimensional BPS particles carry integer charges 
$(k,d)\in \IZ$ with 
respect to these gauge fields. Geometrically, $k,d$ 
are the M2-brane multiplicity on the disc ${\sf D}$, respectively the zero section $C_0$.
The five dimensional $SO(4)$ little group is broken 
to $U(1)\times U(1)$ by the M5-brane, the 
first factor being the little group in the M5 three 
dimensional effective theory. The second factor 
is generated by rotations in the two transverse
dimensions. 
Therefore the BPS spectrum is graded by two 
spin quantum numbers $\sigma, j \in \IZ+1/2$. 
 The degeneracies of BPS states will 
 be denoted accordingly by $N_{k,d,\sigma,j}$.
   By analogy with \cite{Mtop}
a Schwinger computation shows that the couplings $F_{g,1}(t,V)$ 
are given by 
\be\label{eq:openBPSA}
F_{g,1}(t,V) =  \sum_{d\geq 0} 
\sum_{\sigma\in \IZ} {N_{d,\sigma}\over 2\sin(g_s/2)} 
e^{-dt+i\sigma g_s} V^{} 
\ee
where $N_{d,\sigma}$ is a BPS index 
given by 
\[
N_{d,\sigma} =\sum_{j\in \IZ+1/2} (-1)^{2j+1} 
N_{1,d,\sigma,j}.
\]
In particular note that the coefficients 
 $N_{d,\sigma}$ have to be integral but not necessarily positive.
 
The above expression is the restriction of 
\cite[Eqn. 4.4]{knot-top} to open string amplitudes of winding number 
$1$, therefore the sum over the integer $n\geq 1$ corresponding to 
degree $n$ multicover contributions collapses to a single term, $n=1$. 
A more convenient form of equation \eqref{eq:openBPSA} 
can be obtained by a change of variables 
$q=e^{ig_s}$, $Q=e^{-t}$, 
and a redefinition of the spin quantum number, $\sigma = s +1/2$, 
$s\in \IZ$. Then equation \eqref{eq:openBPSA} becomes 
\be\label{eq:openBPSC}
F_{g,1}(t,V) =  \sum_{d\geq 0} 
\sum_{\sigma\in \IZ} {N_{d,s}q^sQ^d\over 1-q}  V. 
\ee
In order to compute the BPS numbers $N_{d,s}$, one has to count spin 
$s+1/2$ bound states of an open M2-brane wrapping the singular holomorphic disc ${\sf D}$ and $d$ M2-branes wrapping the zero 
section $C_0$.   Note that using the Large N duality this implies that the expectation value of the Wilson loop
in the fundamental representation of the knot, which is known as the HOMFLY polynomial of the knot, is given by this expression
(recalling that ${\rm Tr}U$ is paired up with ${\rm Tr}V$):
\be
\langle { Tr} U \rangle =\sum_{d\geq 0} 
\sum_{\sigma\in \IZ} {N_{d,s}q^sQ^d\over 1-q}
\ee

It is known \cite{gauge-gravity,Mtop} that $d$ M2-branes wrapping the 
compact curve $C_0$ form supersymmetric bound states only if $d=1$, in which case there is a single spin 0 state. Therefore the main problem 
is to understand bound state counting for open M2-branes wrapping 
the singular disc ${\sf D}$. 
This is most efficiently done reducing the 
problem to counting D2-D0 bound states in a suitable weakly 
coupled  Type IIA limit. 
More precisely, one can choose the M-theory circle such that 
that the M5-brane is 
mapped again to a D4-brane, but the M2-branes yield open 
D2-branes with boundary on the D4-brane. 
Furthermore the $d=0$ truncation of the right hand side of equation \eqref{eq:openBPSC} 
is interpreted as a partition function of the form 
\be\label{eq:openBPSB} 
\sum_{n\geq 0} C_n q^n 
\ee
counting 
supersymmetric
states of open D2-branes wrapping ${\sf D}$ 
bound to an arbitrary number $n$ of D0-branes
\cite{Mtop,spinning}. The coefficients  $C_{n}$ 
in equation \eqref{eq:openBPSB} are BPS indices 
counting 
states weighted by a sign determined by their spin. 
This index can be exactly computed in the semiclassical limit, in
which case $C_{n}$ equals the Euler character of 
the moduli space of supersymmetric D-brane configurations. 
In order to understand the structure of such moduli spaces, 
it is helpful to consider first configurations of $n$ D0-brane 
bound to a D2-brane wrapping a fixed 
compact holomorphic curve $Z$ in some 
Calabi-Yau threefold.  
According to \cite{spinning, vortex, stabpairs-I},
such configurations are 
mathematically modeled by an abelian vortex configuration of degree 
$n$ on $Z$. The basics of this formalism will be reviewed in some
detail below.

\subsection{D2-D0 bound states, vortices, and stable pairs}

A degree $n$ abelian vortex is a triple 
 pair $(\CL,A,s)$ where $\CL$ is a complex line bundle on $Z$ with first 
 Chern class $c_1(\CL)=n$, 
 $A$ is a $U(1)$ 
connection on  $\CL$ and 
$s$ is a section of $\CL$ satisfying  $D_As=0$. 
This naturally captures the dates of the choice of the gauge field on the D2 brane,
as well as the geometry of D0 brane, which can be identified with $s$.  It corresponds
to the bifundamental field charged under the D2-brane $U(1)$ stretched between D0-brane
and D2-brane.
The moduli space of triples $(\CL,A,s)$ modulo unitary 
gauge transformations is isomorphic to the moduli 
space of pairs $(\CL,s)$ modulo complexified gauge transformations, where $\CL$ is a holomorphic 
line bundle on $Z$ and $s\in H^0(Z,\CL)$ is a 
nontrivial holomorphic section. 
The relation between differential geometric and algebraic geometric data follows as usual 
observing that any connection $A$ on a $C^\infty$ 
complex line bundle $\CL$ determines a Dolbeault 
operator 
 ${\overline \partial}_A$. 
 
 In the algebraic formulation, 
 note that the zero locus of $s$ is a degree $n$ effective divisor $
(s)$ on $Z$,
that is a formal linear combination of points $\sum_{i=1}^k 
n_ip_i$ where $n_i\geq 1$ and $\sum_{i=1}^k n_i=n$. The points 
$p_1,\ldots, p_k$ represent the locations of the D0-branes and 
the integers $n_i\geq 1$, $i=1,\ldots, k$ the D0-brane multiplicity 
at each point.
Assigning to each pair $(\CL,s)$ the divisor $(s)$ 
yields an isomorphism between the 
 moduli space of  isomorphism classes of pairs $(\CL,s)$ and the symmetric product $S^n(Z) = Z^n/\CS_n$, where $\CS_n$ is the permutation group on letters. 
  
From the sheaf theoretic point of view, 
 a pair $(\CL,s)$ can be uniquely characterized up to 
gauge transformations by specifying the germs of local holomorphic 
sections of $\CL$ near each point $p$ of $Z$. 
The simplest case is the trivial vortex configuration, when $\CL$ is isomorphic to the trivial line bundle $\CO_Z$ on $Z$ 
and $s$ is constant. 
The germs of local sections of $\CO_Z$ near each point $p\in Z$ 
are simply germs of local holomorphic functions 
with no 
restrictions on the vanishing order at $p$. 
For a configuration $(\CL,s)$ with $n>0$ the same holds locally near 
any point $p\in Z$, $p\neq p_i$, $i=1,\ldots, k$. 
Near one of the points $p_i$, the germs of holomorphic sections 
of $\CL$ are identified with
germs of meromorphic functions with at most a pole of order 
$n_i$ at $p_i$. In this local picture the section $s$ corresponds to the natural  
inclusion of the local sections of $\CO_Z$ in the 
set of local sections of $\CL$. Note that the complement is 
a finite dimensional vector 
space of dimension $n_i$. In terms of a local coordinate $z_i$ centered 
at $p_i$, this vector space is generated by sections 
of the form $\{z_i^{-l}\}$, $l=1,\ldots, n_i$. More abstractly, 
the local sections of $\CL$ near $p_i$ 
form a rank 1 module over the local ring of functions $\CO_Z$ generated by $\{z_i^{-n_i}\}$.

It may be also helpful to note that there is a dual mathematical model for D2-D0 configurations. In the dual model, $n_i$ 
D0-branes located at $p_i$ are described by the set of local 
holomorphic functions which vanish at least to order $n_i$ at $p_i$. This set is the ideal generated by $z_i^{n_i}$ 
in the ring of local functions near $p_i$. The geometric object 
characterized by this local behavior is the dual line bundle $\CL^{-1}$, which is a sub-sheaf of the trivial line bundle $\CO_Z$. In more abstract language, 
$\CL^{-1}\subset \CO_Z$ is the defining ideal sheaf of the 
effective divisor $(s)=\sum_{i=1}^k n_i p_i$. 

Similar considerations apply \cite{stabpairs-I,stabpairs-III} to a 
singular curve $Z$,
abelian vortices being generalized to stable pairs.
  This essentially means 
that one has to allow the gauge field $A$ to develop singularities 
at the singular points of the curve $Z$. While a complete analytic
treatment of such singularities would be quite difficult, the sheaf 
theoretic point of view discussed above leads to an efficient construction of the moduli space. 
A single D0-brane supported at 
 a smooth point $p$ 
was previously identified with the module of local meromorphic functions 
 with at most a simple pole at $p$. 
 If $p$ is a singular point of $Z$, a single D0-brane at $p$ 
is still defined by a module of local meromorphic 
functions, but this module may have more than one generator. 
Conceptually, this may be easily understood employing the dual 
model. Consider for example the plane curve singularity 
$x^3=y^2$. A D0-brane with multiplicity $1$ located at the 
singular point $x=y=0$ cannot be described as the zero locus 
of a single local holomorphic function. If one simply sets $x=0$ 
or $y=0$ the defining equation of the curve reduces to $y^2=0$, 
respectively $x^3=0$. According to the previous paragraph 
this is in fact a D0-brane configuration with multiplicity $2$, 
respectively $3$. A single D0-brane is the zero locus of 
two local functions, $(x,y)$ which generate an ideal
in the ring of local holomorphic functions. 
The dual stable pair is given by a local module 
over the ring of local functions generated by two elements. 

Informally, the main idea 
of this construction is that at a singular point $p$ of $Z$ the rank of the Chan-Paton 
line bundle $\CL$ on $Z$ is allowed to jump in a controlled way, 
depending on the analytic type of the singularity at $p$. 
 Effectively, the single D2-brane on $Z$ 
behaves locally at $p$ 
as a stack of D2-branes with higher multiplicity $m\geq 1$.
For a fixed number of D0-branes $m$ may take finitely many values determined by $n$ and the singularity type. 
 This point of view 
will be very useful in understanding bound state formation for 
D2-branes wrapping different holomorphic curves with transverse intersection.

A consequence of the above discussion 
is that the moduli space of D2-D0-brane configurations 
supported on $Z$ is no longer isomorphic to the symmetric product 
$S^n(Z)$.  It has been shown in \cite{stabpairs-III} 
that the moduli space of D2-D0-brane configurations supported on $Z$ is 
in this case isomorphic to the Hilbert scheme $\CH^n(Z)$ 
of $n$ points on $Z$. The rigorous definition of the Hilbert scheme 
is not needed for the purpose of the present discussion, but it may be helpful 
to note that there is a natural map $\pi:\CH^n(Z) \to S^n(Z)$
forgetting the extra algebraic structure associated to each 
singular point. From a physical point of view this means that the 
D0-branes are treated simply as non-interacting particles ignoring 
interactions due to 
open string effects. 

Analogous considerations hold for D2-branes 
wrapping a smooth holomorphic disc ${\sf D}$ with boundary on a lagrangian cycle. The holomorphic line bundle $\CL$ must now be 
equipped with a trivialization on the boundary of the disc
$\partial {\sf D}\simeq S^1$, which is part of the boundary conditions 
on the D2-brane fields. Complex line bundles on the disc 
with boundary trivialization are topologically classified by 
the first Chern class, which takes values in the  
relative homology group $H_2({\sf D},\partial{\sf D})\simeq\IZ$. 
Moreover, since the section $s\in H^0(\CL)$ must be compatible 
with the trivialization, the number of zeroes of $s$, counted with multiplicity must equal the first Chern class $n$. 
Summing over all Chern classes yields the partition 
function 
\be\label{eq:smoothdisc}
\sum_{n\geq 0}C_n q^n = \sum_{n\geq 0}
\chi(S^n({\sf D})) q^n = {1\over 1- q} 
\ee
since the symmetric power $S^n({\sf D})$ is contractible 
for any $n\geq 0$. Note that this result 
is the same as the winding number one partition  function 
 of a single lagrangian brane 
in $\IC^3$ given \cite{topvert} by the topological vertex 
$C_{\emptyset,\emptyset,{\Box}}(q)$.   
As observed in \cite[Sect 4]{IKV}, the above formula 
can be alternatively interpreted as the Hilbert 
series of the ring $\IC[t]$ of polynomial functions 
on the complex line $\IC$. By definition, the Hilbert 
series of a polynomial ring $R$ is 
\[
H_R(q) = \sum_{n\geq 0} c_n(R) q^n  
\]
where $c_n(R)$ is the number of degree $n$ monomials 
in $R$. Obviously, $H_{\IC[t]}(q)$ 
is equal to the above partition function. 
It was explained in \cite[Sect 4]{IKV}, that 
$H_{\IC[t]}(q)$ can be also interpreted as a counting function of states in the Hilbert space $\CH$ of a single 
quantum harmonic oscillator. 

Next suppose ${\sf D}$ has singular points away 
from the boundary. Without any loss of essential 
information, one may assume that ${\sf D}$ has 
only one singular point $p$. Several singular points 
may be treated analogously with no new conceptual 
issues. 

In this case the Chern class of a singular vortex configuration 
admits a splitting, $n+l$, where $n\in \IZ$ is determined as above 
by the trivialization on the boundary, and $l\in \IZ$ is a contribution 
supported at the singular point. A rigorous account of this splitting 
is provided at the end of section (\ref{mathsummary}), where 
it is also shown that $l$ takes finitely many values. In addition, 
one has to specify the multiplicity $m\geq 1$ 
 of the singular vortex at $p$, as discussed above. 
Therefore the partition function will be 
in general of the form 
\be\label{eq:singdiscA}
\sum_{l\geq 0} \sum_{m\geq 1} f_{l,m}(q) 
\ee
where only finitely many terms are nontrivial. 
Note that each term $f_{l,m}(q)$ is a power series in $q$ 
because for fixed values of $(l,m)$ one has to sum over 
all possible boundary trivializations, as in the smooth case. 
More detailed information on the terms $f_{l,m}(q)$
requires a more involved technical analysis, as shown 
for  specific examples   
in section (\ref{toruscurves}). A more immediate 
task at this stage is however to explain 
how the above general reasoning can be 
applied to more general M2-brane 
configurations supported on intersecting curves. 

\subsection{Intersecting M2-brane bound states}
The relevant intersecting curve configurations for large  $N$ duality consist of a 
singular holomorphic 
disc ${\sf D}$ as above meeting a smooth 
$(-1,-1)$ rational  curve $C_0$ at 
the singular point $p$. 
One then has to count bound states of 
$k=1$ open M2-branes wrapping ${\sf D}$ and 
$d$ closed M2-branes wrapping the zero section $C_0$.
As shown in \cite{gauge-gravity,Mtop}, 
 M2-branes wrapping a $(-1,-1)$ curve  
 $C_0$ with multiplicity $d\geq 1$ 
 form bound states only for $d=1$, in which 
case the spectrum consists of one BPS state of spin 0.

In addition, when an  M2-brane wrapping ${\sf D}$ is added 
to the system one can 
form new bound states binding a membrane wrapping $C_0$ 
to the membrane wrapping ${\sf D}$. 
If ${\sf D}$ were smooth, the intersection between the two 
M2-branes would be modeled by a curve with a simple nodal singularity $xy=0$. This configuration can be viewed as a limit 
of a single M2-brane wrapping the smooth curve $xy=\epsilon$ 
as $\epsilon \to 0$. Therefore two intersecting M2-branes 
form in this case a single bound state. 

However in the case of interest here ${\sf D}$ is a singular disk, 
which has local multiplicity $m\geq 1$ at the singular point $p$, 
even though its generic multiplicity $1$.  
  In other words
$m$ counts the number of `points' at $p$.  What this means is that
if we were to consider an annulus which ends on one end on the D2-D0 brane bound
state on one side, and on a transverse
D-brane intersecting the curve at $p$ on the other, $m$ counts the Witten index for it.
Therefore 
a membrane wrapping $C_0$ may bind to the singular membrane 
in $m$ distinct ways, depending on which local branch it is attached 
to. This results into a spectrum of $m$ BPS states in the low energy effective action. More generally, $d$ membranes wrapping 
$C_0$ can bind in 
$\binom{m}{d}$ 
 distinct ways to the singular open membrane, resulting 
in as many BPS particles. In particular, if $d>m$ no irreducible bound state may be constructed.
Therefore the partition function for such configurations 
must take the general form 
\be\label{eq:genform}
\sum_{l\geq 0}\sum_{m\geq 1} \sum_{d=0}^m 
\binom{m}{d}(- Q)^d f_{l,m}(q) = 
\sum_{m\geq 1} f_m(q) (1-Q)^m 
  \ee
where 
\[
f_m(q) = \sum_{l\geq 0} f_{l,m}(q).
\]
Here we used the fact, already seen for the unknot, that
the fermion parity of the M2 brane wrapping ${\bf P}^1$ is -1 leading
to $-Q$ for each such state in the above formula.
Moreover, equation \eqref{eq:openBPSA} 
predicts that 
\[
f_m(q) = {g_m(q)\over 1-q} 
\]
with $g_m(q)$ a polynomial with integral coefficients. 
These predictions will be confirmed by explicit 
computations for plane curves of the form 
\[
x^r-y^s=0
\]
in the next section.

For completeness, it is worth noting that the combinatorial 
factors $\binom{m}{d}$ admit 
a geometric interpretation in the 
a weakly coupled IIA limit mapping M2-branes to D2-branes. Then  
the  massless spectrum of open string stretching between 
a D2-brane on ${\sf D}$ and $d$ D2-branes on $C_0$ 
consists of an $N=2$, $d=4$ hypermultiplet reduced to 
one dimension. The bosonic components are
two  complex scalar fields 
$\phi,\psi$ transforming in the bifundamental representation 
of the D-brane gauge fields and its dual. 
Again, the singular D2-brane has effectively multiplicity 
 $m$ at the singular point even if it is generically of rank one.
Therefore $\phi,\psi$ may be identified with linear 
maps $\phi:\IC^m \to \IC^n$, 
$\psi:\IC^n\to \IC^m$ respectively.
Then the 
F-term equations are simply 
\[
\psi \circ \phi =0, \qquad \phi\circ \psi=0.
\]
This implies that the moduli space of flat directions modulo 
gauge transformations is isomorphic to a moduli space of 
representations of a quiver of the form 
\[
\xymatrix{
\IC^m \ar@/^/[r]^\phi & \IC^n \ar@/^/[l]^\psi \\}
\]
subject to the F-term equations. The stability conditions 
are determined as usual by the D-term equations, 
\[
|\phi|^2 -|\psi|^2 = \xi.
\]
The subtle aspect here is that 
even though the singular D2-brane has multiplicity $m$ at $p$, 
one should only mod out by $U(1)\times U(d)$ gauge transformations since the brane has  
generic of rank 1. Moreover, since the diagonal 
$U(1)$ subgroup acts trivially on 
$\phi,\psi$,  it suffices to 
mod out by $U(d)$ gauge transformations. 

A straightforward 
analysis of the resulting stability condition 
shows that $\phi=0$ and $\psi$ must be surjective if 
$\xi<0$ and $\psi=0$ and $\phi$ must be surjective for 
$\xi>0$. Therefore if $\xi>0$, the moduli space of 
stable representations modulo $U(d)$ gauge transformations 
is isomorphic to the grassmannian $G(m,d)$ of $d$-dimensional 
quotients of $\IC^m$ if $d\leq m$ and 
empty if $d>m$. If $\xi<0$ the moduli space is just a point
if $d\geq m$ and empty if $d<m$. 

In string theory FI term $\xi$ is determined by the expectation 
value of the background fields on $Y$, such as the metric and B-field. The previous paragraph implies that for any choice of background fields such that $\xi>0$ the weakly coupled 
IIA analysis agrees with M-theory considerations. 
Namely, the moduli space of expectation values of open string 
modes is isomorphic to the grassmannian $G(m,d)$ which has 
Euler character $\binom{m}{d}$. This is precisely the number of 
bound states predicted by M-theory arguments.

\subsection{Curves of type $(r,s)$}\label{toruscurves}
Returning to the setup described at the beginning 
of this section, consider a singular curve $C$ of the form
\[
x^r-y^s=0, \qquad \lambda =0.
\]
 in a 
resolved conifold $Y$. 
 Here $(r,s)$ are coprime positive integers and it will 
be assumed that $r>s\geq 1$. 
Note that $C$ has only one 
singular point $p$ given by $x=y=0$, $\lambda=0$. 
The construction of section (\ref{topres}) 
produces a lagrangian cycle  $M\subset Y$
which intersects $C$ along a smooth connected one cycle $\eta$. 
Therefore $C$ is divided into two connected components, the holomorphic disc ${\sf D}$ being 
the component containing the singular point $p$.
Note that ${\sf D}$ 
is preserved by the circle 
action 
\[
(x,y,\zeta) \mapsto (e^{-is\varphi}x, e^{-ir\varphi}y, 
e^{i(r+s)\varphi}\zeta)
\]
which fixes only the singular point $p$. 
This action yields a natural action on the moduli space
of vortices, and Euler character computations 
localize to the fixed point set. As shown at the end of 
section (\ref{mathsummary}), 
the fixed point set in the moduli space of vortices 
is discrete and consists of vortex  configurations 
centered at  the singular 
point $p$.  Since $p$ is away from the boundary of 
${\sf D}$, the localization computation of 
the partition function \eqref{eq:singdiscA} 
yields the same answer as the localization computation
for vortices on the open curve $C$. Therefore for 
computational purposes one may work with stable 
pairs on $C$. This yields an explicit computational algorithm for the terms $f_{l,m}(q)$ 
in  \eqref{eq:singdiscA} which is summarized below. 

The first term $f_{0,1}(q)$ in \eqref{eq:singdiscA} 
represents the contribution of topologically trivial 
gauge field configurations. All terms $f_{0,m}(q)$, 
$m\geq 2$ are obviously zero since the trivial line 
bundle has multiplicity $1$ at $p$. 
Just as in the smooth case, $f_{0,1}(q)$
is given by the Hilbert series of the ring $R_C$ of regular functions on $C$. Since $(r,s)$ are coprime, 
the curve $C$ may be given in parametric form 
as $(x,y)=(t^s,t^r)$. Therefore 
$R_C$ is isomorphic to the subring 
$\IC[t^r,t^s]\subset \IC[t]$ 
spanned polynomials of the form 
$p(t^r,t^r)$ with $p(x,y)$ an 
arbitrary polynomial of two variables.  It will be convenient to identify the set of monomials 
$t^n\in \IC[t^r,t^s]$ with the set of exponents 
$n\in \IZ_{\geq 0}$, which will be denoted by 
$\Lambda(r,s)$.  Note that the complement $\Xi(r,s)= \IZ_{\geq 0}\setminus 
\Lambda(r,s)$ is a finite set. 
Therefore $f_{0,1}(q)$ can be identified with the germ
turn is generated by $x,y$ with weights $s,r$ respectively, modulo a relation of degree $sr$:
\be\label{eq:sindiscB} 
f_{0,1}(q) = \sum_{n\in \Lambda(r,s)} q^n = {(1-q^{rs})\over (1-q^r)(1-q^s)}=
{1\over 1-q} -\sum_{n\in \Xi(r,s)} q^n.
\ee
By comparison with the formula \eqref{eq:smoothdisc}
it follows that the effect of the singularity in the topologically trivial sector is to remove 
the states in the Hilbert space of the harmonic oscillator with quantum numbers $n\in \Xi(p,q)$. 

For concretness, suppose $(r,s)=(4,3)$. Then 
$\Lambda(4,3)$ is the set 
\[
0,\qquad 3, 4, \quad 6,7,\cdots 
\]
and the complement $\Xi(3,4)$ is the finite set 
\[
1,2,\qquad 5.
\]
Therefore in this case 
\[
f_{0,1}(q) = {1\over 1-q} - (q+q^2+q^5) 
= {1-q+q^3 -q^5 + q^6\over 1-q}.
\]

The terms $f_{l,m}(q)$ corresponding to topologically 
nontrivial sectors are constructed in a similar manner
in terms of partial fillings of $\Lambda(r,s)$. 
A partial filling of $\Lambda(r,s)$ is a subset 
\[
\Lambda(r,s) \subseteq \Lambda'(r,s) \subseteq \IZ_{\geq 0} 
\]
with the property that if $\Lambda'(r,s)$ contains 
some $n'\in \Xi(r,s)$, then it must contain all 
its translates $n'+n$ by arbitrary elements 
$n\in \Lambda(r,s)$.  Each partial filling is obtained 
by adding finitely many elements in $\Xi(r,s)$ to 
$\Lambda(r,s)$ subject to this selection rule. 
For example all possible partial 
fillings in the case $(r,s)=(4,3)$ 
are 
\[
\bal 
& \Lambda'(4,3)_{(1)}: 
\qquad {\bf 0},{\underline {\bf 1}},\quad3,4,{\underline 5},6,7,\ldots \\
& \Lambda'(4,3)_{(2)}: 
\qquad {\bf 0},\quad {\underline {\bf 2}},3,4,{\underline 5},6,7,\ldots \\
& \Lambda'(4,3)_{(3)}: 
\qquad {\bf 0},\qquad 3,4,{\underline {\bf 5}},6,7,\ldots \\
& \Lambda'(4,3)_{(4)}: \qquad 
{\bf 0},{\underline {\bf  1}},{\underline {\bf 2}},3,4,{\underline 5},6,7,\ldots 
\eal 
\]
the extra elements being underlined in each case.   What this 
means is the for example for $\Lambda'(4,3)_{(1)}$ the line bundle has one additional
section $s'$ represented by ${\underline {\bf 1}}$, which does not vanish at the origin as we put $x=y=0$.  The
additional element in the ring given by $s'y$ given by ${\underline 5}$, does vanish at the origin, as it vanishes
as we set $y=0$.  Similar considerations apply to the rest.
A disallowed filling is for example 
\[
0,1,\quad3,4,\quad,6,7,\ldots 
\]
since the translation of $1$ by $4$ is $1+4=5$, 
which is missing in the above sequence.   This is consistent with the fact
that we can multiply a section by the holomorphic functions of $x,y$ and still get a section
of the same bundle, and so $5$ should also have been in the sequence of the sections of the line bundle.

Note that 
any 
partial filling $\Lambda'(r,s)$ 
contains a unique finite subset $\Gamma(r,s)$ 
consisting of all elements $n'$ which cannot be 
decomposed as 
\[
n'= n''+n
\]
with $n''\in \Lambda'(r,s)$ and 
$n\in \Lambda(r,s)$, $n\neq 0$. 
Moreover it is easy to show that any element $n'
\in \Lambda'(r,s)$ can be written as $n'=n''+n$ with 
$n''\in \Gamma(r,s)$ and $n\in \Lambda(r,s)$. 
The elements of $\Gamma(r,s)$ will be called the generators of $\Lambda'(r,s)$. In the above example the generators are marked in
each case with boldface characters. 

The first Chern class  $l$ of the vortex corresponding
to $\Lambda'(r,s)$ 
is the number of elements in the complement 
$\Lambda'(r,s)\setminus \Lambda(r,s)$, which is the same as the number of additional
sections we have introduced while
the multiplicity $m$ is the number of generators, which is also the number
of sections which do not vanish at $p$ as we set $x=y=0$.
In the above example, 
\[
l_{(1)} = l_{(2)}=2, \qquad l_{(3)}=1,\qquad 
l_{(4)} =3. 
\]
and 
\[
m_{(1)}=m_{(2)}=m_{(3)}=2, \qquad m_{(4)}=3.
\]
Note that $0$ is always a generator, and never an extra 
element. The pair $(l,m)$ assigned to a partial filling 
$\Lambda'(r,s)$ will be called below the type of the 
partial filling. 

The terms $f_{(l,m)}(q)$ are then obtained 
by summing the Hilbert series of all 
modules associated to partial fillings 
$\Lambda'(r,s)$ of fixed type
$(l,m)$. That is 
\be\label{eq:singdiscC} 
f_{(l,m)}(q) = q^l \sum_{\Lambda'(r,s)\ \mathrm{of\ type}\ (l,m)}  \ \ 
\sum_{n\in \Lambda'(r,s)} q^{n} 
\ee
The factor $q^l$ reflects the fact that all such configurations have first Chern class $l$. 
For $(r,s)=(4,3)$ the resulting contributions are 
\[
\bal
f_{1,2}(q) & = q\sum_{n\in \Lambda'(4,3)_{(3)}} q^n 
= q\bigg({1\over 1-q} -q-q^2\bigg)\\ 
f_{2,2}(q) & = q^2 \sum_{n\in \Lambda'(4,3)_{(1)}} q^n + q^2 \sum_{n\in \Lambda'(4,3)_{(2)}}q^n 
=
\bigg({1\over 1-q} - q^2\bigg) 
 + q^2 \bigg({1\over 1-q} - q\bigg) \\
f_{3,3}(q) & = q^3 
\sum_{n\in \Lambda'(4,3)_{(4)}} q^n = 
{q^3\over 1-q} \\
\eal
\]
all other terms being trivial. Then the coefficients 
$f_m(q)$ in equation \eqref{eq:genform} are 
\[
\bal 
f_1(q) & =  {1-q+q^3 -q^5 + q^6\over 1-q}\\
f_2(q) &  = {q+q^2 -q^3 +q^4
+q^5 \over 1-q}\\
f_3(q) & = {q^3\over 1-q} \\
\eal 
\]

The 
 HOMFLY polynomial of the $(r,s)$ torus knot is given by 
 \be\label{eq:torushomflyA} 
 H_{(r,s)}(q,Q) = \left({Q\over q}\right)^{(r-1)(s-1)/2} 
 {1\over 1-q^r} \sum_{j=0}^{r-1} 
 {q^{sj+(r-1-j)(r-j)/2}\over [j]![r-1-j]! }
 \prod_{i=j+1-r}^j (q^i-Q).
 \ee
  where $[0]!=1$ and $[j]!=(1-q^j)[j-1]!$ for all $j\geq 1$. 
  Then a straightforward computation yields 
\[ 
\sum_{m=1}^3 (1-Q)^m f_m(q) = \bigg({q\over Q}\bigg)^6 
H_{(4,3)}(q,Q),
\]
confirming large $N$ duality for the $(4,3)$ torus knot.
 Note that  \cite[Thm. 19]{hilbert-links} 
 proves the agreement between formula \eqref{eq:torushomflyA} and 
 the 
 stable pair localization computation for all $(r,s)$. The 
 examples considered in this section
  are meant to explain the 
 localization computation in a physical context. 
 
 The next case treated explicitly here is   
  $(r,s)=(2,2k+1)$, $k\geq 1$. Then the HOMFLY polynomial 
  \eqref{eq:torushomflyA}   
 reduces to 
 \be\label{eq:torushomflyB}
 \bal
 H_{(2,2k+1)}(q,Q) & =  \left({Q\over q}\right)^{k} {1-Q\over 1-q} 
 {1-  q^{2k+2} -qQ(1 -q^{2k})\over 1-q^2} \\
 & =  \left({Q\over q}\right)^{k}{1-Q\over 1-q}  
 \bigg[1+(q-Q)\sum_{j=0}^{k-1} q^{2j}\bigg]. 
 \eal
 \ee
  The subset 
 $\Lambda(2k+1,2)\subset \IZ_{\geq 0}$ consists of the following elements 
 \[
 0, \quad 2,\quad \cdots \quad 2k, \ 2k+1, \ 2k+2, \cdots 
 \]
 its  complement $\Xi(2k+1,2)$ being 
 \[
 1,\quad 3,\quad \cdots \quad 2k-1.
 \]
 There are $k+1$ partial fillings 
 $\Lambda'(2k+1,2)_{(j)}$, $0\leq j \leq k+1$ 
 as follows 
 \[
 \bal 
\Lambda'(2k+1,2)_{(0)} & =\Lambda(2k+1,2),\\
\Lambda'(2k+1,2)_{(1)} & =\Lambda(2k+1,2)\cup 
\{1,\ldots, 2k-1\},\\
& \ \, \vdots \\
\Lambda'(2k+1,2)_{(j)} & =\Lambda(2k+1,2)\cup 
\{2j-1,\ldots, 2k-1\},\\
& \ \, \vdots \\
\Lambda'(2k+1,2)_{(k)} & =\Lambda(2k+1,2)\cup 
\{2k-1\}.\\
\eal 
\]
Each $\Lambda'(2k+1,2)_{(j)}$, $1\leq j \leq k$ 
has two generators, $0, 2j-1$ and the complement 
of $\Lambda(2k+1,2)$ contains $k-j+1$ 
elements.
Therefore 
\[
 l_{(j)} = k-j+1, \qquad m_{(j)} = 2
 \]
 for all $1\leq j\leq k$. Obviously, $l_{(0)}=0$, 
 $m_{(0)}=1$.   
 Therefore  
\[
{f_1(q)} = \sum_{n\in \Lambda(2k+1,2)} q^n = 
{1\over 1-q} -\sum_{j=1}^k q^{2j-1} = {1+q^{2k+1}\over 1-q^2} 
\]
\[
\bal
{f_2(q)} &  = \sum_{j=1}^k q^{k-j+1} \bigg({1\over 1-q} - \sum_{i=1}^{j-1} q^{2i-1}\bigg) \\
& = \sum_{j=1}^k q^{k-j+1} {1+q^{2j-1}\over 1-q^2} 
 = {q(1-q^{2k})\over (1-q)(1-q^2)}.\\
\eal 
\]
Note that 
\[
f_1(q) + f_2(q) = {1-q^{2k+2}\over (1-q)(1-q^2)}
\]
Then a straightforward computation yields 
\[
\bal 
\sum_{m=1}^2(1-Q)^mf_m(q) = \bigg({q\over Q}\bigg)^k
H_{(2,2k+1)}(q,Q)
\eal 
\]

\section{A summary for mathematicians}\label{mathsummary}

This section recapitulates the previous, in language perhaps more amenable to
mathematicians.  As before, the goal
is to explain how a conjecture of Oblomkov and the second author 
\cite{hilbert-links} is related to a certain series of string dualities.  
On the one hand, this provides a physics
proof of the conjecture.  On the other, the conjecture was proven (in the mathematical sense) 
for torus knots in \cite{hilbert-links}.  This may then be viewed as confirmatory
evidence for the string dualities which occur in the discussion below.

We recall the conjecture in question.  Let $C$ be a curve in $\IC^2$,
say given by $f(x,y) = 0$.  Assume $C$ passes through the origin.
Then the intersection of $C$ with the boundary
of a small ball around the origin gives a link in the $3$-sphere.  
Note this link has a natural orientation since it bounds a complex variety,
and in fact a natural framing (though we will not use this).  
Recall
that the HOMFLY polynomial is an invariant of links which assigns to a link $L$
a certain rational function $H(L)$ in the variables $q^{\pm 1/2}, Q^{\pm 1/2}$,
characterized by the following skein relation:

\begin{eqnarray}
  \label{eq:skein1}
  Q^{1/2} \, H(\undercrossing) -  Q^{-1/2} \, 
  H(\overcrossing) & = & (q^{1/2} - q^{-1/2})  
  \, H(\smoothing) \\
  Q^{1/2} - Q^{-1/2} & = & (q^{1/2}-q^{-1/2})H(\bigcirc)
\end{eqnarray}

On $C$, we consider the moduli space $C^{[n]}$ parameterizing pairs $(F,s)$,
where $F$ is a torsion free sheaf, $s$ is a section $s:\CO_C \to F$, 
and $\dim F/s\CO_C = n$.  Note that in \cite{hilbert-links} 
the same notation was used
for the Hilbert scheme of $n$ points on $C$; as shown in \cite{stabpairs-III} 
these spaces are isomorphic for Gorenstein (and in particular planar) 
curves $C$.  By $C^{[n]}_0$ we
denote the space of such pairs in which the section vanishes only at
the origin.  By $C^{[n];m}_0$ we denote the locus where 
$m = \dim_\IC F / (x,y)F$.  Let 
$\mu = \dim \IC[[x,y]]/(\partial_x f, \partial_y f)$
be the Milnor number of the singular point.  We can now state:

\begin{conj} \label{conj:OS} \cite{hilbert-links}.
  \[H(\mbox{the link of }C) = \left(\frac{Q}{q}\right)^{\mu - 1} \!\!\!\!\!
  \sum_{n,m} q^n (1-Q)^m \chi(C^{[n];m}_0)\]
\end{conj}

One contribution of the present article is to explain how a certain
sequence of string dualities connects the left
of the conjecture to the right.  The HOMFLY polynomial enters physics 
through Witten's observation \cite{wit1} that it computes
the expectation value of the knot viewed as a Wilson line in the Chern-Simons
gauge theory on the three sphere.  Here the gauge group is $U(N)$, and its
holonomy around the knot is traced in the fundamental representation.  
Witten later \cite{CSstring} explained that this theory was equivalent to the type IIA topological 
string theory
on $T^* S^3$, with $N$ lagrangian D-branes on $S^3$.   The Wilson loop
expectations (and hence the HOMFLY polynomial) are reproduced 
by introducing \cite{knot-top} the conormal bundle of the knot
as a lagrangian brane $N^* K \subset T^* S^3$ and
counting open strings with one end on $N^* K$ and the other on $S^3$.

As $N$ grows large one may take the view \cite{gauge-gravity} that 
the $S^3$ shrinks and the  space $X= T^* S^3$ is ultimately replaced by 
the small resolution of the conifold, i.e., by the total space $Y$ of 
the bundle $\CO(-1) \oplus \CO(-1)$ over $\IP^1$. 
The D-branes on $S^3$ vanish along with the $S^3$.
Attempting to follow $N^* K$ through the conifold transition is problematic, since
it meets the collapsing $S^3$.  Instead, it is better to first deform it
off the $S^3$.  In the case that the knot $K$ arose as an algebraic knot, 
we have explicitly constructed such a deformation in section \ref{lagalgknots},
and followed it through the conifold transition.  
The essential feature of the resulting $L_K \subset Y$ 
is that it intersects the fiber over infinity in a single circle; the unique
holomorphic curve passing through this circle is the singular curve itself.  
Conjecturally this curve is in fact the {\em only} irreducible curve with boundary
on $Y$; this is proven in the next section in the case where $K$ arose from
the curve $x^r = y^s$. 

At this stage we see that the HOMFLY polynomial of $K$ should be computed by 
counting curves in $Y$ with boundary on $L_K$.  The mathematical foundations of open 
Gromov-Witten theory are not presently available, but nonetheless for torus
knots it is possible to describe the inevitable result of torus localization 
of the virtual class as in \cite{KL}.  This is done in the subsequent section, 
yielding agreement with known formulas for the HOMFLY polynomial in these cases.

According to \cite{Mtop, knot-top,LMV,framedknots}, we may
lift to $M$-theory.  Indeed, the topological string computes certain supersymmetric
quantities in the full type IIA string theory on
$Y \times \IR^{3,1}$, which in turn 
is viewed as a limit of M-theory on $Y \times \IR^{4,1}$. 
The variables work out so that the coefficient of 
$q^r Q^s$ in $(1-q) H$ counts certain M2-branes.  More precisely, 
one considers an M5-brane $L \times \IR^{2,1}$ for some 
$\IR^{2,1} \subset \IR^{4,1}$.  Note that this brane breaks the symmetry 
group of the $\IR^{4,1}$ to $Spin(2,1) \times Spin(2)$; we will only be interested in the 
$Spin(2) \times Spin(2)$ action.  This group acts on all spaces of BPS states of M2 branes
with boundary on this M5 brane, so
these acquire a bigrading by the characters $q,t$ of the group.  (Here $t$ is 
the character of the rotation transverse to the Lagrangian.) 
The M2 brane states also carry two additional gradings, corresponding to the 
class of the brane in $H_2(Y,L) = \IZ \IP^1 \oplus \IZ {\sf D}$, where ${\sf D}$
is the class of the singular disc bounding the lagrangian.  Writing
$N_{d', d, \sigma, j}$ for the space of states of character $q^\sigma t^j$
and homology class $d \IP^1 + d' {\sf D}$, the prediction of the 
above dualities is that the HOMFLY polynomial is (upto an appropriate $q^\cdot Q^\cdot$) 
given by 
\be \label{eq:homflyMm}
H(q,Q) = \frac{1}{1-q} \sum N_{1, d, \sigma, j} Q^d q^\sigma (-1)^{2j+1}
\ee

The geometry of $M2$-branes is not well enough understood that the 
$N_{1,d,\sigma,j}$ may be computed directly.  However, according to 
\cite{Mtop, knot-top,LMV,framedknots} the above index
may be computed in {\em a different} type IIA limit of the M-theory, in
which one of the dimensions of the $\IR^4$ is compactified on a circle, 
and the different momenta modes  around this circle are converted into bound states
of D0-branes to D2-branes.  The D2-branes must of course still
have boundary along the Lagrangian. 
(The reason one is free to compute in any limit one
likes is that all the states in question are BPS.)  In the large volume limit, 
the space of D2-D0 branes is understood to be mathematically modelled
by the space of stable pairs \cite{spinning, vortex, stabpairs-I}, and the index 
above is just its (appropriately weighted) Euler characteristic.  
These spaces {\em are not} 
identical to the spaces $C^{[n];m}_0$.  In an upcoming paper of the first
author, they will be shown to be related 
by wall crossing, and as a consequence conjecture \ref{conj:OS} 
will be deduced from \eqref{eq:homflyMm}.

Here we argue instead at the level of M-theory.
Since the coefficient of $Q^d$ is counting bound states formed by
one M2-brane wrapping
{\sf D} and $d$ M2 branes wrapping $\IP^1$, we must analyse how this
configuration may occur.  It is already known that M2 branes wrapping the
$\IP^1$ may not bind to each other; this for instance follows by running
through the above series of dualities for the partition function of Chern-Simons
theory itself; the consequence being that in the absence of Lagrangians 
there is a unique BPS state consisting of a single M2 brane wrapping the $\IP^1$. 
Thus each of the M2 branes wrapping the $\IP^1$ must bind to {\sf D}.  

Let $g_m(q)$ be the generating polynomial of which the coefficient of $q^n$ is
the number of spin $n$ M2 branes wrapping {\sf D} which
can bind up to $m$ M2 branes on $\IP^1$.  Then, since the M2 branes on
$\IP^1$ are indistinguishable and have fermion number $-1$, the generating
polynomial of bound states of such a brane to some number of branes wrapping
$\IP^1$ is just $(1-Q)^m g_m(q)$.  Thus \eqref{eq:homflyMm} can be rewritten as

\be 
H(q,Q) = \frac{1}{1-q} \sum_m g_m(q) (1-Q)^m
\ee

To relate this to conjecture \ref{conj:OS}, it remains only to explain why
\be
  \frac{g_m(q)}{1-q} = \sum_n q^n \chi(C^{[n];m}_0)
\ee 
i.e., why the BPS M2-branes which can bind exactly $m$ M2 branes on $\IP^1$
may be computed by stable pairs $s:\CO_C \to F$ where $\dim F/(x,y)F= m$. 
As mentioned above, the relation to stable pairs and the appearance of the 
$1/(1-q)$ is standard: the M2 brane has momentum modes around the circle
which become D2-D0 bound states, and the particular index being computed
becomes the Euler number of the stable pairs space.
One might worry about the appropriate 
boundary conditions for the sheaf and the section. 
But whatever boundary conditions
are chosen we will surely want all zeroes of the section $s$ to lie
in the connected component $\sf D$ of $C \setminus L$ containing the origin. 
Assuming we choose $L$ sufficiently near the origin that $\sf D$ is 
contractible, then the space of such pairs contracts to $C^{[n]}_0$.  

The essential thing to explain is what
binding to $m$ branes on $\IP^1$ has to do with the number of generators 
$\dim F/(x,y)F= m$. To count the number of ways an M2 brane on $\IP^1$ may bind to 
a given M2 brane on {\sf D}, we first pass to the IIA theory and compute instead
the number of ways the D2 brane on $\IP^1$ may bind to a D2 brane $F$.  
The heuristic given in the previous section is that the virtual 
number of points on $F$ at the origin which are available for the 
branes on $\IP^1$ is just $\dim F/(x,y)F$.  To elaborate on this slightly,
the ``open strings are Ext'' philosophy here specializes to the statement
that the space of open strings from $F$ to a brane $\CO_{\IP^1}$ 
is $\mathrm{Ext}^1_Y(F, \CO_{\IP^1})$.  This immediately localizes
to the intersection of $\IP^1$ and ${\sf D}$; since this is a point the local
to global spectral sequence collapses and we are reduced to computing Ext 
of modules in the complete local ring.  Let us give coordinate $z$ to the 
$\IP^1$ direction; then we are computing
\[\mathrm{Ext}^1_{\IC[[x,y,z]]}(F,\IC[[z]]) = 
 \mathrm{Hom}_{\IC[[x,y]]}(F, \IC) = (F/(x,y)F)^\vee
\]
One can also compute open strings in the other direction,
$\mathrm{Ext}^1_Y(\CO_{\IP^1},F)$; the result is that 
this canonically parameterizes the nontrivial 
syzygies of the completion of $F$ as a module over $\IC[x,y]$.  Because
the curve is planar, this space has the same dimension as that
parameterizing the generators. 

\vspace{4mm}
We make a quick note about the sample computations
in the previous section of the right hand side of conjecture \ref{conj:OS}. 
The main point is a certain invariant introduced there did not in fact
require the existence of a torus action. 
Let $R$ be the complete local ring of
the singularity; then the Jacobian factor parameterizes $R$-modules $M$ such that
$\IC[[t]] \supset M \supset R$.  Note that given a stable pair $(M,s)$ one has an abstract
isomorphism $M \otimes_R \IC((t)) = \IC((t))$; requiring $s \mapsto 1$ fixes the isomorphism.  
In other words a stable pair with quotient supported at $0$ is equivalent data to 
a rank one $R$-submodule of $\IC((t))$.  Let $M$ be such a module, then 
$M \IC[[t]] = t^{-k} \IC[[t]]$ for some $k$, and $\IC[[t]] \subset t^k M \subset R$. 
Thus there is a map from the space of stable pairs to the Jacobian factor.  The fiber over
some module $M$ is just the set of elements in $M$, up to constant multiple.  It is straightforward
to see that the space of elements with leading term $t^a$ is a vector space and hence has Euler
characteristic one.  On the other hand $\dim_\IC M/t^a R  = a + \dim_\IC M/R$.  Thus the contribution
of $M$ to the Euler numbers of pairs spaces is $q^{\dim_\IC M/R} \CH_M(q)$, where $\CH_M(q)$ is
the Hilbert function of $M$.  Note that $\dim_\IC M/R$ is the ``l(M)'' of the previous section. 
Let us also write $m(M)$ for the number of generators, and $\CJ^{l;m}$ for 
the locus in the Jacobian factor of modules with $m$ generators.  Then the $f_m$ of
the previous section are 

\[f_{m}(q) = \sum_n q^n \chi(C^{[n];m}_0) = \int_{\CJ^{;m}} q^{\ell(M)} \CH_M(q) d \chi(M) \]

The integral is with respect to Euler characteristic, and has the meaning that we sum 
possible $\CH_M(q)$ weighted by the Euler characteristic of the locus of modules with this
Hilbert series.  One may as desired further stratify by $l$, and introduce

\[f_{l;m} = \int_{\CJ^{l;m}} \CH_M(q) d \chi(M) \]

in order to write

\[\sum_{m,n} q^n (1-Q)^m \chi(C^{[n];m}_0) = \sum_{l,m} q^l (1-Q)^m f_{l,m}(q) \]

\section{Large $N$ duality and topological 
amplitudes for torus knots}\label{torusknots}

The main goal of this section is to generalize the 
large $N$ duality results for the unknot 
reviewed in section (\ref{unknot})  to arbitrary $(s,r)$ 
torus knots. The open topological {\bf A}-model amplitudes 
for lagrangian cycles associated to torus knots 
 will be explicitly computed on both sides of the transition employing an equivariant virtual localization approach analogous to \cite{KL}. 
 Note that the mirror topological {\bf B}-model has been studied 
 in \cite{torusknotsmirror}, reproducing the HOMFLY polynomials 
 via a matrix model approach.

The first task, carried out in section (\ref{pqlagcycles}),
is to write down an explicit analytic presentation 
the lagrangian cycles $L_\mu \subset X_\mu$, 
 constructed in section 
(\ref{lagtrans}) 
and show  that they are preserved by 
the circle action 
\be\label{eq:circactA}
(x,y,z,w) \mapsto (e^{is\varphi}x, e^{ir\varphi}y, 
e^{-is\varphi}z, e^{-ir\varphi}w).
\ee
Then it shown in section (\ref{opdef}) that 
on the deformation side the open string instanton corrections to Chern-Simons 
theory are encoded in a formula of the form 
\eqref{eq:opinstA}. 
The next section (\ref{GWres}) contains the 
virtual localization computation  of Gromov-Witten 
invariants on the resolution $Y$ 
with lagrangian boundary conditions 
on $M_\epsilon$. 
In particular it is shown that $M_\epsilon$ is preserved 
by the circle induced by \eqref{eq:circactA} 
and the disc ${\sf D}_\epsilon$ obtained in section 
(\ref{topres}) is the only circle invariant 
Riemann surface 
in $Y$ with boundary in $M_\epsilon$. 
The tangent-obstruction complex for circle 
invariant stable maps 
with lagrangian boundary conditions on $M_\epsilon$ 
is derived by linearizing the defining equations 
of $M_\epsilon$ near the one-cycle $\eta_\epsilon = \partial 
{\sf D}_\epsilon$. A by-product of this computation 
is an argument proving that the disc ${\sf D}_\epsilon$
is rigid as a Riemann surface with boundary on $M_\epsilon$, even in the absence of the circle action.  
The final details of the localization 
computation are given in section (\ref{comphomfly}). 
The main result is that the winding number one  {\bf A}-model partition function of the lagrangian cycle $M_\epsilon$ is in agreement with the HOMFLY polynomial 
of the $(s,r)$ torus knot up to an overall sign depending 
on orientations.  The proof is  
essentially an open {\bf A}-model reflection of 
the Chern-Simons $S$-matrix formula  \cite{toruslinks, torusknotsmirror} relating the HOMFLY polynomial 
of  a torus knot to  the colored invariants of the unknot 
\cite{toruslinks, torusknotsmirror}.

\subsection{Lagrangian cycles for torus knots}\label{pqlagcycles}
 Lagrangian cycles for torus knots are obtained as a special 
 case of the  construction explained in section 
 (\ref{lagalgknots}) for general algebraic knots. 
 Consider the family of the curves $Z_\mu \subset X_\mu$ 
\be\label{eq:rscurveA}
Z_\mu:\qquad x^r-\alpha^r y^s=0, \qquad z^r-\alpha^r (-w)^s =0
\ee
where $(r,s)$ are coprime integers with $r>s\geq 1$ or $r=s=1$, 
and 
$\alpha \in \IR\setminus \{0\}$ is a fixed nonzero 
real number. As explained in section (\ref{lagalgknots}), the specialization 
of $Z_\mu$ at $\mu=0$ has $r-s+1$ connected components 
classified by the distinct roots of the equation 
$\eta^r-\alpha^{2r}(-\eta)^s=0$. 
The component corresponding to $\eta=0$ is a union of two 
irreducible components $C^\pm$ given by 
\[
x^r-\alpha^r y^s=0, \qquad z=w=0,
\]
respectively 
\[
z^r-\alpha^r (-y)^s =0, \qquad x=y=0.
\]
The remaining $r-s$ are disjoint smooth components isomorphic to 
$\IC^\times$. Let $\gamma^+:S^1\to X$ be a parametric presentation 
of the intersection $\phi_0(C^+\setminus \{0\})\cap P_a$, 
where $P_a\subset X$ is the sphere bundle $|{\vec v}|=a$. 
Its inverse image $\phi_0^{-1}\circ \gamma^+:S^1\to X_0$ 
has a parametric presentation of the form  
\be\label{eq:intcompB}
\bal 
(x,y,z,w) = (\alpha b_1^se^{is\theta}, b_1^re^{ir\theta}, 0, 0) 
\eal 
\ee
where $b_1$ must be a solution of the equation 
\be\label{eq:boundaryBA}
b_1^{2r}+ \alpha^2 b_1^{2s} = 4a.
\ee
Some elementary real analysis shows that this 
equation has a unique positive real solution 
for any fixed $\alpha\neq 0, a>0$. 
Then the
 construction of section (\ref{lagsect}) then yields a lagrangian cycle 
$L_{\gamma^+}\subset X$. It will be checked below that 
$L_{\gamma^+}$ does not intersect the zero section, hence 
its inverse image $L_0=\phi_0^{-1}(L_{\gamma^+})$ is a 
lagrangian cycle on $X_0$ supported away from the conifold singularity.

For $\mu>0$ the connected components of $Z_\mu$ are in one-to-one 
correspondence with distinct roots of the equation $\eta^r-\alpha^{2r}(\mu-\eta)^s=0$. Each such component is given by 
\[
(x,y,z,w) = (t^s,t^r,\eta t^{-s}, (\mu-\eta)t^{-r}).
\]
In particular for sufficiently small $\mu>0$ there exists a continuous 
family 
$\eta(\mu)$ of roots specializing to $\eta=0$ at $\mu=0$. 
Let $C_\mu\subset X_\mu$ be the corresponding components of 
$Z_\mu$. Each connected component of the 
intersection $C_\mu\cap \phi_\mu^{-1}(P_a)$
must be an orbit of the circle 
action \eqref{eq:circactA} since both $C_\mu$, 
$\phi_\mu^{-1}(P_a)\subset X_\mu $ are invariant cycles.
Taking into account the parametric presentation of $C_\mu$, each intersection component must be of the form 
\be\label{eq:intcomp}
\bal 
(x,y,z,w) = (\alpha b_1^se^{is\theta}, b_1^re^{ir\theta}, 
\alpha b_2^se^{-is\theta}, 
-b_2^re^{-ir\theta}) 
\eal 
\ee
with $\theta$ an angular coordinate on $S^1$ and $b_1,b_2\in \IR_{>0}$. 
The  parameters $b_1,b_2$ must satisfy the condition
\be\label{eq:boundaryA} 
\alpha^2 (b_1^s+b_2^s)^2 + (b_1^r+b_2^r)^2 = 2(\mu +\sqrt{\mu^2 + 4a^2})
\ee
which follows from the defining equation of 
$\phi_{\mu}^{-1}(P_a)\subset X_\mu$, 
and 
\be\label{eq:boundaryB} 
\alpha^2 (b_1b_2)^s+(b_1b_2)^r= \mu,
\ee
which follows from $xz-yw=\mu$. 
By continuity,
 for sufficiently small $\mu>0$ the intersection of 
$\phi_\mu(C_\mu)$ with $P_a$ 
consists of 
two connected one-cycles $\gamma_\mu^\pm$ 
conjugate under the 
antiholomorphic involution $\tau_\mu$ defined in \eqref{eq:antiholinv}.
Moreover $\gamma_\mu^+$ specializes to the 
cycle $\gamma^+$ constructed above, while $\gamma_\mu^-$ 
specializes to its conjugate, which is the intersection of 
$\phi_0(C^-\setminus\{0\})$ with $P_a$. 
In fact this picture can be confirmed by detailed analytic 
computations which will be omitted in the interest of brevity.

Applying the 
construction of section (\ref{lagalgknots}) 
to the one-cycles $\gamma^+_\mu:S^1\to X$ yields a family of 
lagrangian cycles $L_{\gamma^+_\mu}\subset X$. 
The lagrangian 
cycles $L_\mu\subset X_\mu$ are  the inverse 
images, $L_\mu=\phi_\mu^{-1}({L_{\gamma^+_\mu}})$, via the symplectomorphisms $\phi_\mu:X_\mu\to X$. 

The next task is to check that the lagrangian cycles 
$L_\mu$, $L_0$ are invariant under 
the circle action \eqref{eq:circactA} 
and do not intersect the zero section. 
Since the arguments are very similar, it suffices to 
present the details in one case only, say $L_\mu$. 
The explicit 
form of the circle action on $X$ is  
\be\label{eq:circactB}
\bal 
& \left[\begin{array}{c} u_1\\ u_2 \end{array}\right]
\mapsto R(s\varphi)  \left[\begin{array}{c} u_1\\ u_2 
\end{array}\right], \qquad 
\left[\begin{array}{c} u_3\\ u_4 \end{array}\right]
\mapsto R(r\varphi)  \left[\begin{array}{c} u_3\\ u_4 
\end{array}\right]\\
& \left[\begin{array}{c} v_1\\ v_2 \end{array}\right]
\mapsto R(s\varphi)  \left[\begin{array}{c} v_1\\ v_2 
\end{array}\right], \qquad \,
\left[\begin{array}{c} v_3\\ v_4 \end{array}\right]
\mapsto R(r\varphi)  \left[\begin{array}{c} v_3\\ v_4 
\end{array}\right]\\
\eal
\ee
where 
\[
R(\varphi) = \left[\begin{array}{cc}
\cos(\varphi) & -\sin(\varphi) \\ \sin(\varphi) & \cos(\varphi) \\
\end{array}\right].
\]
According to equations \eqref{eq:lagcycleA}, 
the defining equation of $L_{\gamma_\mu^+}$ is 
\[
{\vec u} = {\vec f}(\theta),\qquad 
{\dot {\vec f}}(\theta)\cdot ({\vec v} - {\vec g}(\theta))=0, 
\]
where the functions ${\vec f}(\theta), {\vec g}(\theta)$ are determined by equation 
\eqref{eq:intcomp}. One then  finds
\be\label{eq:toruslagA}
\bal
& \left[\begin{array}{c} 
f_1(\theta) \\ f_2(\theta) \\ \end{array}\right] =\alpha {b_1^s+b_2^s\over {{c}}} R(s\theta)
\left[\begin{array}{c} 
1 \\ 0 \\ \end{array}\right] \qquad 
\left[\begin{array}{c} 
f_3(\theta) \\ f_4(\theta) \\ \end{array}\right] = -{b_1^r+b_2^r\over {{c}}} R(r\theta)
\left[\begin{array}{c} 
1 \\ 0 \\ \end{array}\right] 
\eal 
\ee
and 
\be\label{eq:toruslagB}
\bal
& \left[\begin{array}{c} 
g_1(\theta) \\ g_2(\theta) \\ \end{array}\right] = - \alpha 
{b_1^s-b_2^s\over 4}{c} R(s\theta)
\left[\begin{array}{c} 
0 \\ 1 \\ \end{array}\right] \qquad 
\left[\begin{array}{c} 
g_3(\theta) \\ g_4(\theta) \\ \end{array}\right] = {b_1^r-b_2^r\over 4}{c} R(r\theta)
\left[\begin{array}{c} 
0 \\ 1 \\ \end{array}\right]
\eal 
\ee
with $c =\sqrt{2(\mu + \sqrt{\mu^2 + 4a^2})}$. Then it is straightforward to check that 
$L_{\gamma_\mu^+}$ is preserved by the torus action using the 
elementary identity $R({\varphi})R({\varphi'})=R({\varphi}+{\varphi}')$.
In fact $L_{\gamma_\mu^+}$ admits a parameterization of the form 
 \[
 {\vec u}= {\vec f}(\theta), \qquad  
  \left[\begin{array}{c} v_1 \\ v_2 \end{array}
 \right] = R(s\theta) 
  \left[\begin{array}{c} v_{10} \\ v_{20} \end{array}
 \right] \qquad 
 \left[\begin{array}{c}v_3 \\ v_4 \end{array}
 \right] = R(r\theta) 
  \left[\begin{array}{c} v_{30} \\ v_{40} \end{array}
 \right],
 \]
 where $(v_{10}, \ldots, v_{40})$ are real parameters satisfying 
 \be\label{eq:toruslagC}
 \bal 
 \alpha(b_1^s+b_2^s)v_{10} - (b_1^r+b_2^r) v_{30} & =0\\
  s\alpha(b_1^s+b_2^s) v_{20} - r(b_1^r+b_2^r) v_{40} & = 
  -{c\over 4} 
 \big[\alpha s(b_1^{2s}-b_2^{2s}) + r (b_1^{2r}-b_2^{2r}) \big].\\
 \eal
  \ee
  The first equation in \eqref{eq:toruslagC} 
  follows from the defining equation  ${\vec u}\cdot {\vec v}=0$ of $X$, 
  and the second from the equation 
 ${\dot{\vec f}}(\theta)\cdot ({\vec v} -{\vec g}(\theta))=0$.
  Note that equations 
 \eqref{eq:toruslagC} define a real 2-plane in the fiber of 
 $T^*S^3$ over the point ${\vec u}_{0}={\vec f}(0)$.  
 The points in this plane are in one-to-one correspondence with 
 orbits of the circle action on the lagrangian cycle.
 
Note also that the intersection of $L_{\gamma_\mu^+}$ with the zero section ${\vec v}=0$ is determined by the equation 
\[
 {\dot {\vec f}}(\theta)\cdot  {\vec g}(\theta)=0
\]
which yields 
\be\label{eq:interszerosect} 
s\alpha^2 (b_1^{2s}-b_2^{2s}) + r(b_1^{2r}-b_2^{2r}) =0. 
\ee
Since $b_1,b_2$ satisfy simultaneously 
equations \eqref{eq:boundaryA}-\eqref{eq:boundaryB}, 
equation \eqref{eq:interszerosect} 
will have no solutions for generic values of $\mu,a>0$.
Therefore in the generic case, 
this intersection with the zero section is empty.

\subsection{Open string {\bf A}-model on the deformation}\label{opdef}
Now consider an open {\bf A}-model with target space $X_\mu$ and
lagrangian branes on  the lagrangian cycles 
$ L_\mu,S_\mu$, where $L_\mu$ is defined in
equations \eqref{eq:toruslagA}, \eqref{eq:toruslagB} 
and $S_\mu\simeq S^3$ is the fixed point set of the antiholomorphic 
involution \eqref{eq:antiholinv} on $X_\mu$. 
Note that both cycles are preserved by the circle 
action \eqref{eq:circactA}. 
Moreover, 
$L_\mu$ intersects an irreducible component of the 
curve \eqref{eq:rscurveA} 
along an orbit \eqref{eq:intcomp} of the $S^1$-action. 
Then it follows that the 
holomorphic 
cylinder ${\sf C}_\mu$ given by 
\be\label{eq:rsholcylinder}
(x,y,z,w) = (\alpha b_1^st^s, b_1^rt^r, \alpha 
b_2^st^{-s}, -b_2^rt^{-r}),
\ee
with 
\[
\sqrt{{b_2\over b_1}} \leq |t|\ \leq 1,
\]
has boundary components on $S_\mu$, $L_\mu$ respectively. 
Obviously, ${\sf C}_\mu$ is preserved by the circle 
action \eqref{eq:circactA}. 
Set $\alpha=1$ in the following. 

Equation \eqref{eq:rsholcylinder} describes a 
circle invariant genus zero stable map 
to $X_\mu$ with two boundary components 
mapped to   $S_\mu,L_\mu$. According to 
\cite{CSstring} such instantons are expected to generate 
Wilson loop corrections to the Chern-Simons action. 
If ${\sf C}_\mu$ is the only torus invariant holomorphic 
cylinder in $X_\mu$ with boundary components on 
$L_\mu$, $S_\mu$, 
 these corrections can be easily evaluated 
by a virtual localization computation analogous to 
\cite{KL}. Such a computation has been carried out 
for example in \cite{geom-delpezzo} in a similar context. 
As required in section (\ref{largeNunknot}), 
the final formula for the instanton series is of the form 
\be\label{eq:opinstC} 
Z_{op-inst}(g_s,{\sf q}) =\sum_{n\geq 1} 
{e^{-{t}_{\sf C}} \over n} \mathrm{Tr}(U^n) 
\mathrm{Tr}(V^n) 
\ee
where $U,V$ are the holonomy of the gauge fields 
on $S_\mu,L_\mu$ about the boundary components of 
${\sf C}_\mu$, and $t_{\sf C}$ is the symplectic 
area of ${\sf C}_\mu$. 

It remains to show that ${\sf C}_\mu$ is indeed the unique 
torus invariant cylinder in $X_\mu$ with boundary components 
in $S_\mu,L_\mu$ respectively. The argument is analogous 
with \cite{geom-open}, although more technically involved 
since the present torus action allows continuous families 
of invariant curves on $X_\mu$. The main steps will be
summarized below omitting many computational details. 

   First note that 
   any invariant map $\IC^\times \to X_\mu$  
 must be of the form 
 \be\label{eq:invcurveA}
t\mapsto (x,y,z,w)=(\alpha_1t^s,\alpha_2t^r,\alpha_3t^{-s},-\alpha_4t^{-r})
\ee
where 
$\alpha_1,\ldots, \alpha_4$ are constant parameters 
satisfying 
\be\label{eq:alphamurel}
\alpha_1\alpha_3+\alpha_2\alpha_4=\mu.
\ee
Let $C_{{\vec \alpha}}\subset X_\mu$, ${\vec \alpha} =(\alpha_1, 
\ldots, \alpha_4)$ denote the image of this map. 
Since $L_\mu,S_\mu$ are preserved by the circle action, any connected of the intersection of $C_{\vec \alpha}$ with 
the lagrangian cycles must be an orbit of the form 
$t=\rho e^{i\theta}$, with $\rho\in \IR_{>0}$. 
Then equations \eqref{eq:toruslagA}-\eqref{eq:toruslagC} for 
 for $L_\mu$ imply the following conditions 
 \be\label{eq:interscondA} 
 \rho_L^s \alpha_1 + \rho_L^{-s} \balpha_3 = {A\over c} (b_1^s+b_2^s), \qquad 
 \rho_L^r \alpha_2 + \rho_L^{-r} \balpha_4 = {A\over c} 
 (b_1^r+b_2^r).
 \ee
 where $\rho_L$ is the radius of a 
 component of the intersection $C_{\vec \alpha} \cap L_\mu$ 
 and 
 $A=\sqrt{(|\alpha_1\rho_L^s+\balpha_3 \rho_L^{-s}|^2 +
|\alpha_2\rho_L^r+\balpha_4\rho_L^{-r}|^2)/2}$. 
At the same time any connected component 
of the intersection $C_{\vec \alpha} \cap L_\mu$
must satisfy 
\be\label{eq:interstaufixedB}
\alpha_1\rho_S^{2s} =\balpha_3, \qquad 
\alpha_2\rho_S^{2r} = \balpha_4,
\ee
where $\rho_S$ denotes again the radius of the orbit. 
Equations \eqref{eq:interscondA}, 
\eqref{eq:interstaufixedB} imply that $\alpha_i$, $i=1,\ldots,4$ must be non-zero real numbers, if ${C}_{\vec \alpha}$ 
intersects both $S_\mu,L_\mu$ nontrivially. For example, 
if $\alpha_1=0$, it follows easily that all the remaining 
coefficients $\alpha_2,\ldots, \alpha_4$ must be also trivial, 
which contradicts relation \eqref{eq:alphamurel}. 
Then by a reparametrization of the domain, the 
 map \eqref{eq:invcurveA} can be set in the form 
\be\label{eq:invcurveC}
t\to (\alpha \beta_1^s t^s, \beta_2^r t^r, \alpha \beta_2^s t^{-s}, 
-\beta_2^r t^{-r}) 
\ee
with $\beta_1,\beta_2\in \IR_{>0}$, 
$\alpha\in \IR\setminus\{0\}$, 
the intersections with $S_\mu$, 
$L_\mu$ 
being given by 
\[
|t|=\sqrt{\beta_2\over \beta_1}, \qquad |t|=1
\]
respectively.
Using again equations \eqref{eq:toruslagA}-\eqref{eq:toruslagC} for 
 for $L_\mu$, one finds the following nonempty intersection 
 conditions 
  \be\label{eq:interscondG} 
{\alpha\over c'} (\beta_1^s+\beta_2^s) = {1\over c}(b_1^s+b_2^s),
\qquad 
{1\over c'}(\beta_1^r+\beta_2^r) = {1\over c}(b_1^r+b_2^r).
\ee
\be\label{eq:interscondH} 
-{\alpha c'\over 4} (\beta_1^s-\beta_2^s) = v_{20},
\qquad 
{ c'\over 4}(\beta_1^r-\beta_2^r) = v_{40}.
\ee
\be\label{eq:interscondI} 
\bal 
& {s}(b_1^s+b_2^s)v_{20} 
-{r}(b_1^r+b_2^r) v_{40} = -{c\over 4}
\big[ s(b_1^{2s}-b_2^{2s}) + r(b_1^{2r}-b_2^{2r}) \big]
\\
& \qquad \qquad \qquad 
v_{10}=v_{30}=0,\qquad v_{20}^2 + v_{40}^2 = a'^2.\\
\eal
\ee
where ${\vec v}_0$ is a vector in the plane \eqref{eq:toruslagC} parametrizing a common
circle orbit of $L_\mu$ and $C_{\vec \alpha}$, 
and 
\[
a'=\sqrt{a^2+|{\vec v}_0|^2}, \qquad
c' =\sqrt{2(\mu + \sqrt{\mu^2 + 4a'^2})}.
\] 
Recall that the coefficients $b_1,b_2$ are given functions 
of $(\mu,a)$ determined by equations 
\eqref{eq:boundaryA}-\eqref{eq:boundaryB}, as 
explained in section (\ref{pqlagcycles}). 

Next note that it suffices to show that the orbit parametrized by the 
vector ${\vec v}_0$ coincides with the boundary of the built in 
cylinder ${\sf C}_\mu$, since then the two cylinders must coincide by holomorphy. This follows by an elementary but 
fairly tedious computation in real analysis. The strategy is
to solve for $(v_{20}, v_{40})$ in equations 
\eqref{eq:interscondI} and substitute the solutions in equations 
\eqref{eq:interscondG}-\eqref{eq:interscondH}. Then one 
solves for $(\alpha\beta_1^s, \alpha\beta_2^s)$ respectively 
$(\beta_1^r,\beta_2^r)$ in the resulting equations 
imposing at the same time the positivity conditions $\beta_1, \beta_2>0$. Note that this will yield a priori independent expressions of the form 
\be\label{eq:betasolA}
\beta_i^r = F_i(c',b_1,b_2),\qquad 
\alpha \beta_i^s = G_i(c',b_1,b_2),
\ee
$i=1,2$, where $F_{i}(c',a,\mu)$, $G_i(c',a,\mu)$, $i=1,2$ 
are explicit functions of $(c',a,\mu)$. 
Moreover the expressions \eqref{eq:betasolA} 
must satisfy the obvious compatibility condition
\be\label{eq:matchingA} 
F_1(c',b_1,b_2)^s G_2(c',b_1,b_2)^r = 
F_2(c',b_1,b_2)^s G_1(c',b_1,b_2)^r. 
\ee
A straightforward but fairly long computation 
shows that the matching condition \eqref{eq:matchingA}
is equivalent to
\[
F_+(\eta)^sG_+(\eta)^r = F_-(\eta)^sG_-(\eta)^r
\]
where 
$\eta=c'^2$, 
\[
F_\pm(\eta) = \bigg(1+{r^2D^2\over s^2}\bigg) {D\eta\over 2c} 
 \mp {r\over s}{BD} \pm
\bigg[ \bigg(1+{r^2D^2\over s^2}\bigg)(1+D^2)
\bigg({\eta^2\over 4c^2} - {\widetilde \mu} \eta \bigg) 
-B^2\bigg]^{1/2}
\]
\[
G_\pm(\eta) = \bigg(1+{r^2D^2\over s^2}\bigg) {\eta\over 2c} 
 \pm {B} \pm {rD\over s}
\bigg[ \bigg(1+{r^2D^2\over s^2}\bigg)(1+D^2)
\bigg({\eta^2\over 4c^2} - {\widetilde \mu} \eta \bigg) 
-B^2\bigg]^{1/2}
\]
and 
\[
\bal 
D = {b_1^r+b_2^r\over b_1^s+b_2^s}\qquad
B  ={c\over 2} \bigg({b_1^s-b_2^s\over b_1^s+b_2^s} +{r\over s} 
D{b_1^r-b_2^r\over b_1^s+b_2^s}\bigg)\\
\eal 
\]
On then has to analyze the monotonicity properties of the 
functions $F_\pm(\eta)$, $G_\pm(\eta)$ on the intervals 
where $\beta_1,\beta_2>0$. Suppressing the details, which 
are quite elementary, it follows 
that for sufficiently small $\mu>0$ 
equation \eqref{eq:matchingA} admits only the 
solution $c'=c$, if $a>0$ is in addition bounded 
above by a constant $a_0(r,s)$ depending only on $r,s$. Returning  to the expressions \eqref{eq:betasolA}, this implies in turn that $\beta_i=b_i$ for 
$i=1,2$. Therefore the two orbits indeed coincide. 
 
\subsection{Open Gromov-Witten invariants on the 
resolution}\label{GWres}
The goal of this section is to compute the Gromov-Witten invariants for stable maps 
$f:\Sigma \to Y$ with lagrangian boundary conditions on the
cycle $M_\epsilon$ constructed in section (\ref{topres}) for a polynomial $f(x,y)$ of the form 
\[
f(x,y)=x^r-y^s,
\]
with $r>s\geq 1$ coprime. These invariants will be computed 
assuming the existence of a virtual fundamental cycle and a virtual localization result for the moduli space of such maps, 
by analogy with \cite{KL}. 

Recall that the main steps in the construction of 
$M_\epsilon \subset Y$ are as follows. 
Let $C^+\subset X_0$ be the plane curve 
determined by 
\[
f(x,y)=0, \qquad z=w=0
\]
in the singular conifold $X_0$. 
Let $\gamma^+:S^1\to X=T^*S^3$ be the one-cycle obtained by intersecting the 
sphere bundle $P_a$, $a>0$ with the  image $\phi_0(C^+)$, where $\phi_0:X_0\to X$ is the symplectomorphism constructed below 
equation \eqref{eq:sympmorphismA}. Let $\gamma^+_\epsilon = \phi_0\circ \varrho_\epsilon\circ \phi_0^{-1}\circ \gamma^+$ be the 
the dilation of $\gamma^+$ via the radial map $\varrho_\epsilon: 
X_0\setminus\{0\}\to X_0(\epsilon)$, 
\be\label{eq:dilationA}
\varrho_{\epsilon}(x,y,z,w) = \bigg(x,{\sqrt{|z|^2+|y|^2+\epsilon^2}
\over \sqrt{|z|^2+|y|^2}}y, {\sqrt{|z|^2+|y|^2+\epsilon^2}
\over \sqrt{|z|^2+|y|^2}}z,w\bigg).
\ee
Applying the construction in section (\ref{lagsect}) to $\gamma^+_\epsilon$ yields a lagrangian cycle 
$L_\epsilon \subset X$. As shown in equation \eqref{eq:lagcycleres}, 
$M_\epsilon$ is the inverse image $\sigma^{-1}\circ
\varrho_\epsilon^{-1} \circ \phi_0^{-1}(L_\epsilon)$.

The cycle $L_\epsilon\subset X$ admits an explicit parametric 
presentation analogous to the presentation of
the cycles $L_\mu\subset X_\mu$ in 
section (\ref{pqlagcycles}). 
 Note that the one-cycle 
$\phi_0^{-1}(\gamma^+_\epsilon) = \phi_0^{-1}(
\varrho_\epsilon \circ \gamma^+)$ is parametrically given by 
\be\label{eq:gammaepsilonA}
(x,y,z,w) = (b_1^se^{is\theta}, \sqrt{b_1^{2r}+\epsilon^2}\, 
e^{ir\theta}, 0, 0),
\ee
where $b_1=b_1^+(a)$ is the unique positive real solution 
of the equation 
\[
b_1^{2s}+b_1^{2r} = 4a.
\]
Then $L_\epsilon\subset X$ is given by equations of the form  
\be\label{eq:lageqA}
{\vec u} = {\vec f}(\theta), \qquad 
{\dot {\vec f}}(\theta)\cdot ({\vec v} - {\vec g}(\theta)) =0
\ee
where 
\[
\bal
& \left[\begin{array}{c} 
f_1(\theta) \\ f_2(\theta) \\ \end{array}\right] = 
{b_1^q\over c } R(s\theta)
\left[\begin{array}{c} 
1 \\ 0 \\ \end{array}\right] \qquad
\left[\begin{array}{c} 
f_3(\theta) \\ f_4(\theta) \\ \end{array}\right] = 
-{\sqrt{b_1^{2r}+\epsilon^2}\over c} 
R(r\theta)
\left[\begin{array}{c} 
1 \\ 0 \\ \end{array}\right] 
\eal 
\]
\[
\bal
& \left[\begin{array}{c} 
g_1(\theta) \\ g_2(\theta) \\ \end{array}\right] = 
{b_1^s\over 4}c R(s\theta)
\left[\begin{array}{c} 
0 \\ 1 \\ \end{array}\right] \qquad
\left[\begin{array}{c} 
g_3(\theta) \\ g_4(\theta) \\ \end{array}\right] = 
-{{\sqrt{b_1^{2r}+\epsilon^2}}\over 4}c
R(p\theta)
\left[\begin{array}{c} 
0 \\ 1 \\ \end{array}\right] 
\eal 
\]
and $c={\sqrt{4a+\epsilon^2}}$. Using the above formulas, 
it is straightforward to show that $L_\epsilon$ is invariant 
under the circle action \eqref{eq:circactB}.

Now recall 
that the defining equations of $Y$ in $\IC^4\times \IP^1$ are 
\[ 
x\lambda = w\rho, \qquad y\lambda = z\rho
\]
where $[\lambda,\rho]$ are homogeneous coordinates on $\IP^1$.
There are two affine coordinate patches on $Y$, 
$U$ given by 
$\rho\neq 0$ with coordinates 
\[
x,\ y, \ \zeta ={\lambda\over \rho},
\]
 and $U'\subset y$ given by $\lambda\neq 0$,
  with 
coordinates 
\[
z,\ w,\ \zeta'={\rho\over \lambda}.
\]
Obviously, the  transition functions are 
 \[
 w=x\zeta, \qquad z=y\zeta, \qquad \zeta'={1\over \zeta}.
 \]
The strict transform $C\subset Y$ of $C^+$ is
contained in the first patch and has defining equations 
\[
f(x,y)=0, \qquad \zeta=0.
\]

Moreover, note that equation \eqref{eq:circactA} 
also defines a circle action on singular threefold $X_0$ which preserves  $C^+$. This lifts to a circle action $S^1\times Y\to Y$,
\be\label{eq:circactD} 
(x,y,z,w)\times [\lambda, \rho]
 \mapsto (e^{is\varphi}x, e^{ir\varphi}y, e^{-iq\alpha}z, e^{-ip\alpha}w)\times 
 [e^{-i(p+q)\alpha}\lambda, \rho],
\ee
which preserves $C$. 
Since the blow-up map 
$\sigma :Y \to X_0$ and the dilation map \eqref{eq:dilationA} 
are  equivariant,  
it follows that the action \eqref{eq:circactD} 
preserves $M_\epsilon$.
Therefore it also preserves the singular holomorphic disk
${\sf D}_\epsilon$ with boundary on $M_\epsilon$
obtained by intersecting $M_\epsilon$ and $C$. Note that 
${\sf D}_\epsilon$ is given in parametric form by 
\be\label{eq:holdiskA}
(x,y,\zeta) = \big(t^s,t^r,0\big), \qquad |t|\leq b_1.
\ee

Next one has to show that \eqref{eq:holdiskA} 
is the unique torus invariant disk instanton 
$f:\Delta \to Y$ 
with lagrangian boundary conditions on $M_\epsilon$. 
Using equations \eqref{eq:lageqA} it is straightforward to check 
that the only coordinate hyperplane in $Y$ intersecting 
$M_\epsilon$ nontrivially is $\lambda=0$, in which case 
the intersection is the one-cycle $\eta_\epsilon={\sf D}_\epsilon$. 
All other coordinate hyperplanes, $x=0$, $y=0$, $\rho=0$ do not intersect $M_\epsilon$. In particular this implies the image 
$f(\Delta)$ 
of such a map cannot be contained in the surface $\rho=0$. 
Then torus invariance implies that $f(\Delta)$ is either 
disjoint from the surface $\rho=0$, or intersects it transversely 
at the torus fixed point $z=w=0$, $\rho=0$. 
In the first case the fixed point $t=0$ in the domain must be 
mapped to the fixed point $x=y=0$, $\lambda =0$ in the target. 
Moreover, 
in both cases, the restriction of the map $f$ to the punctured 
disk $\Delta\setminus \{0\}$ must be of the form 
\be\label{eq:holmapA}
(x,y,\zeta) = (\alpha_1t^{\pm s}, \alpha_2t^{\pm r}, 
\alpha_3t^{\mp(r+s)})
\ee
for some complex parameters $(\alpha_1,\alpha_2,\alpha_3)$. 

If the first case holds, the map $f$ must be of the 
form 
\[
(x,y,\zeta) = (\alpha_1t^{s}, \alpha_2t^{ r}, 0)
\]
or 
\[
(x,y,\zeta) = (0,0,\alpha_3 t^{(r+s)})
\]
since $f(0)=(0,0,0)$.
The second subcase is ruled out because $M_\epsilon$ 
does not intersect the zero section $x=y=0$. In the first subcase 
the image $f(\Delta)$ is contained in the surface $\lambda=0$
which intersects $M_\epsilon$ along the boundary of ${\sf D}_\epsilon$. 
Therefore $f(\Delta)$ and ${\sf D}_\epsilon$ must have common boundary, 
which implies they must coincide. 

The second case can hold only if $\alpha_3\neq 0$, which implies 
that the image $f(\Delta)$ cannot be contained in the surface 
$\lambda=0$. Then torus invariance implies that $f(\Delta)$ 
must be disjoint from the surface $\lambda=0$ since any common point would have to be a fixed point of the torus action. At the 
same time the only fixed point in the domain is mapped to the 
fixed point $z=w=0$, $\rho=0$. Therefore $f(\Delta)$ is contained 
in the coordinate chart $U'$. 
In terms of the coordinates $(z,w,\zeta')$, 
equation \eqref{eq:holmapA} reads 
\be\label{eq:holmapB}
(z,w,\zeta') = (\alpha_1\alpha_3 t^{\mp r}, \alpha_2\alpha_3
t^{\mp s}, \alpha_3^{-1} t^{\pm(r+s)}).
\ee
Since $\alpha_3\neq 0$, the condition $f(0)=(0,0,0)$ implies that 
$\alpha_1=\alpha_2=0$. This is again ruled out since $M_\epsilon$ 
does not intersect the zero section. 

In conclusion, ${\sf D}_\epsilon$ is indeed the unique torus invariant holomorphic 
disc on $Y$ with boundary in $M_\epsilon$. Then 
the computation of Gromov-Witten invariants reduces to the computation
of 
 multicover contributions of ${\sf D}_\epsilon$ via a virtual localization 
 theorem. 
One then requires an explicit form of lagrangian boundary conditions for 
an $S^1$-invariant 
 stable map $f:\Sigma \to Y$ which factors through the disc 
${{\sf D}_\epsilon}\subset Y$. Let 
${Ann}(M_\epsilon)\subset T^*Y|_{M_\epsilon}$ 
be the subbundle of the cotangent bundle of $Y$ which annihilates 
the tangent bundle $TM_\epsilon \subset TY|_{M_\epsilon}$.  
The boundary conditions are determined by a framing of 
${Ann}(M_\epsilon)|_{\eta_\epsilon}$, that is three sections 
of $T^*Y|_{\eta_\epsilon}$ 
which form a basis of ${Ann}(M_\epsilon)$ at 
any point on $\eta_\epsilon =\partial{\sf D}_\epsilon$. 
This computation reduces basically to the linearization of the defining equations of
$M_\epsilon$ in $Y$, which is standard differential geometry. 
Omitting the intermediate steps, the resulting generators are,
in local coordinates $(x,y,\zeta)$,
\be\label{eq:realgen}
\bal 
\alpha& = 
b_1^s\big[2AC+(s-r)(b_1^{2r}+\epsilon^2)B+ (s-r)^2 b_1^{2s}(b_1^{2r}+\epsilon^2)\big](e^{-is\theta}dx + e^{is\theta}d\bx)\\
&\ \ \, + b_1^{2r}\big[BC - (s-r)b_1^{2s}A + (s-r)^2 b_1^{2s}(b_1^{2r}+\epsilon^2)\big](e^{-ir\theta}dy + e^{ir\theta}d\by).\\
\eal 
\ee
\be\label{eq:cpxgen}
\bal
\beta= & \ 
 e^{-i(r+s)\theta} d{\overline \zeta} + 
  {b_1^s\sqrt{b_1^{2r}+\epsilon^2}\over C}\bigg[{B\over 2c^2b_1^s} 
e^{-is\theta}dx +{(s-r)b_1^s\over 2c^2}e^{is\theta} d\bx\\
&\ -{1\over 4c^2b_1^r} \bigg(
{2b_1^{2r}+\epsilon^2 \over b_1^{2r}+\epsilon^2}A + (s-r)\epsilon^2\bigg)e^{-ir\theta} dy \\
&\ + 
{1\over 4c^2b_1^r}\bigg(
{\epsilon^2 A\over b_1^{2r}+\epsilon^2} + (s-r)(2b_1^{2r}+\epsilon^2)\bigg) e^{ir\theta}d\by\bigg]
\\
\eal
\ee
where 
\[
\bal 
A = 2sb_1^{2s}+(r+s)(b_1^{2r}+\epsilon^2), 
\qquad 
B = (r+s)b_1^{2s}+2r
(b_1^{2r}+\epsilon^2).\eal
\]
\[
C = sb_1^{2s} + r(b_1^{2r}+\epsilon^2).
\]
In particular, $\alpha$ is real and $\beta$ is complex.

\subsubsection{Deformation theory}\label{deftheory}
Let $\Delta\subset \IC$ be the disk $|t|\leq b_1$. 
Let $f:\Delta \to Y$ be the map 
\be\label{eq:diskmap}
t\mapsto (x,y,\zeta) = (t^s,t^r,0).
\ee
Obviously $f$ factors through the disk ${\sf C}\subset Y$ 
mapping the boundary of the disk, 
$|t|=b_1$ to the boundary $\eta_\epsilon =\partial {\sf C}\subset  M_\epsilon$. 
Let $f_\partial$ denote the restriction of $f$ to the boundary.
Let $\CT_{(\Delta, f)}$ denote the sheaf of germs of holomorphic sections of the bundle $f^*T_Y$ satisfying the boundary conditions 
\be\label{eq:bdcondA} 
f_\partial^*(\alpha)\big(s|_{\partial \Delta}\big) =0 ,\qquad 
f_\partial^*(\beta)\big(s|_{\partial \Delta}\big) =0 ,\qquad 
\ee
where $\alpha,\beta$ are the generators of the annihilator 
sub-bundle ${Ann}(M_\epsilon)|_{\eta_\epsilon}$
given in equations \eqref{eq:realgen}-\eqref{eq:cpxgen}.
Let $\CT_\Delta$ be the sheaf of germs of holomorphic sections 
of the tangent bundle $T_\Delta$ satisfying the boundary condition 
\be\label{eq:bdcondB} 
\gamma|_{\partial \Delta} \big(s|_{\partial \Delta}\big) =0
\ee
where $\gamma = td{\overline t} + {\overline t}dt$. 

The deformation complex of the stable map $(\Delta,f)$ with 
lagrangian boundary conditions along $M_\epsilon$ is 
\be\label{eq:defcomplex} 
\bal
0 & \to H^0(\Delta, \CT_\Delta) \to 
H^0(\Delta, \CT_{(\Delta, f)}) \to Def(\Delta, f) \\
& \to H^1(\Delta, \CT_\Delta) 
\to H^1(\Delta, \CT_{(\Delta, f)}) \to Obs(\Delta,f) \to 0.\\
\eal
\ee
In particular one has to compute the ${\check {\rm C}}$ech
cohomology groups 
$H^k(\Delta, \CT_{(\Delta, f)})$ with $k=0,1$. This will be done below using 
the following open cover of $\Delta$
\[
 U = \{t\, |\, 0< |t|\leq b_1\},\qquad
U'=\{t\, |\, 0\leq |t|<b_1\}.
\]
Local sections over $U,U'$ are of the form 
\[
\bal 
s= \sum_{n\in \IZ} \big(a_n t^n \partial_x + b_n t^n \partial y + 
c_n t^n \partial_\zeta\big)\\
s' = \sum_{n\geq 0} \big(a'_n t^n \partial_x + b'_n t^n \partial y + 
c'_n t^n \partial_\zeta\big)\\
\eal
\]
The coefficients $(a_n,n_n,c_n)$, $n\in \IZ$, are subject to boundary
conditions of the form 
\[\bal
\oc_{-n} & = A_1 a_{n-r} + A_2 \oa_{r+2s-n} + B_1 b_{n-s} 
+B_2 \ob_{2r+s-n} \\
a_{s+n}+{\oa}_{s-n} & = C_1(b_{r+n}+\ob_{r-n}).\\
\eal
\]
Changing the variable $n$ to $n+r+s$ in the first equation  yields 
the equivalent formulation 
\be\label{eq:bdcondC} 
\bal 
\oc_{-(n+r+s)} & = A_1 a_{s+n} + A_2 \oa_{s-n} + B_1 b_{r+n} 
+B_2 \ob_{r-n}\\
a_{s+n}+{\oa}_{s-n} & = C_1(b_{r+n}+\ob_{r-n}).\\
\eal
\ee
These conditions are derived from \eqref{eq:bdcondA}, the coefficients 
$A_1,A_2,B_1,B_2,C_1$ being determined from the explicit expressions of $\alpha',\beta'$. The resulting functions of $(p,q,b_1,
\epsilon)$ are fairly complicated, but explicit formulas for these coefficients will not be needed in the following. It suffices to note that the following conditions are satisfied 
\be\label{eq:nondegcond} 
\bal 
A_1^2-A_2^2 \neq 0, & \qquad  A_1C_1+B_1\neq 0, \qquad 
A_2C_2+B_2\neq 0 \\
& \ \ A_1C_1+B_1+ A_2C_2+B_2\neq 0. 
\eal 
\ee
for sufficiently generic 
values of $\epsilon>0$.
This will be assumed from now on. 

The cohomology group $H^0(\Delta, \CT_{(\Delta, f)})$ is isomorphic 
to the kernel of the ${\check {\rm C}}$ech differential, which consists of 
sections 
\[
\bal 
s= \sum_{n\geq \IZ} \big(a_n t^n \partial_x + b_n t^n \partial y + 
c_n t^n \partial_\zeta\big)\\
\eal
\]
where $(a_n,b_n,c_n)$ are subject to the boundary conditions 
\eqref{eq:bdcondB}, and 
\[
a_n=0, \qquad b_n =0, \qquad c_n =0 
\]
for all $n<0$. The behavior of these equations depends on the value 
of $n$, resulting in several different cases. Recall that under the current assumption $r>s\geq 1$. 

 $a)\ n>r.$ The boundary conditions \eqref{eq:bdcondC} reduce to 
 \[
 A_1a_{n+s} + B_1 b_{n+r} =0, \qquad 
 a_{n+s} = C_1 b_{n+r}.
 \]
If $B_1\neq A_1C_1$, these equations admit only the trivial solution, 
hence $a_{n+s}=b_{n+r}=0$ for all $n>r$. 

$b)\ s<n\leq r.$ The boundary conditions \eqref{eq:bdcondC} reduce to 
\[
 A_1a_{n+s} + B_1 b_{n+r} + B_2 \ob_{r-n} =0, \qquad 
 a_{n+s} = C_1( b_{n+r} +\ob_{r-n}),
 \]
which are equivalent to 
\be\label{eq:bdcondD}
(A_1C_1+B_1)b_{n+r} + (A_1C_1+B_2)\ob_{r-n}=0,\qquad 
a_{n+s} = C_1( b_{n+r} +\ob_{r-n}).
\ee

$c)\ -r\leq n<-s.$ The boundary conditions become 
\[
A_2 \oa_{s-n} + B_1 b_{n+r} + B_2\ob_{r-n} =0,
\qquad 
\oa_{s-n} = C_1(b_{n+r}+\ob_{r-n}),
\]
which are equivalent to 
\be\label{eq:bdcondE}
(A_2C_1+B_1)b_{n+r} + (A_2C_1+B_2)\ob_{r-n}=0,\qquad 
\oa_{s-n} = C_1( b_{n+r} +\ob_{r-n}).
\ee
Now note that the first equations in 
\eqref{eq:bdcondE} yield 
\be\label{eq:bdcondF} 
(A_2C_1+B_1)\ob_{r-n} + (A_2C_1+B_2)b_{n+r}=0
\ee
for all $s<n\leq r$, by complex conjugation and changing 
$n$ into $-n$. Therefore for any $s<n\leq r$ the following equations 
must hold simultaneously 
\[
\bal 
(A_1C_1+B_1)b_{n+r} + (A_1C_1+B_2)\ob_{r-n}&=0\\
(A_2C_1+B_2)b_{n+r}+(A_2C_1+B_1)\ob_{r-n} &=0\\
\eal 
\]
This implies $b_{r+n}=b_{r-n}=0$ in this range, provided that 
\[
(B_1-B_2)((A_1+A_2)C_1+B_1+B_2)\neq 0,
\]
which is the case for generic $\epsilon$. The remaining equations 
in \eqref{eq:bdcondD}, \eqref{eq:bdcondE} then imply that $a_{n+s}=0$ as well if $s<n\leq q$. 

Therefore it has been proven so far 
that 
\[
a_n =0, \qquad \mathrm{for}\quad  n>2s
\]
and 
\[
b_n =0, \qquad \mathrm{for}\quad  n>r+s \quad \mathrm{or} \quad 
 0\leq n < r-s.
\]

$d) -(r+s)<n<-r.$ Then the boundary condition yield
\[
A_2 \oa_{s-n} + B_2 \ob_{r-n} =0,\qquad \oa_{s-n} = C_1\ob_{r-n}. 
\]
As long as $A_2C_1+B_2\neq 0$, it follows that $b_{r-n}=0$, $a_{s-n}=0$ 
in this range. These results duplicate those obtained in case $(a)$. 

$e)\ -s\leq n \leq s.$ In this case the resulting equations are 
\be\label{eq:bdcondG}
\bal 
 A_1 a_{n+s} + A_2 \oa_{q-n} + B_1 b_{n+r} 
+B_2 \ob_{r-n}& =0\\
a_{n+s}+{\oa}_{s-n} & = C_1(b_{n+r}+\ob_{r-n}).\\
\eal 
\ee
Note that for any $-s\leq n\leq s$ the second equation in \eqref{eq:bdcondG} is invariant 
under complex conjugation, followed by the reflection $n\to (-n)$. 
This is expected since it originates in the boundary condition 
given by the real one-form $\alpha'$. For sufficiently generic 
coefficients, the first equation does not have this property. 
In fact, using this transformation, the first set of equations 
in \eqref{eq:bdcondG} is  equivalent to 
\[
\bal 
A_1 a_{n+s} + A_2 \oa_{s-n} + B_1 b_{n+r} 
+B_2 \ob_{r-n}& =0, \qquad 0< n \leq s\\
A_2 a_{n+s} + A_1 \oa_{s-n} + B_2 b_{n+r} 
+B_1 \ob_{r-n}& =0, \qquad 0< n \leq s\\
A_1a_s+A_2\oa_s + B_1b_r + B_2 \ob_r & =0,\\
\eal 
\]
Assuming again $\epsilon$ to be sufficiently generic such that$A_1^2\neq A_2^2$, these equations determine $a_n$, $0\leq n\leq 2s$ uniquely in terms of the variables $b_n$, $(r-s)\leq n\leq 
r+s$ as follows:
\be\label{eq:acoeff}
\bal
a_{n+s}& = {(A_2B_2-A_1B_1)b_{n+r}+(A_2B_1-A_1B_2)\ob_{r-n}
\over A_1^2-A_2^2},\qquad 0\leq n\leq q\\
\oa_{s-n}& = {(A_2B_1-A_1B_2)b_{n+r}+
(A_2B_2-A_1B_1)\ob_{r-n}\over A_1^2-A_2^2},\qquad 
0<n\leq s.
\\
\eal
\ee
 Substituting in the second set of equations in \eqref{eq:bdcondG}
 yields 
 \[
 \left((A_1+A_2)C_1+{B_1+B_2}\right) (b_{n+r} + 
 \ob_{r-n}) =0
 \]
 for $-s\leq n\leq s$. Since 
 \[ 
 C_1(A_1+A_2)+B_1+B_2\neq 0
 \]
generically, this implies the reality condition 
\be\label{eq:bcoeff}
b_{n+r} + 
 \ob_{r-n} =0.
 \ee
Therefore the space of solutions can be parameterized by the independent variables $b_n$ with $r-s\leq n\leq r$, where 
$b_n\in \IC$ for $n\neq r$ and $b_r\in i\IR$. 
The last case is 

$f)\ n \leq -(r+s).$ Using the previous cases, the boundary conditions 
reduce to $c_n=0$ for all $n\geq 0$ and no additional conditions 
on $a_n,b_n$. 

In conclusion the cohomology group 
$H^0(\Delta, \CT_{(\Delta, f)})$ is isomorphic to the space of sections of the form 
\be\label{eq:defspace} 
s= \sum_{n=0}^{2s} a_n t^n \partial_x + \sum_{n=r-s}^{r+s} 
b_n t^n \partial_y 
\ee
where the coefficients $a_n,b_n$ are subject to conditions \eqref{eq:acoeff}, \eqref{eq:bcoeff}. 

Note that the above computation implies 
that the holomorphic cylinder ${\sf D}$ admits no finite deformations 
in $Y$ as a Riemann surface with boundary on $M_\epsilon$. The argument 
relies on the fact that the coefficients 
$c_n$ are zero for all infinitesimal deformations of the map 
$f:\Delta\to Y$. This implies that the disk ${\sf D}$ cannot
be deformed in the normal directions to the plane $\zeta=0$. Any such 
deformation would yield by linearization an infinitesimal deformation 
with some $c_n\neq 0$. Therefore the disk ${\sf D}$ may admit 
only deformations in the plane $\zeta=0$. However note that the 
lagrangian cycle $M_\epsilon$ intersects the plane $\zeta=0$ 
along the boundary $\eta_\epsilon$ of ${\sf D}$. Hence any deformation 
of ${\sf D}$ would have to intersect $M_\epsilon$ along the same 
cycle $\eta_\epsilon$. Then the claim follows noting that any 
two irreducible 
holomorphic curves passing through the same circle must coincide. 

The cohomology group $H^1(\Delta, \CT_{(\Delta, f)})$
is isomorphic to the cokernel of the ${\check {\rm C}}$ech 
differential. The image of the differential map consists of 
sections of the form 
\[ 
s= \sum_{n\in \IZ} \big(a_n t^n \partial_x + b_n t^n \partial_y + 
c_n t^n \partial_\zeta\big)-
\sum_{n\geq 0} \big(a'_n t^n \partial_x + b'_n t^n \partial_y + 
c'_n t^n \partial_\zeta\big)
\]
on $U\cap U'=\{t\, |\, 0<|t|<b_1\}$, where the coefficients 
$(a_n,b_n,c_n)$, $n\in \IZ$,
 are subject to the conditions \eqref{eq:bdcondC}. 
The coefficients $(a_n',b_n',c'_n)$, $n\in \IZ_{\geq 0}$ are arbitrary. 
In order to determine the cokernel, consider the equation 
\be\label{eq:image}
s = \sum_{n\in \IZ} \big(\alpha_n t^n \partial_x + 
\beta_n t^n \partial_y + 
\gamma_n t^n \partial_\zeta\big)
\ee
in the variables $(a_n,b_n,c_n)$, $n\in \IZ$, $(a'_n,b'_n,c'_n)$, 
$n\in \IZ_{\geq 0}$, 
where $(\alpha_n,\beta_n,\gamma_n)$, $n\in \IZ$ are arbitrary coefficients. This implies $$(a_n,b_n,c_n) = 
(\alpha_n,\beta_n,\gamma_n)$$ for all $n<0$
and 
\[
(a_n-a'_n, b_n-b'_n,c_n-c'_n) = 
(\alpha_n,\beta_n,\gamma_n)
\]
for all $n \geq 0$. The effect of boundary conditions must be analyzed again on a case by case basis, depending on the value of 
$n$.

$a')\ r<n<r+s.$ The boundary equations become
\[
\bal
A_1a_{n+s} + A_2\balpha_{s-n}+B_1b_{n+r}+B_2\bbeta_{r-n} & =
\bgamma_{-(n+r+s)} \\
a_{n+s}+\balpha_{s-n} & = C_1(b_{n+r} + \bbeta_{r-n}),\\
\eal
\]
which are equivalent to 
\be\label{eq:bdcondH}
\bal 
(A_1C_1+B_1) b_{n+r} + (A_2-A_1)\balpha_{s-n} + 
(A_1C_1+B_2)\bbeta_{r-n} & = \bgamma_{-(n+r+s)}\\
a_{n+q} +\balpha_{s-n}- C_1(b_{n+r} + \bbeta_{r-n}) & =0.\\
\eal
\ee

$b')\ -(r+s)<n<-r.$ In this case the boundary conditions read
\[
\bal
A_1\alpha_{n+s} + A_2\oa_{s-n}
+B_1\beta_{n+r}+B_2\ob_{r-n} & =
\bgamma_{-(n+r+s)} \\
\alpha_{n+s}+\oa_{s-n} & = C_1(\beta_{n+r} + \ob_{r-n}),\\
\eal
\]
and are equivalent to 
\be\label{eq:bdcondI}
\bal 
(A_2C_1+B_2)\ob_{r-n} +(A_1-A_2)\alpha_{n+s}+(A_2C_1+B_1)
\beta_{n+r}& =\bgamma_{-(n+r+s)} \\
\oa_{s-n}+\alpha_{n+s}-C_1(\beta_{n+r} + \ob_{r-n}) & =0\\
\eal
\ee
By complex conjugation and reflection, $n\to (-n)$, equations 
\eqref{eq:bdcondI} yield 
\be\label{eq:bdcondJ}
\bal 
(A_2C_1+B_2)b_{n+r} +(A_1-A_2)\balpha_{s-n}+(A_2C_1+B_1)
\bbeta_{r-n}& =\gamma_{n-(r+s)} \\
a_{n+s}+\balpha_{s-n}-C_1(\bbeta_{r-n} + b_{n+r}) & =0\\
\eal
\ee
for all $r<n < r+s$. If 
\[
A_1C_1+B_1\neq 0, \qquad A_2C_1+B_2\neq 0, 
\]
equations \eqref{eq:bdcondI}, \eqref{eq:bdcondJ} admit solutions 
if and only if the linear relation 
\be\label{eq:linobs}
\bal 
& (A_1C_1+B_1)(A_2C_1+B_2) (\gamma_{n-(r+s)}-
\bgamma_{-(n+r+s)}) =\\
& (B_1+B_2+C_1(A_1+A_2))((A_1-A_2)\balpha_{s-n} + (B_1-B_2)\bbeta_{r-n}).\\
\eal
\ee
holds.

$c')\ n\geq r+s.$ The boundary conditions are identical to case 
$(a')$ above. 

$d')\ n\leq -(r+s).$ The boundary conditions  are very similar to case 
$(b')$, except the first equation in \eqref{eq:bdcondI} now reads 
\[
(A_2C_1+B_2)\ob_{r-n} +(A_1-A_2)\alpha_{n+s}+(A_2C_1+B_1)
\beta_{n+r} =\oc_{-(n+r+s)}. 
\]
By complex conjugation and reflection, $n\to (-n)$, this becomes 
\[
(A_2C_1+B_2)b_{n+r} +(A_1-A_2)\balpha_{s-n}+(A_2C_1+B_1)
\bbeta_{r-n} =c_{n-(r+s)}. 
\]
This equation is very similar to the first equation in \eqref{eq:bdcondJ}, 
except the right hand side is $c_{n-(r+s)}$ instead of 
$\gamma_{n-(r+s)}$. As a result the resulting system of linear equations in 
$b_{n+r}$, $c_{n-(r+s)}$, 
\[
\bal 
(A_1C_1+B_1) b_{n+r} + (A_2-A_1)\balpha_{s-n} + 
(A_1C_1+B_2)\bbeta_{r-n} & = \bgamma_{-(n+r+s)}\\
(A_2C_1+B_2)b_{n+r} +(A_1-A_2)\balpha_{s-n}+(A_2C_1+B_1)
\bbeta_{r-n} -c_{n-(r+s)} & =0, \\
\eal 
\]
admits solutions for any values of $\alpha_{s-n}$, $\beta_{r-n}$. 

$e')\ s<n\leq r.$ In this case the boundary equations read
\[
\bal 
A_1a_{n+s} + A_2\balpha_{s-n} + B_1 b_{n+r} +B_2 \ob_{r-n} 
& = \bgamma_{-(n+r+s)}\\
a_{n+s}+\balpha_{s-n} &= C_1(b_{n+r}+\ob_{r-n})\\
\eal
\]
and are equivalent to 
\be\label{eq:bdcondK}
\bal 
(A_1C_1+B_1)b_{n+r} + (A_1C_1+B_2)\ob_{r-n} +
(A_2-A_1)\balpha_{n-s} & = \bgamma_{-(n+r+s)}\\
a_{n+s} - C_1(b_{n+r}+\ob_{r-n})+\balpha_{s-n}&=0.\\
\eal
\ee

$f')\ -r\leq n<-s.$ 
\[
\bal
A_1\alpha_{n+s} + A_2\oa_{s-n} + B_1 b_{n+r} +B_2 \ob_{r-n} 
&= \bgamma_{-(n+r+s)}\\
\alpha_{n+s} + \oa_{s-n}& = C_1(b_{n+r}+\ob_{r-n})\\
\eal 
\]
Again, by complex conjugation and reflection these equations
become 
\[
\bal
A_1\balpha_{s-n} + A_2a_{n+s} + B_1 \ob_{r-n} +B_2 b_{n+r} 
&= \gamma_{n-(r+s)}\\
\balpha_{s-n} + a_{n+s}& = C_1(\ob_{r-n}+b_{n+r})\\
\eal 
\]
with $s<n\leq r$. They are equivalent to 
\be\label{eq:bdcondL} 
\bal
(A_2C_1+B_2)b_{n+r} + (A_2C_1+B_1)  \ob_{r-n} + (A_1-A_2)
\balpha_{s-n} & = \gamma_{n-(r+s)}\\
 a_{n+s}- C_1(\ob_{r-n}+b_{n+r}) +\balpha_{s-n} &=0.\\
 \eal
 \ee 
 Since $a_{n+s}$, $b_{n+r}$, $b_{r-n}$ are independent variables, 
 equations \eqref{eq:bdcondK}, \eqref{eq:bdcondL} admit solutions 
 for any values of $\alpha_{r-n}$, $\gamma_{-(n+r+s)}$, 
 $\gamma_{n-(r+s)}$. 
 The remaining case is 
 
 $g')\ -s\leq n\leq s.$ In this range the boundary conditions read 
 \be\label{eq:bdcondM}
 \bal 
A_1 a_{n+s} + A_2 \oa_{s-n} + B_1 b_{n+r} 
+B_2 \ob_{r-n}& =\bgamma_{-(n+r+s)}\\
a_{n+s}+{\oa}_{s-n}-C_1(b_{n+r}+\ob_{r-n})&=0.\\
\eal
\ee
By complex conjugation and reflection, the first set of these equations is 
equivalent to 
\[
\bal 
A_1 a_{n+s} + A_2 \oa_{s-n} + B_1 b_{n+r} 
+B_2 \ob_{r-n}& =\bgamma_{-(n+r+s)}, \qquad 0< n \leq s\\
A_2 a_{n+s} + A_1 \oa_{s-n} + B_2 b_{n+r} 
+B_1 \ob_{r-n}& =\gamma_{n-(r+s)}, \qquad 0< n \leq s\\
A_1a_s+A_2\oa_s + B_1b_r + B_2 \ob_r & =\bgamma_{-(r+s)},\\
\eal 
\]
 If $A_1^2\neq A_2^2$, these equations yield
\be\label{eq:acoeffB}
\bal
a_{n+s} = & \
 {(A_2B_2-A_1B_1)b_{n+r}+(A_2B_1-A_1B_2)\ob_{r-n}
\over A_1^2-A_2^2},\qquad 0\leq n\leq \\
&\ + {A_1\bgamma_{-(n+r+s)}-A_2\gamma_{n-(r+s)}\over 
A_1^2-A_2^2}\\
\oa_{s-n} =& \ {(A_2B_1-A_1B_2)b_{n+r}+
(A_2B_2-A_1B_1)\ob_{r-n}\over A_1^2-A_2^2},\qquad 
0<n\leq s.\\
&\ + {A_1\gamma_{n-(r+s)}-A_2\bgamma_{-(n+r+s)}\over 
A_1^2-A_2^2}\\
\eal
\ee
 Substituting in the second set of equations in \eqref{eq:bdcondM}
 yields 
 \[
 \left((A_1+A_2)C_1+B_1+B_2\right) (b_{n+r} + 
 \ob_{r-n}) = \gamma_{n-(r+s)}+ \bgamma_{-(n+r+s)}.
   \]
 for $-s\leq n\leq s$. Since $b_{n+r}$, $b_{r-n}$ are independent variables, these equations admits solutions for any values of 
 $\gamma_{n-(r+s)}$, $\bgamma_{-(n+r+s)}$ provided that 
 $(A_1+A_2)C_1+B_1+B_2\neq 0$.  
 
 Summarizing the above results, it follows that for sufficiently generic 
 $\epsilon$ equation 
 \eqref{eq:image} admits solutions if and only 
 if the coefficients $(\alpha_n,\beta_n,\gamma_n)$, $n\in \IZ$, 
 satisfy the linear relations  
 \eqref{eq:linobs}.  
 This implies that the cohomology group 
 $H^1(\Delta, \CT_{(\Delta,f)})$ is an $(s-1)$-dimensional 
 complex vector space  
 which can be identified with the space of sections 
 of the form 
 \be\label{eq:obsspace}
 s= \sum_{n=1-s}^{-1} \gamma_n t^n \partial_\zeta  
 \ee
 on $U\cap U'$. In particular, if $s=1$, this space is trivial.
 
 For completeness, note that the computation of the cohomology groups $H^i(\Delta, \CT_\Delta)$, $i=0,1$ is entirely analogous, 
 and technically much simpler. One finds that $H^0(\Delta, \CT_\Delta)$ is generated by sections of the form 
 \[
 a_{-1} \partial_t + a_0 t \partial_t + a_1 t^2 \partial_t
 \]
 with 
 \[
 a_{-1}+ \oa_1 =0, \qquad a_0 +\oa_0 =0.
 \] 
 The obstruction space $H^1(\Delta, \CT_\Delta)$ is trivial. 
  
 \subsubsection{Virtual localization}\label{virtloc}
  Now let ${\overline\CM}_{g,1}(Y,M_\epsilon;d,1)$ be 
 the moduli space of 
genus $g\geq 0$ stable maps with $h=1$ boundary 
components mapped to $M_\epsilon$, in the relative homology class 
$ d[C_0]+[ {\sf C}]\in H_2(Y,M_\epsilon)$, 
$d\in \IZ_{\geq 0}$.  The circle action 
  \[
 (x,y,\zeta) \mapsto (e^{-is\varphi}x,e^{-ir\varphi}y,e^{i(r+s)\varphi}\zeta) 
 \]
 on $Y$ preserves $M_\epsilon$, hence it induces an action on the 
 moduli space of stable maps. 
  Let also 
${\overline M}_{g,1}(Y,d)$ denote the moduli space of 
genus $g$ stable maps to $Y$ with one marked point in the homology class 
$d[C_0]\in H_2(Y)$. This moduli space is equipped with a natural 
evaluation map at the marked point, 
$ev: {\overline M}_{g,1}(Y,d)\to Y$.  

A map $f:\Sigma \to Y$ 
 determines a circle fixed point in the moduli space \\
 ${\overline\CM}_{g,1}(Y,M_\epsilon;d,1)$ if 
 and only if there exists a circle action on the domain $\Sigma$ such 
 that $f$ is equivariant. 
 This implies that domain must be a union $\Sigma=\Sigma_0\cup_\nu \Delta$ where $\Sigma_0$ is a closed nodal Riemann surface 
 without boundary which intersects the disk $\Delta$ 
 at a single point $\nu$, which is a simple node of $\Sigma$.
 Moreover the image of the restriction $f|_{\Delta}$ must coincide 
 with the holomorphic disc ${\sf D}$, which has been shown 
 below \eqref{eq:holdiskA} to be the unique torus invariant 
 disc in $Y$ with boundary in $M_\epsilon$. In more detail, 
 the following conditions must hold 
 \begin{itemize}
 \item 
 $\Delta$ admits a parameterization 
 $\Delta = \{|t|\leq r_1\}$ such that $\nu$ is identified with the point $t=0$ and 
 \[
 f|_{\Delta}(t) = (t^s,t^r,0).
 \]
 The circle action on $\Delta$ is given by $t\mapsto e^{-i\varphi}t$. 
  \item Note that there is an algebraic torus actions
 ${\IC}^\times \times Y\to Y$ which agrees with above real torus 
 action by restriction to the unit circle.  Then  the data $(\Sigma_0, f_0,\nu)$, with 
 $f_0=f|_\Sigma$ must be a $\IC^\times$-invariant stable map to $Y$ such that 
 $f_0(\nu) = p$, where $p\in Y$ is the point $x=y=\zeta=0$. 
 \end{itemize}
 These conditions imply that the fixed locus ${\overline\CM}_{g,1}(Y,M_\epsilon; d,1)^{S^1}$  is isomorphic to the fixed subspace 
$$ev_\nu^{-1}(p)^{\IC^\times}\subset {\overline M}_{g,1}(Y,d)^{\IC^\times}.$$ 
 
The deformation complex of a fixed stable map $(\Sigma,f)$ is  
  \be\label{eq:defcomplexB} 
\bal
0 & \to Aut(\Sigma) \to 
Def(f)\to Def(\Sigma, f) \\
& \to Def(\Sigma)
\to Obs(f) \to Obs(\Sigma,f) \to 0.\\
\eal
\ee
 where the notation is self-explanatory. All terms carry natural circle 
 actions since $(\Sigma,f)$ is a circle invariant map. The fixed part 
 of the deformation complex determines the virtual fundamental cycle on 
 the fixed locus, while the moving part determines the virtual 
 normal bundle to the fixed locus. 
 Each term will be analyzed below assuming that $\Sigma_0$ is 
 nonempty. In the special case $\Sigma_0=\emptyset$ the deformation 
 complex \eqref{eq:defcomplexB} reduces to \eqref{eq:defcomplex} 
 analyzed in the previous subsection. 
 
  Given the structure of fixed maps explained above, there is an  
 exact sequence
  \be\label{eq:normseqA} 
  \bal 
 0 & \to Def(f) \to H^0(\Delta, \CT_{(\Delta, f|_\Delta)})
 \oplus Def(f_0)  \to T_pY \\
 & \to Obs(f) \to 
 H^1(\Delta, \CT_{(\Delta, f|_\Delta)}) \oplus Obs(f_0) 
 \to 0.\\
 \eal
 \ee
 This yields the following relations in the representation ring of 
 the circle 
 \be\label{eq:normrelA} 
 \bal 
 Obs(f)^{f}-Def(f)^{f} & = H^1(\Delta, \CT_{(\Delta, f|_\Delta)})^{f} 
 - H^0(\Delta, \CT_{(\Delta, f|_\Delta)})^{f}\\
 & \ \ \,   +Obs(f_0)^f -
 Def(f_0)^f\\
  Obs(f)^{m}-Def(f)^{m} & = H^1(\Delta, \CT_{(\Delta, f|_\Delta)})^{m} - H^0(\Delta, \CT_{(\Delta, f|_\Delta)})^{m}  \\
  & \ \ \, +Obs(f_0)^m -
 Def(f_0)^m  + T_pY.\\
 \eal
 \ee
 Moreover standard arguments imply 
 \be\label{eq:normrelB}
 \bal 
 Aut(\Sigma)^{f,m} & = Aut(\Sigma_0,\nu)^{f,m}+
 Aut(\Delta,0)^{f,m}\\
 Def(\Sigma)^{f} & = Def(\Sigma_0,\nu)^{f}\\
   Def(\Sigma)^{m} & = Def(\Sigma_0,\nu)^{m}+ 
   T_\nu\Sigma_0 \otimes T_0\Delta\\ 
 \eal 
 \ee
 while the cohomology groups 
 $H^0(\Delta, \CT_{(\Delta, f|_\Delta)})$ have been 
 determined in equations \eqref{eq:defspace}, \eqref{eq:obsspace}. 
 There is however a discrete ambiguity
  in reading off their equivariant content, 
 reflecting a choice of orientation on the moduli space of 
 stable maps with lagrangian boundary conditions \cite{KL}.
 As explained in \cite{KL},  the difference between these choices is 
 encoded in an overall sign which cannot be fixed in the absence of a rigorous 
 construction of the moduli space equipped with a virtual cycle. 
 Therefore 
 the present computation will be a test of large $N$ duality up to sign. 
Given equations \eqref{eq:acoeff}, \eqref{eq:bcoeff}, the 
deformation space \eqref{eq:defspace} is isomorphic to a 
vector space of the form 
\[
\IR\langle \partial_y\rangle\oplus \bigoplus_{n=r-s}^{r-1} \IC\langle 
t^n\partial_y\rangle.
\]
At the same time, the obstruction space \eqref{eq:obsspace} is 
naturally identified with the complex vector space 
\[
\bigoplus_{n=1-s}^{-1}\IC\langle t^n \partial_\zeta\rangle.
\]
This yields the following relations in the representation ring of $S^1$ 
  \be\label{eq:equivinfobs}
 H^0(\Delta, \CT_{(\Delta, f|_\Delta)})^m  = \sum_{n=1}^{s} 
 R^{n} ,\qquad 
 H^1(\Delta, \CT_{(\Delta, f|_\Delta)})^m  = \sum_{n=1}^{s-1}
 R^{-(r+n)},
 \ee
 \[
  H^0(\Delta, \CT_{(\Delta, f|_\Delta)})^f 
  = \IR, \qquad H^1(\Delta, \CT_{(\Delta, f|_\Delta)})^f =0,
  \]
     where $R$ is the canonical representation of $S^1$ on $\IC$, 
     and $\IR$ denotes the trivial real representation.  
 Note also that $Aut(\Delta)$ is isomorphic to the space of sections 
 of $\CT_\Delta$ of the form $a\partial_t + bt\partial_t$ 
 with $a\in \IC$, $b\in i\IR$. Therefore 
 $$ Aut(\Delta)^f = \IR,\qquad Aut(\Delta)^m = R.$$
 The subgroup of automorphisms preserving the origin, 
 $Aut(\Delta,0)$ is generated by $t\partial_t$ over $\IR$, therefore 
 it has only a fixed part $Aut(\Delta,0)^f=\IR$. 
 
 Collecting all the above results one obtains
 \be\label{eq:fixedpart} 
 \bal
  Obs(\Sigma,f)^f - Def(\Sigma,f)^f & = Obs(f_0)^f-Def(f_0)^f 
  +Aut(\Sigma_0,\nu)^f - Def(\Sigma_0,\nu)^f\\
  & = Obs(\Sigma_0,f_0)^f-Def(\Sigma_0,f_0)^f.\\
  \eal 
  \ee
  \[
  \bal
   Obs(\Sigma,f)^m - Def(\Sigma,f)^m &   = 
   Obs(f_0)^m-Def(f_0)^m +Aut(\Sigma_0,\nu)^m - 
   Def(\Sigma_0,\nu)^m \\
   & \ \ \, + \sum_{n=1}^{s-1}
 R^{-(r+n)} -\sum_{n=1}^{s} 
 R^{n} + T_pY -  T_\nu\Sigma_0 \otimes T_0\Delta\\
   \eal
   \]
  This implies that the virtual fundamental cycle of the fixed locus 
  is the restriction of the natural virtual cycle of the fixed locus 
  $[{\overline M}_{g,1}(Y,d)^{\IC^\times}]^{vir}$ with the subspace 
  $ev^{-1}(p)^{\IC^\times}$. 
  The equivariant K-theory class of the 
  virtual normal bundle is given by 
  \[
  \bal 
  N^{vir} &  = N^{vir}_{{\overline M}_{g,1}(Y,d)^{\IC^\times}/
  {\overline M}_{g,1}(Y,d)} -T_pY + R{\mathbb L}^{-1}\\
     & \ \ \, + \sum_{n=1}^{s-1}
 R^{-(r+n)} -\sum_{n=1}^{s} 
 R^{n} 
 \eal 
   \]
  where ${\mathbb L}$ is the tautological line bundle 
  on ${\overline M}_{g,1}(Y,d)$  associated to the marked point. 
       Then the residual formula for 
  open Gromov-Witten invariants is 
  \be\label{eq:openGW} 
  \bal 
 &  GW_{g,1}(d,1) = (-1)^{s-1}  
  {\prod_{n=1}^{s-1}(r+n)\over s!} 
  {e_{\IC^\times}(T_pY)\over \alpha} \\
 & \int_{[{\overline M}_{g,1}(Y,d)^{\IC^\times}_p]^{vir}} 
  {1\over 
  e_{\bf \IC^\times}
  \big(N^{vir}_{{\overline M}_{g,1}(Y,d)^{\IC^\times}/
  {\overline M}_{g,1}(Y,d)}\big)^{-1} (\alpha - \psi)} \\
    \eal
 \ee
 where ${\overline M}_{g,1}(Y,d)^{\IC^\times}_p$ denotes 
 the union of connected components of the fixed locus 
contained in $ev^{-1}(p)$. 
  Standard  formal manipulations show that this 
 formula is equivalent to 
\be\label{eq:openGWB} 
\bal 
 &  GW_{g,1}(d,1) = (-1)^{s-1}
  {\prod_{n=1}^{s-1}(r+n)\over s!} 
 \int_{[{\overline M}_{g,1}(Y,d)]_{\IC^\times}^{vir}} 
 {ev^*\phi_{\IC^\times}(p)\over \alpha(\alpha - \psi)} \\
 \eal
 \ee 
 where $[{\overline M}_{g,1}(Y,d)]_{\IC^\times}^{vir}$ denotes 
 the equivariant virtual cycle of the moduli space, $\phi_{\IC^\times}(p)\in H^*_{\IC^\times}(Y)$ is the equivariant Thom class of $p\in Y$, 
  and $\alpha = \ch(R)$.



 \subsection{Comparison with HOMFLY 
 polynomial}\label{comphomfly}
The goal of this section is to compare the generating function 
for the open Gromov-Witten invariants\\
 $GW_{g,1}(Y,M_\epsilon; d)$
with the HOMFLY polynomial of $(s,r)$-torus knots. It will be shown that large $N$ duality for $(s,r)$ torus knots follows from known 
results on large $N$ duality for the unknot. 
The manipulations of enumerative 
invariants justifying this statement parallel similar manipulations in 
Chern-Simons theory relating invariants of $(s,r)$ torus knots 
to colored invariants of the unknot 
\cite{toruslinks, torusknotsmirror}.

The main observation is that 
 the Gromov-Witten invariants given in \eqref{eq:openGWB} 
 for some coprime $(r,s)$ can be expressed in terms of analogous invariants  invariants determined by the curve  
 \be\label{eq:unknotcurve}
x=z=w=0
\ee
in $X_0$ and the associated lagrangian cycles. 
In order to emphasize the dependence on $(r,s)$, the lagrangian cycles used in the above 
construction 
will be denoted by $M_\epsilon^{(s,r)}$, and the 
corresponding invariants by $GW_{g,1}^{(s,r)}(d,1)$. 

Consider the construction of lagrangian cycles carried out in sections (\ref{lagsect}) -- (\ref{topres}) for a curve $C$ 
of the form \eqref{eq:unknotcurve}. 
By analogy with section (\ref{pqlagcycles}) one can easily check 
that the lagrangian cycle $M^{(1,0)}_\epsilon$ obtained in this case 
is preserved by any circle action on $Y$ of the form 
\be\label{eq:circactM}
(x,y,\zeta) \mapsto (e^{-is\varphi}x, e^{-ir\varphi}y, 
e^{i(r+s)\varphi}\zeta)
\ee
with $r,s\in \IZ$. 
Moreover, $M^{(1,0)}_\epsilon$ intersects the strict transform 
of $C$ along an orbit of the torus action, obtaining a unique 
holomorphic circle invariant 
disk ${\sf D}_0$ on $Y$ with boundary 
on $M^{(1,0)}_\epsilon$. In this case  ${\sf D}_0$
is smooth and
Gromov-Witten invariants with boundary conditions on 
$M^{(1,0)}_\epsilon$ can be constructed in close analogy with 
\cite{KL}. Let ${\overline\CM}_{g,1}(Y,M_\epsilon;d,k)$ be 
 the moduli space of 
genus $g\geq 0$ stable maps with $h=1$ boundary 
components mapped to $M^{(1,0)}_\epsilon$, in the relative homology class 
$ d[C_0]+k[{\sf D}_0]\in H_2(Y,M_\epsilon)$, 
$d\in \IZ_{\geq 0}$, $k\in \IZ_{>0}$.  
In contrast with the previous section, the winding number 
 $k$ will be allowed 
to take arbitrary values in the present context. 
Then  there is a 
residual formula of the form 
\be\label{eq:openGWC}
GW_{g,1}^{(1,0)}(d,k) = (-1)^{k-1}{\prod_{n=1}^{k-1}(r k + n) \over 
(k-1)!} 
 \int_{[{\overline M}_{g,1}(Y,d)]_{\IC^\times}^{vir}} 
 {ev^*\phi_{\IC^\times}(p)\over k\alpha(k\alpha - s\psi)} \\
\ee
In particular, setting $k=s$ in equation \eqref{eq:openGWC}, 
it follows that 
\be\label{eq:openGWD} 
GW_{g,1}^{(s,r)}(d,1) = s GW_{g,1}^{(1,0)}(d,s).
\ee
Now define the generating functions with fixed winding numbers $1$, 
respectively $s$,
\[
F_{1}^{(s,r)}(g_s,Q, V) =\sum_{g\geq 0} \sum_{d\geq 0} g_s^{2g-1} Q^{d} GW^{(s,r)}_{g,1}(d,1) \mathrm{Tr}(V) 
\]
\[
F_{s}^{(1,0)}(g_s,Q, V) = \sum_{g\geq 0} \sum_{d\geq 0} g_s^{2g-1} Q^{d} GW^{(1,0)}_{g,1}(d,s) \mathrm{Tr}(V^s) 
\]
where the open  string Gromov-Witten are defined by residual formulas \eqref{eq:openGWB}, \eqref{eq:openGWC} with respect to a 
circle action of the form \eqref{eq:circactM}. 
 
Large $N$ duality for the unknot yields the following identity 
\cite[Eqn. (5.6)]{framedknots} 
\be\label{eq:largeNunknotA} 
F_{q}^{(1,0)}(g_s,Q, V) = {(-1)^{s-1}\over s} \sum_{R} \chi_R(C_{{(s)}})
e^{i(r/s)\kappa_R g_s/2}  W_R^{(1,0)}(q,Q)\mathrm{Tr}(V^s) ,
\ee
the terms in the right hand side being explained below. 
\begin{itemize}
\item The sum in the right hand side of \eqref{eq:largeNunknotA}
is over all Young diagrams $R$ and 
$\chi_R(\CC_{(s)})$ denotes the character of the conjugacy class 
determined by the vector ${\vec k}=(k_j)_{j\geq 1}$, with 
$k_j=1$ if $j=s$and $k_j=0$ otherwise in the 
representation determined by $R$. See 
\cite[Sect. 4.1]{framedknots} for more details. 
\item $W_R^{(1,0)}(q,Q)$ is the HOMFLY polynomial 
colored by the representation $R$ of $U(N)$, expressed as a 
function of the large $N$ Chern-Simons theory on $S^3$, 
\[
g_s=\left({2\pi \over k+N}\right), \qquad 
\lambda = \left({2\pi N \over k+N}\right), \qquad q=e^{ig_s} 
\qquad Q=e^{i\lambda}.
\]
Up to a normalization factor, $W_R^{(1,0)}(q,Q)$ is given by the 
quantum dimension of $R$,  
\[
W_R^{(1,0)}(q,Q) = Q^{-|R|/2} \mathrm{dim}_q(R).
\]
where $|R|$ is the total number of boxes in the Young diagram $R$. 
\item For any Young diagram $R$, the number $\kappa_R$ 
is defined by 
\[
\kappa_R = |R| + \sum_{i=1}^{l_R} (l_i^2 -2il_i)
\]
where $l_R$ is the number of rows of $R$ and $l_i$ is the length 
of the $i$-th row, $i=1,\ldots, l_R$. 
\end{itemize}

As explained in 
\cite[Sect. 3.2]{framedknots} the factor 
 $e^{im\kappa_R g_s/2}$ 
encodes the framing dependence of colored HOMFLY polynomials, 
 $m$ being the framing of the knot with respect to the canonical framing.  
The expression $e^{i(r/s)\kappa_Rg_s/2} W_R^{(1,0)}(q,Q)$ 
in the right hand side 
of equation \eqref{eq:largeNunknotA} must therefore be interpreted as a 
a colored HOMFLY polynomial with fractional framing. The relation  
between quantum knot invariants with fractional framing and 
residual open string  Gromov-Witten invariants  
has been observed in a similar context in \cite{largeN,locgluing}. 

Formula \eqref{eq:largeNunknotA} was initially tested in specific examples 
 for the free term in the $\lambda$-expansion  of 
$W_R^{(1,0)}(q,Q)$. The higher order terms were implicitly
tested in \cite{locgluing} in the process of finding an enumerative interpretation of the topological  vertex \cite{topvert}. 
In fact formula \eqref{eq:largeNunknotA} follows rigorously 
using more recent results in the mathematical literature
 \cite{proofMV,unknotHodge,two-part} on one and two-partition Hodge integrals. Details will be omitted because this is a standard 
 virtual localization computation.  

 The important fact for the present goal is to note that 
 equations \eqref{eq:openGWD}, \eqref{eq:largeNunknotA} yield an identity of the form 
\be\label{eq:torusknotsA} 
F_1^{(s,r)}(g_s,q,Q) = (-1)^{s-1}\sum_{R} \chi_R(C_{{(s)}})
e^{i(r/s)\kappa_R g_s/2}  W_R^{(1,0)}(q,Q).
\ee
Now recall that according to \cite[Sect 3.3]{toruslinks}, 
\cite[Eqn (2.43)]{torusknotsmirror}, the HOMFLY 
polynomials of $(s,r)$ torus knots is expressed in terms of 
colored HOMFLY polynomials of the unknot as follows 
\be\label{eq:torusknotsB} 
W^{(q,p)}_{\Box} (q,Q)
= \sum_{R} \chi_R(\CC_{(s)}) e^{2\pi i (r/s) h_R} \mathrm{dim}_q(R). 
\ee 
Next note that 
\[
h_R = {N|R|\over 2(k+N)} + {\kappa_R \over 2(k+N)}
\]
which implies 
\[ 
e^{2\pi i (r/s) h_R} = e^{i(r/s) \lambda |R|/2} e^{i(r/s)\kappa_R g_s}.
\] 
Since only diagrams $R$ with $q$ boxes contribute to the right hand 
side of \eqref{eq:torusknotsA}, \eqref{eq:torusknotsB}, it follows that 
\[
F_1^{(s,r)}(g_s,q,Q) = q^{-r/2} (-1)^{s-1}W^{(s,r)}_{\Box} (q,Q) \mathrm{Tr}(V).
\]
This is the expected large $N$ duality prediction for torus 
knots. The factor $(-1)^{s-1}$ reflects a specific choice of 
orientation of the moduli space of stable maps with lagrangian 
boundary conditions, as explained above. 
 
 \bibliography{newref.bib}
 \bibliographystyle{abbrv}
\end{document}